\documentclass[aps,prd,amsmath,floats,floatfix,
twocolumn, superscriptaddress,nofootinbib,showpacs]{revtex4-1}
\usepackage[colorlinks, pdfborder={0 0 0}, plainpages=false]{hyperref}
\usepackage{euscript,graphicx}
\usepackage{amsthm,dsfont,amsfonts,amsmath,amssymb}
\usepackage{euscript,color,fontenc,textcomp,relsize}
\usepackage{bm,url,float,cleveref}
\usepackage[caption=false]{subfig}
\usepackage{lipsum}
\usepackage{array}
\usepackage{setspace}
\usepackage[normalem]{ulem}
\usepackage{xcolor}
\newcolumntype{P}[1]{>{\centering\arraybackslash}p{#1}}
\newcommand{\appropto}{\mathrel{\vcenter{
  \offinterlineskip\halign{\hfil$##$\cr
    \propto\cr\noalign{\kern2pt}\sim\cr\noalign{\kern-2pt}}}}}
\newcommand{\vek}[1]{\boldsymbol{#1}}

\newcommand{\IISERP}{\affiliation{Indian Institute of Science Education \& Research, Dr. Homi Bhabha Road, Pashan, Pune - 411008, India}}%
\newcommand{\Syracuse}{\affiliation{Department of Physics, Syracuse University, Syracuse NY 13244, USA}}%
\newcommand{\IUCAA}{\affiliation{Inter University Centre for Astronomy \& Astrophysics, Ganeshkhind, Pune - 411007, India}}%
\newcommand{\CITA}{\affiliation{Canadian Institute for Theoretical Astrophysics, University of Toronto, Toronto, ON M5S 3H8, Canada}}%
\newcommand{\SNU}{\affiliation{Seoul National University, Seoul 151-742, Korea}}
\newcommand{\CORNELL}{\affiliation{Cornell Center for Astrophysics and Planetary Science, Cornell University, Ithaca, New York 14853, USA}}
\newcommand{\CALTECH}{\affiliation{Theoretical Astrophysics 350-17, California Institute of Technology, Pasadena, CA 91125, USA}}
\newcommand{\CIAR}{\affiliation{Canadian Institute for Advanced Research, 180 Dundas St. West, Toronto, ON M5G 1Z8, Canada}}
\newcommand{\JPL}{\affiliation{Jet Propulsion Laboratory, California Institute of Technology, 4800 Oak Grove Drive, Pasadena, CA 91109, USA}}
\newcommand{\AEI}{\affiliation{Max Planck Institute for Gravitational Physics (Albert Einstein Institute), Am M¨uhlenberg 1, 14476 Potsdam-Golm, Germany}}
\newcommand{\PENN}{\affiliation{Department of Physics, The Pennsylvania State University, University Park, PA 16802, USA}}
\newcommand{\CalState}{\affiliation{Gravitational Wave Physics and Astronomy Center, California State University Fullerton, Fullerton, California 92834, USA}}

\begin{document}

\date{\today}

\title{Detection and characterization of spin-orbit resonances in the advanced gravitational wave detectors era}

\author{Chaitanya Afle}\IISERP\Syracuse
\author{Anuradha Gupta}\PENN\IUCAA
\author{Bhooshan Gadre}\IUCAA
\author{Prayush Kumar}\CORNELL\CITA
\author{Nick Demos}\CalState
\author{Geoffrey Lovelace}\CalState
\author{Han Gil Choi}\SNU
\author{Hyung Mok Lee}\SNU
\author{Sanjit Mitra}\IUCAA
\author{Michael Boyle}\CORNELL
\author{Daniel A. Hemberger}\CALTECH
\author{Lawrence E. Kidder}\CORNELL
\author{Harald P. Pfeiffer}\CITA\CIAR\AEI
\author{Mark A. Scheel}\CALTECH
\author{Bela Szilagyi}\CALTECH\JPL

\begin{abstract}
Spin-orbit resonances have important astrophysical implications as the evolution and subsequent coalescence of supermassive black hole 
binaries in one of these configurations may lead to low recoil velocity of merger remnants. It has also been shown that black hole spins in comparable mass stellar-mass black hole binaries could preferentially lie in a resonant plane when their gravitational waves (GWs) enter the
advanced LIGO frequency band~\cite{Gerosa13}. Therefore, it is highly desirable to investigate the possibility of  detection and subsequent characterization of such GW sources in the advanced detector era, which can, in turn, improve our perception of their high mass counterparts. The current detection pipelines involve only non-precessing templates for compact binary searches whereas parameter estimation pipelines can afford to use approximate precessing templates.  
In this paper, we test the performance of these templates in detection and characterization of spin-orbit resonant binaries. We use fully precessing time-domain SEOBNRv3 waveforms as well as {\it four} numerical relativity (NR) waveforms to model GWs from spin-orbit resonant binaries and filter them through IMRPhenomD, SEOBNRv4 and IMRPhenomPv2 approximants. We find that the non-precessing approximants IMRPhenomD and SEOBNRv4 recover only $\sim70\%$ of injections with fitting factor (FF) higher than 0.97 (or 90\% of injections with ${\rm FF} >0.9$). This loss in signal-to-noise ratio is mainly due to the missing physics in these approximants in terms of precession and non-quadrupole modes. However, if we use a new statistic, i.e., maximizing the matched filter output over the sky-location parameters as well, the precessing approximant IMRPhenomPv2 performs magnificently better than their non-precessing counterparts with recovering $99\%$ of the injections with FFs higher than 0.97. Interestingly, injections with $\Delta \phi = 180^{\circ}$ have higher FFs ($\Delta \phi$ is the angle between the components of the black hole spins in the plane orthogonal to the orbital angular momentum) as compared to their $\Delta \phi =0^{\circ}$ and generic counterparts. This is because $\Delta \phi=180^{\circ}$ binaries are not as strongly precessing as $\Delta \phi =0^{\circ}$ and generic binaries. This implies that we will have a slight observation bias towards $\Delta \phi=180^{\circ}$ and away from $\Delta \phi=0^{\circ}$ resonant binaries while using non-precessing templates for searches.
Moreover, all template approximants are able to recover most of the injected NR waveforms with FFs $>0.95$. For all the injections including NR, the systematic error in estimating chirp mass remains below $<10\%$ with minimum error for $\Delta \phi = 180^{\circ}$ resonant binaries. The symmetric mass ratio can be estimated with errors below $15\%$. The effective spin parameter $\chi_{\rm eff}$ is measured with maximum absolute error of 0.13.
The in-plane spin parameter $\chi_p$ is mostly underestimated indicating that a precessing signal will be recovered as a relatively less precessing signal.
Based on our findings, we conclude that we not only need improvements in waveform models towards precession and non-quadrupole modes but also better search strategies for precessing GW signals.
\end{abstract}


\maketitle

\section{Introduction}
The advanced LIGO (aLIGO) detectors have successfully 
completed their second observing run (O2) and have detected GWs from five binary black holes (BBHs) so far \cite{detection, GW151226, GW170104, GW170608, GW170814}. The advanced Virgo detector \cite{virgo2} joined aLIGO during the end of O2 and the first ever three-detector observation of GWs was made from GW170814 \cite{GW170814}. Not to mention that the first ever detection of binary neutron star merger, GW170817 \cite{GW170817}, has opened the window of a long-awaited multi-messenger astronomy through which we are now able to ``hear'' and ``see'' the sources together.
In the coming years, KAGRA \cite{KS11} and LIGO-India \cite{LIGO_I_UNNI} will join aLIGO and Virgo detectors and help us reveal the exciting physics and astrophysics  of GW sources.

In practice, the associated weak GW signals from binary coalescence, buried in the noisy interferometric data, are extracted by employing the {\it matched filtering} technique \cite{SathyaDhurandhar1991,SathyaDhurandhar1994}. This is an optimal technique if and only if one can construct search templates that accurately model the GW signals. However, the general relativity (GR) based modeling of a binary coalescence is tricky as it happens in three physically distinct phases, namely inspiral, merger and ringdown. One requires different analytical and/or numerical schemes to describe each of these phases. For example, the post-Newtonian (PN) approximation to GR accurately
models the inspiral part whereas numerical relativity (NR) simulations are employed to model the other two phases \cite{NR}. That is why PN and NR waveforms are stitched together to construct the full inspiral-merger-ringdown (IMR) waveforms for coalescing binaries \cite{Ajith2007}.

The black holes (BHs) in a binary system are expected to carry some intrinsic spin which adds many interesting features
to the emitted GWs. In GR, the spin angular momentum of a compact object is defined as $\vek S=G\,m^2\,\chi\, \vek s/c$, where $\chi$ is the dimensionless Kerr
parameter, $m$ is BH mass and $\vek s$ denotes the direction of $\vek S$. The BHs can have their Kerr parameter in the range $[0, 1]$ ($\chi=0$ means non-spinning while $\chi=1$ means the maximally spinning BH). Interestingly, when the BH spins are not aligned (or anti-aligned) to the orbital angular momentum, the orbital plane as 
well as the spin vectors precess around the total angular momentum of the system. This precession of spins and orbital plane introduces modulations into the GW amplitude and phase. Therefore, it is important to include all possible spin and precession effects while modeling GWs from such systems in order to characterize them uniquely. A precessing BBH on quasi-circular inspiral is fully characterized by 15 parameters. These include 8 intrinsic ones: masses $m_1$, $m_2$; magnitude of the spins $|\vek S_1|=G\,m_1^2\,\chi_1 \, |\vek s_1|/c$ and $|\vek S_2|=G\,m_2^2\,\chi_2 \, |\vek s_2|/c$, and angles to define their orientation $\theta_1$, $\phi_1$, $\theta_2$, $\phi_2$ in the orbital triad ($\vek s_{1,2}$ are the unit vector along $\vek S_{1,2}$). The other 7 parameters are extrinsic and include the luminosity distance $D_L$, right ascension $\alpha$, declination $\delta$, orbital inclination $\iota$, polarization angle $\psi$, time of 
arrival $t_0$ and initial phase $\Phi_0$.
While we could not constrain the spins of either BHs in GW150914, GW170104, GW170608 and GW170814, one of the components in GW151226 certainly had non-zero spin \cite{GW151226, O1BBH}.
However, none of the GW events showed any clear sign of precession and 
GW community is hopeful to detect precessing GW signal in the coming years. 

Apart from being the most general and interesting case of binary coalescence, precessing BBHs have additional astrophysical implications as well. One of them is the spin-orbit resonances pointed out by Schnittman \cite{JS}. 
The binaries in such configurations  have their two spin vectors and the orbital
angular momentum vector lying in the same plane - {\it the resonant plane}. Schnittman showed that the spin-orbit resonances are more prominent in comparable mass highly spinning binaries and later on
Gerosa et al. \cite{Gerosa13} demonstrated that unless the tides are in-efficient the BH spins in comparable mass stellar mass BBHs would  preferentially lie in a resonant plane when their GWs enter the aLIGO frequency window. 
Refs.~\cite{KSB09,BKS12} showed that the spins of comparable mass supermassive BBHs can get aligned prior the merger due to the spin precession if the spin of primary BH was initially partially aligned with the orbital angular of the system. This alignment of spins can significantly reduce the recoil velocity of the final BH formed from the merger. Moreover, Schnittman found that the partially aligned spins (with orbital angular momentum) during the inspiral of the binary are strongly influenced by the presence of spin-orbit resonances. This implicates that spin-orbit resonances have important effects on determining the amount of recoil the final BH will experience after the merger of BBH.   
Therefore, it is important to explore the possibility of detection and characterization of such stellar mass sources using aLIGO detectors which can, in turn, improve our understanding of their high mass counterparts. Previous studies of precessing (generic) binaries have been focused at neutron star - black hole systems~\cite{Tito, Ian}. For those systems, Ref.~\cite{Tito} has shown that an aligned-spin search pipeline will be able to detect at least $50\%$ of the precessing neutron star - black hole binaries. In this paper we will instead focus on the a subset of the BBH population that has entered the spin-orbit resonance configuration.

On the waveform modeling side, most of the effort has been put into the modeling of GWs from non-spinning binaries and binaries which do not precess; and very less on 
the precessing binaries. That is why we have several phenomenological IMR waveform models for non-spinning (IMRPhenomA \cite{Ajith2007}) 
and  non-precessing binaries (IMRPhenomB, IMRPhenomC, IMRPhenomD \cite{Ajith_prl, Santamaria:2010yb, Khan:2015jqa, Husa:2015iqa}) 
but only one model for spinning, precessing binaries (IMRPhenomP \cite{Hannam2014}).
All these models are based on PN description of the early inspiral.
Additionally, there is an effective-one-body (EOB) approach to tackle the inspiral phase of BBH coalescence. Based on EOB approach, we have
other IMR models: SEOBNRv2 \cite{Taracchini:2013rva}, SEOBNRv4 \cite{SEOBNRv4} and SEOBNRv3 \cite{Pan2013}. The first two models are for non-precessing systems while the third one additionally accounts for spin-precession. Detection searches on O1 data used SEOBNRv2 waveform templates 
\cite{O1data} while detailed
parameter estimation of detected sources was performed with IMRPhenomP, SEOBNRv2
and SEOBNRv3 \cite{GW150914props, PE_precessing, O1BBH, SEOBNRv3_PE}.
Note that these are the currently available waveform models we have so far to detect and characterize GW signals from BBHs.

In this paper, we investigate the performance of currently available state-of-the-art waveform models in detection and characterization of GW signals from BBH in spin-orbit resonant configuration. We employ two non-precessing models, namely IMRPhenomD and SEOBNRv4, and one precessing model IMRPhenomPv2 as our templates. The time-domain IMR waveform approximant SEOBNRv3 is used to model GW from fully precessing resonant binaries.\footnote{In this paper, we term binaries in spin-orbit resonances as resonant binaries.} As Schnittman \cite{JS} demonstrated that the spin-orbit resonances are effective only in comparable mass highly spinning binaries, we restrict the mass ratio $q=m_1/m_2$ and BH spin magnitudes $\chi_{1,2}$ of our injections in the range $[1,3]$ and $[0.5, 1]$, respectively. We impose the resonance requirement on binary spin vectors and restricting the angle between their azimuthal projections onto the orbital plane $\Delta\phi$ to be either $0^{\circ}$ or $180^{\circ}$ (see Sec.~\ref{Sec_SOR} for detail). For comparison, we also consider generic precessing binaries which may or may not satisfy resonant condition(s). In addition, we also perform full numerical relativity (NR) simulations of three spin-orbit resonant binaries
using the Spectral Einstein Code \cite{sxs}, as well as one using a code developed by
the Korean Gravitational-Wave group (KGWG). We use these NR waveforms to study
the approximants considered in a model-independent way.

We compute fitting factor (FF) for both types of injections (SEOBNRv3 and NR) using the above mentioned template approximants and results are summarized in Figs.~\ref{template_comparison_fig_v3} to \ref{FF_NR}. 
We find that the  non-precessing template approximants, IMRPhenomD and SEOBNRv4, recover $\sim70\%$ of injections (all three kinds, $\Delta \phi=0^{\circ}$, $\Delta \phi=180^{\circ}$ and generic) with FF$>0.97$. Moreover, binary injections with $\Delta \phi = 180^{\circ}$ are recovered with higher FFs as compared to $\Delta \phi = 0^{\circ}$ and generic cases. Binaries that are close to edge-on have somewhat
lower FFs but the overall trend persists for all orbital inclinations considered. Note that the SEOBNRv3 injections have orbital inclination such that $\cos \iota \in [-1, 1]$.
This relative loss in signal-to-noise ratio (SNR) for binaries that are neither face-on nor face-off is
primarily due to the exclusion of higher harmonics in the template approximants. This is because our signal
approximant (SEOBNRv3) contains $(l, m)=(2, \pm2), (2, \pm1)$ GW modes whereas the template approximants IMRPhenomD and SEOBNRv4 only include $l = |m| = 2$ modes in the co-precessing frame. This causes additional mismatch between the signal and template waveforms and hence low FF values. Even though the effect of inclination angle (or non-availability of non-quadrupole modes) is dominant in FF loss, we find that our template waveforms also fall short in capturing full precessional effects present in the signal even if we restrict ourselves in the comparable mass regime with $q\in[1, 3]$. The non-precessing template approximants yield  low FFs for injections which have high in-plane spin parameter $\chi_{\rm p}$ and negative effective spin parameter $\chi_{\rm eff}$. Notice that large $\chi_{\rm p}$ and negative $\chi_{\rm eff}$ means that the binary is strongly precessing. 
This is the reason behind $\Delta \phi = 0^{\circ}$ resonant binaries having low FFs as compared to $\Delta \phi = 180^{\circ}$ resonant and generic binaries (see the discussion in Sec.~\ref{sec:SEOBNRv3results}).
This implies that the non-precessing IMR waveform models need to be calibrated with NR simulations that cover more asymmetric masses, negative effective spins and high in-plane spin components.  The precessing approximant IMRPhenomPv2, on the other hand, incorporates precession effects (in an effective way) as well as GW modes $(l,m)=(2,\pm2), (2,\pm1)$ in the detector frame. Therefore, to fully exploit its properties, we use a new statistic, proposed in Ref.~\cite{Harryetal}, to compute FFs for IMRPhenomPv2 templates. We compute the so called ``sky-maxed'' FF while maximizing the overlap not only over the intrinsic parameters but also over the sky-location dependent parameters such as $\iota$, $\delta$, $\alpha$ and $\psi$ along with $t_0$, $\Phi_0$ and $D_L$. In Figs.~\ref{template_comparison_fig_v3} to \ref{FF_NR}, we see that IMRPhenomPv2 performs magnificently better than IMRPhenomD and SEOBNRv4, recovering almost all the injections ($\sim 99\%$) with FF$>0.97$. 
Therefore, we recommend to use IMRPhenomPv2 templates and sky-maxed FF statistic for precessing BBH searches.
Finally, we note that our results for NR injections are consistent with those obtained for SEOBNRv3 injections, which is what one expects given the demonstrated reliability of SEOBNRv3 model for near equal-mass binaries~\cite{Babaketal}.

Next, we compute systematic biases in the recovery of resonant binary parameters. Note that we are not quoting statistical errors in this paper which depend on source's SNR and mostly dominate the systematic errors for the current LIGO/Virgo GW events. We find that, as expected, IMRPhenomPv2 performs better in recovering mass parameters, chirp mass $M_c=(m_1\, m_2)^{3/5}/(m_1+m_2)^{1/5}$ and symmetric mass ratio $\eta=m_1\,m_2/(m_1+m_2)^2$, as compared to IMRPhenomD and SEOBNRv4 approximants. For all the three template aproximants the error in $M_c$ recovery remains below $2\%$ for low mass binaries ($M_c<25M_{\odot}$) while increasing for high mass injections for which the same can be as large as $10\%$. On the other hand, we always recover $\eta$ smaller than the injected value with a maximum systematic bias of $15\%$. Similarly, all approximants always recover spin magnitude of heavier BH lower than the injected value with IMRPhenomPv2 yielding smaller errors as compared to the other two approximants. The aligned spin parameters $\chi_{\rm eff}$ and $\hat{\chi}$ (defined in Eq.~(\ref{eq_chi_cap})) are always recovered with relatively positive values unless the injected $\chi_{\rm eff}/\hat{\chi}$ value is close to $+1$, i.e., very high injected positive spins lead to underestimation of $\chi_{\rm eff}$ and $\hat{\chi}$. The overall absolute errors in these spin parameters is $<0.13$. Moreover, the  in-plane spin parameter $\chi_p$ is mostly underestimated and the same can have absolute error as large as $0.5$. Underestimation of $\chi_{\rm p}$ implies that we have bias towards recovering a precessing GW signal as a relatively less precessing signal.
Based on our findings in this paper, we conclude that we are 
presently equipped to detect a good fraction of resonant binaries with non-precessing templates but need precessing models and better detection statistic if we do not want to miss any of them. On the other hand, the characterization of resonant binaries for astrophysical studies requires better precessing waveform models.

Similar studies, i.e., the comparison of different waveform approximants, have been done by several other authors in the past \cite{Prayush2016, Verma, NRsystematics} though most of them were limited to non-precessing binaries. For instance, using a set of 84 non-precessing numerical simulations with mass ratio $1 \leq q \leq 3$ and $\chi_{1, 2}$ up to $0.9$, 
Kumar et al.~\cite{Prayush2016} studied the accuracy of IMRPhenom\{C, D\} and SEOBNRv\{1,2\} approximants. In both the cases they found that the more recent IMRPhenomD and SEOBNRv2 models perform very well at modelling comparable mass binaries. On the other hand, Verma et al. \cite{Verma} studied the effect of non-quadrupole modes, in non-precessing systems with $q\in[1, 10]$ and $\chi_{1,2}$ up to $0.98$, and showed that the sub-dominant modes are important for the detection of aligned-spinning and parameter estimation of anti-aligned-spinning binaries. In another study, Ref.~\cite{NRsystematics}
investigated the effect of waveform model systematics on the characterization of the first GW signal, GW150914. They perform a full Bayesian analysis on mock GW signals from NR simulations with physical parameter similar to that of GW150914 while employing SEOBNRv2, SEOBNRv3 and IMRPhenomPv2 approximants as recovery template. It has been found that all the three approximants give results consistent with the original Bayesian analysis using original GW data \cite{GW150914props,PE_precessing,O1BBH}. In this paper, we build upon these past studies through the following improvements: (a) we neither ignore nor approximate spin-precession effects in our signals, (b) we consider an astrophysically interesting but dynamically distinct class of BBHs, one that has not been studied with the recent improvements in waveform modeling technology.

The paper is organized as follows. In Sec.~\ref{Sec_II}, we briefly review the spin-orbit resonances, their astrophysical implications and GW modeling. 
Section~\ref{Sec_III} gives a brief summary of various waveform models that are currently available in the detection and parameter estimation pipelines. Our results on the detection and characterization of spin-orbit resonances are presented in Sec.~\ref{Sec_IV} and Sec.~\ref{Sec_V}, respectively. Section~\ref{sec:conclusions} summarizes the main conclusions and future work.

\section{Spin-orbit resonances}
\label{Sec_II}
In this section, we describe  Schnittman's spin-orbit resonances and the
equations that govern the dynamics of binaries in such configurations. Later on, we discuss the astrophysical implications 
and GW modeling of resonant binaries.

\subsection{Equilibrium configurations for spin-orbit resonances}
\label{Sec_SOR}
The comparable mass spinning and precessing binaries can reside in an equilibrium configuration 
in which the two spins $\vek S_1$, $\vek S_2$ and orbital angular momentum $\vek L$ remain in a common plane - resonant plane - throughout their inspiral phase. This implies that, by definition, the total angular momentum of the system 
$\vek J=\vek S_1 + \vek S_2 + \vek L$ will also lie in this resonant plane.
In the absence of any GW damping, $\vek S_1$, $\vek S_2$ and $\vek L$ will precess around $\vek J$ at a constant
frequency,  keeping their relative orientations fixed. This is why Schnittman coined such equilibrium states as {\it spin-orbit resonant configuration}
as the precessional frequencies of $\vek S_1$, $\vek S_2$ and $\vek L$  around $\vek J$ are rather same. However, when radiation reaction effects are switched on, the three angular momenta remain co-planer with their relative orientations slowly varying on the radiation reaction time scale. Interestingly, binaries that are not initially in the vicinity of these equilibrium configurations may eventually get captured and librate about them during their inspiral.

In practice, $\vek s_1$ and $\vek s_2$ are freely specified by four angles $(\theta_1, \phi_1, \theta_2, \phi_2)$ in a non-inertial triad whose $\vek z$-axis is along $\vek L$: 
 \begin{subequations}
\label{eq:s1_s2}
\begin{align}
\vek s_1 &= \left ( \sin \theta_1\,\cos \phi_1, \sin \theta_1\,\sin \phi_1, \cos \theta_1 \right )\,,\\
\vek s_2 &= \left ( \sin \theta_2\,\cos \phi_2, \sin \theta_2\,\sin \phi_2, \cos \theta_2 \right )\,,
\end{align}
\end{subequations}
and $\vek k$ - unit vector along $\vek L$ - can be defined as $(0,0,1)$. This means that the orientation of a binary system, characterized by $m_1, m_2, \chi_1, \chi_2$ and $x = (G\,m\, \pi \, f/c^3)^{2/3}$, is described by only 3 angular parameters $(\theta_1, \theta_2, \Delta \phi)$,
where $f$ is GW frequency. Note that all these angles vary on the precessional and radiation reaction time-scales. 

Schnittman's equilibrium configurations are obtained by the requirement that $\vek S_1$, $\vek S_2$ and $\vek L$ remain co-planar throughout 
their evolution. Equivalently,
 \begin{subequations}
\begin{align}
\label{eq:ls1s2_tp}
\vek L \cdot (\vek S_1 \times \vek S_2) = 0\,,\\
\label{eq:ls1s2_tp_d}
\frac{d}{dt}(\vek L \cdot (\vek S_1 \times \vek S_2)) = 0\,.
\end{align}
\end{subequations}
Equation~(\ref{eq:ls1s2_tp}) can be reduced to 
\begin{equation}
\label{eq:SOR_angle}
\sin \theta_1 \, \sin \theta_2\, \sin\Delta \phi = 0 \,,
\end{equation}
which implies that $\Delta \phi =0^{\circ}$ or $\pm 180^{\circ}$, leading to two families of spin-orbit resonances. 
Further, Eq.~(\ref{eq:ls1s2_tp_d}) reduces to
\begin{equation}
\label{eq:SOR_cond}
(\vek \Omega_1 \times \vek S_1)\cdot[\vek S_2 \times(\vek L+\vek S_1)] = (\vek \Omega_2 \times \vek S_2)\cdot [\vek S_1 \times(\vek L +\vek S_2)]\,,
\end{equation}
where we have used precession equations for $\vek S_1$ and $\vek S_2$ while $\vek \Omega_{1,2}$ provide precession frequency of $\vek S_{1,2}$. We employ PN accurate expression for $\Omega_{1,2}$ which can be given, for example, by Eqs.~(2.2) in Ref.~\cite{KSB10}. Equations~(\ref{eq:SOR_angle}) and (\ref{eq:SOR_cond})
together allow us to solve for $\theta_2$ given $\theta_1, \Delta \phi, m_1, m_2, \chi_1, \chi_2$ and $x$. As $x$ changes during the evolution (because of the radiation reaction) of binary, the one-parameter family of spin-orbit resonances sweeps
out a significant portion of $\theta_1-\theta_2$ plane (see, e.g., Figs.~3 \& 4 in Ref.~{\cite{JS}} and Figs.~1 \& 2 in Ref.~\cite{KSB10}). 
If ($\theta_1, \theta_2$) values of a generic precessing binary  during its evolution happen to lie in the neighborhood of any of the solutions,
it will be strongly captured by the nearest equilibrium configuration. This has interesting astrophysical implications, some of which we discuss in the
following section.

\subsection{Astrophysical implications of spin-orbit resonances}

The spin-orbit resonances have important astrophysical implications (see, e.g., Refs.~\cite{JS,KSB09,BKS12, Gerosa13} for detail).
This is because these resonances have the ability to align (anti-align) spins 
of comparable mass supermassive BH binaries prior to their merger \cite{KSB09,BKS12}.
The alignment of spins ensures that BHs formed via BH binary coalescences do not experience large recoil velocities and will be retained in their host galaxies.
Using a toy model for BBH formation, Ref.~\cite{Gerosa13} argued that the BH spins in comparable mass stellar mass BH binaries would preferentially lie in a resonant plane due
the spin-orbit resonances when their GWs enter the aLIGO frequency window \cite{Gerosa13}. This model suggests that binaries belonging to two resonant families can be associated with two different binary formation channels, both of which involve efficient tides. For example, binaries in $\Delta \phi = 0^{\circ}$ configuration are offsprings of the {\it reverse mass ratio} formation channel in which the heavier BH is formed during the second supernova explosion. However, binaries in $\Delta \phi = 180^{\circ}$ configuration are probably formed in {\it standard mass ratio} scenario where the more massive star will evolve to form the more massive component of the BBH.

Refs.~\cite{GG2014, Gerosa2014, Trifiro2016} extensively studied the dynamics of compact binaries in Schnittman's equilibrium configuration and proposed ways to distinguish the two resonant families that will help us constrain the binary formation channels. Ref.~\cite{GG2014} explored the dynamics of comparable mass binaries influenced by spin-orbit resonances in an inertial frame associated with the initial direction of $\vek J$. 
The authors argued that the accurate measurement of the projections of $\vek S_1$, $\vek S_2$ and $\vek L$ along $\vek J$ (at a reference frequency) facilitates the classification of sources between the two spin-orbit resonant families. 
Ref.~\cite{Gerosa2014} computed overlap between waveforms corresponding to the two resonant families and showed that they both exhibit qualitatively and quantitatively different features in their GW emission. Due to these distinctions, binaries (with SNR $\gtrsim10$) belonging to either of the resonant families can be distinguished even if 
the precession induced modulations are minimal (in the case when the line of sight is along the direction of $\vek J$). Ref.~\cite{Trifiro2016} performed the full parameter estimation on resonant binaries using {\scshape lalinference} package of 
{\scshape lalsuite} \cite{lalsuite}. In their analysis, the (conserved) projected effective spin parameter $\xi=(\chi_1\, \cos \theta_1 + q\, \chi_2\, \cos \theta_2)/(1+q)$, angle 
between the line of sight and $\vek J$, and the signal amplitude were varied while keeping the masses and spin magnitudes fixed. It was shown that the two equilibrium configurations can be distinguished for a wide range of binaries 
if the binary is not in a finely tuned highly symmetric configuration. Moreover, Refs.~\cite{Kesdenetal2015, Gerosaetal2015} derive an effective potential to describe the dynamics of such BBHs. Using this effective potential, Refs.~\cite{Kesdenetal2015, Gerosaetal2015} classified BBHs' precession into three spin morphologies: `$\Delta \phi$ liberating around $0^{\circ}$', `$\Delta \phi$ liberating around $180^{\circ}$' and `$\Delta \phi$ circulating between $0^{\circ}$ and $180^{\circ}$'. BBHs during their inspiral may transition from one morphology to other.

Because of the interesting features that spin-orbit resonances offer, it is worthwhile to investigate the possibility of detection and characterization of these binaries. The next section discusses waveform models that can be used to model GWs from resonant binaries.

\subsection{GW models for spin-orbit resonant binaries}
We assume that comparable mass compact binaries which experience spin-orbit resonances would have circularized before entering the aLIGO frequency window.
To simulate GWs emitted by such binaries, we employ the SEOBNRv3 approximant \cite{Pan2014} implemented in the {\scshape lalsimulation} package of {\scshape lalsuite} \cite{lalsuite} and perform NR simulations of {\it four} binaries in resonant configurations. The details of both types of simulated signals are given below.

\begin{table*}[t]
\caption{Injection parameters for SEOBNRv3 approximant used to model GWs from spin-orbit resonances. Note that all the angular parameters (cos$\theta_1$, $\phi_1$,  $\delta$, $\alpha$, $\iota$, $\psi$) are uniformly sampled in their respective ranges while $D_L$ is uniform in volume with radius between 1 to 100 Mpc. Also, all these parameters including spins are defined at a reference frequency of $20$ Hz.}
\begin{center}
\begin{tabular*}{\textwidth}
{m{0.20\columnwidth}m{0.20\columnwidth}m{0.1\columnwidth}m{0.20\columnwidth}m{0.2\columnwidth}m{0.2\columnwidth}m{0.2\columnwidth}m{0.2\columnwidth}m{0.2\columnwidth}m{0.20\columnwidth}m{0.2\columnwidth}}
\toprule
$\chi_1, \chi_2$ & $M (M_{\odot})$ & $q$ & $\theta_1$ & $\phi_1$ & $\delta$ & $\alpha$ & $\iota$ & $\psi$  & $D_L$(Mpc) \\ \hline \\
$[0.5,\,0.99]$ & [$6,\,100$] & [$1,\,3$] & [$0,\,\pi$] & [$0,\,2\pi$] & [$0,\,\pi$] & [$0,\,2\pi$] & [$0,\,\pi$] & [$0,\,2\pi$] & [1, 100] \\
\hline \hline
\end{tabular*}
\end{center}
\label{Tab_injections}
\end{table*}

\subsubsection{SEOBNRv3}
Pan et al. \cite{Pan2013} 
developed an inspiral-merger-ringdown waveform model - also known as SEOBNRv3 - to model GWs from precessing BBHs using the effective-one-body (EOB) approach. SEOBNRv3 is built upon its non-precessing version SEOBNRv2  \cite{Taracchini:2012, Taracchini:2013rva}. We discuss the EOB approach and SEOBNRv2 waveforms in more detail in Sec.~\ref{Sec_SEOBNRv2}. 
The SEOBNRv3 model employs the precessing convention introduced by Buonanno, Chen, and Vallisneri \cite{BCV2003} and uses a non-inertial {\it precessing source frame} 
to describe the dynamics of the system. The $\vek z$-axis of the precessing source frame is along the Newtonian orbital angular momentum $\vek L_{\rm N}$ while $\vek x$ and $\vek y$ axes follow the minimum rotation prescription given in Refs.~\cite{BYPV2004, BOP}.
In the precessing source frame, the precession-induced modulations in phase and amplitude are minimized which help the waveforms to take a simple non-precessing form \cite{BYPV2004, BOP,OVHMS2011,SHHA2011,SHH2012}. In this precessing source frame the dominant $(l, m) = (2,2)$ mode alone is modeled, which upon
transformation back to an inertial frame gives (partial) sub-dominant $(l,m) = (2, \pm 1)$
harmonics. More recently, Ref.~\cite{Babaketal}
has improved the merger-ringdown prescription of SEOBNRv3. In this paper, we use this latter version of SEOBNRv3.

The computation of SEOBNRv3 waveforms involves solving a set of Hamilton's equations (see Eqs. (11) in Ref.~\cite{Pan2013}) as well as ordinary differential equations for $\vek s_1$, $\vek s_2$, $\vek k$ and $x$ (see Eqs. (8) and (9) in Ref. \cite{GG2014} and Eqs. (A1) and (A2) in Ref. \cite{GG2015}) and hence it is  computationally expensive.
Clearly, we require initial conditions for these variables to integrate the system of differential equations. In our analysis, we start our integration from the epoch at which the instantaneous GW frequency of the dominant $(l,m) = (2,2)$ mode is 20 Hz. We sample the values of $m_1, m_2, \chi_1, \chi_2$, $\theta_1$ and $\phi_1$ at 20 Hz from ranges given in Table \ref{Tab_injections}. The initial values of $\theta_2$ and $\phi_2$ are derived 
from Eq. (\ref{eq:SOR_cond}) and the relation $\phi_1-\phi_2 = 0^{\circ}, 180^{\circ}$ for two resonances, respectively. Ranges for the initial values of all other parameters are also listed in Table~\ref{Tab_injections}. 

\subsubsection{Numerical relativity simulations}
\label{Sec_NR_inj}
Numerical relativity waveforms are believed to be the best representation of true GW signals as they involve solving the full general relativistic binary problem in the dynamical and highly non-linear 
regime of the binary's coalescence. However, due to their high computational cost, it is not currently viable to cover the full range of binary parameters with numerical simulations and we have to make do with a selected set of parameter values. In this paper, we use {\it four} NR simulations of BBHs in
spin-orbit resonant configuration, whose physical parameters at their respective initial epochs are
listed in Table~\ref{tab:Injection_NR}. 

\begin{table*}[t]
\caption{Details of the NR waveforms used in this paper. Case 1, 3 and 4 correspond to $\Delta \phi = 0^{\circ}$ resonance while Case 2 corresponds to $\Delta \phi = 180^{\circ}$ resonance. The Cases 1, 2 and 3 are from SXS collaboration with simulation id SXS:BBH:0623, SXS:BBH:0624 and SXS:BBH:0622, respectively.
The values of $\theta_1$, $\theta_2$, $\phi_1$ and $\phi_2$ presented here are in LAL wave frame and at frequency $Mf_0$ after removing the junk radiation.}
\begin{center}
\begin{tabular*}{\textwidth}{@{\extracolsep{\fill}}c c c c c c c c c c }
\toprule \\
Case & $q$ & $\chi_1$ & $\chi_2$ & $\theta_1$ & $\phi_1$ & $\theta_2$ & $\phi_2$ & $N_{cyc}$ & $Mf_0$\\
\hline \\
1 & 1.11 & 0.9 & 0.9 & 82.5$^{\circ}$ & 83.5$^{\circ}$ & 97.9$^{\circ}$ & 82.8$^{\circ}$ & 75 & 0.0033 \\
2 & 1.2 & 0.85 & 0.85 & 44.3$^{\circ}$ & 196.0$^{\circ}$ & 31.0$^{\circ}$ & 10.5$^{\circ}$ & 38 & 0.0057\\
3 & 1.2 & 0.85 & 0.85 & 40.9$^{\circ}$ & 12.3$^{\circ}$ & 59.6$^{\circ}$ & 5.6$^{\circ}$ & 36 & 0.0058\\
4 & 1.2 & 0.85 & 0.85 & 45.2$^{\circ}$ & 256.4$^{\circ}$ & 64.54$^{\circ}$ & 252.7$^{\circ}$ & 21 & 0.0074\\
\hline \hline
\end{tabular*}
\end{center}
\label{tab:Injection_NR}
\end{table*}

\begin{figure}[ht!]
\begin{center}
\includegraphics[width=3.5in]{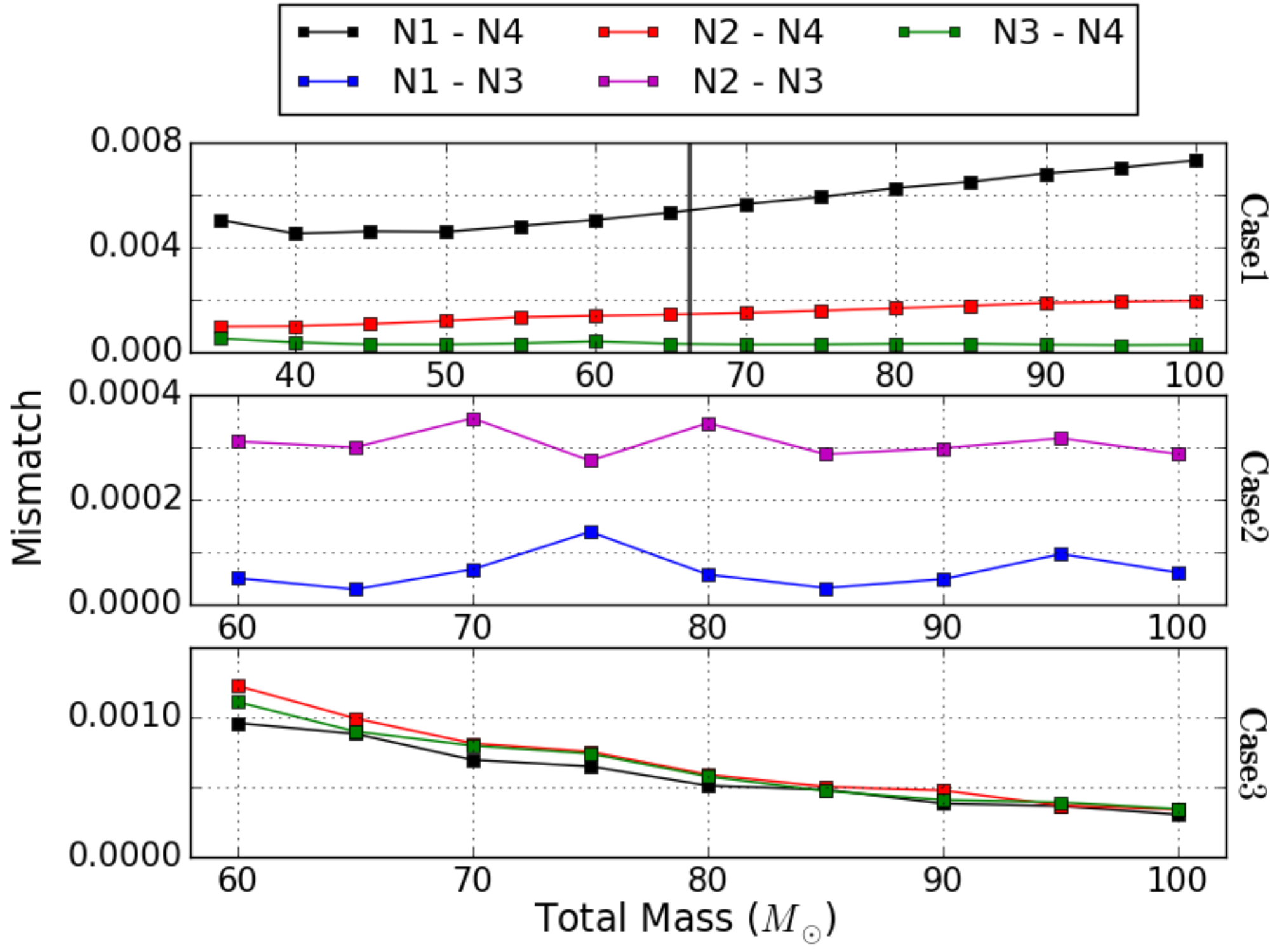}
\end{center}
 \caption{Mismatch between different resolutions (N1,2,3) and the highest available resolution of NR waveforms for Cases 1 (top), 2 (middle) and 3 (bottom), as a function of  the total mass of the binary (shown on the x-axis). 
 We include higher GW modes up to $l=5$ and choose inclination angle $\iota$  and reference phase $\Phi_{\rm ref}$ to be zero in these plots. The solid black line in the topmost panel represents the total mass when the lowermost frequency of the waveform is equal to 10 Hz. Such line for Case 2 and 3 lies beyond total mass of $100 M_{\odot}$. The masses in the three panels are chosen such that the NR waveform completely covers the aLIGO frequency band \cite{Chu:2015kft}.}
\label{NR_mismatch}
\end{figure}

The first {\it three} simulations were performed using the Spectral Einstein Code
(SpEC)~\cite{SpECwebsite}. Quasi-equilibrium initial data for
these configurations was constructed in the extended conformal
thin-sandwich formalism~\cite{Yo2004,Cook2004,
Pfeiffer2003} through the superposition of Kerr-Schild
metrics~\cite{Lovelace2008}. We restrict the initial orbital 
eccentricity below $10^{-3}$ using the iterative 
procedure of~\cite{Mroue:2012kv,Buonanno:2010yk,Buchman:2012dw}.
We evolve the binary on a multi-domain
computational grid that extends from the inner excision boundaries,
which are located slightly inside the apparent horizons,
to an outer spherical boundary with the radius of a few hundred
$M$~\cite{Szilagyi:2014fna}. We use a first-order representation
of the evolution equations~\cite{Friedrich1985,Garfinkle2002,
Pretorius2005c,Lindblom:2007} with a damped (generalized) harmonic
gauge condition~\cite{Szilagyi:2009qz}. The excision boundaries are
dynamically adjusted to track the shapes of the apparent
horizons~\cite{Scheel2009,Szilagyi:2009qz,Hemberger:2012jz} during
the evolution up until merger, at which point we transition to
a grid with one excision boundary~\cite{Scheel2009,Hemberger:2012jz}.
At the outer boundary, we impose constraint-preserving outgoing-wave
boundary conditions~\cite{Lindblom2006,Rinne2006,Rinne2007},
while the inter-domain boundary conditions are enforced with a
penalty method~\cite{Gottlieb2001,Hesthaven2000}. We adaptively
refine the evolution grid based on the truncation error of
evolved fields, local constraint violation magnitude, and the
truncation error of the apparent horizon
finders~\cite{Szilagyi:2014fna}. Finally, waveforms at asymptotic
null infinity are computed from those extracted at finite-radius
spheres using polynomial extrapolation~\cite{Boyle-Mroue:2008},
which has been shown to be sufficiently accurate for LIGO data
analyses~\cite{Taylor:2013zia,Chu:2015kft}.
Our SpEC simulations are performed at multiple resolutions.
They span $\sim 75$ GW cycles for Case 1 which corresponds to a
$q=1.11$ binary with $\chi_{1,2}=0.9$, and $36-38$ cycles
for Cases 2 and 3, both of which have $q=1.2$ and $\chi_{1,2}=0.85$, but belong to different resonant families.
In Fig.~\ref{NR_mismatch} we show the convergence of all three simulations. We find that different resolutions agree well with each other, with the numerical resolution error limited to causing mismatches smaller than $8\times 10^{-3}$.

The NR simulation for Case 4 was performed by KGWG group using Einstein Toolkit \cite{Loffler:2011}. The Einstein Toolkit is an open source code based on CACTUS \cite{Goodale:2002a} which is a modular code, with its modules often referred to as thorns. We use the Carpet \cite{Schnetter2004} thorn for mesh refinement and set the finest resolution \(M/128\) around the horizon with smaller mass covering its diameter with 25  grid points. Outer boundary  of our simulation is at \(120M\).  We use the TwoPunctures \cite{Ansorg2004} thorn to solve constraint equations for initial data and we determine the initial momentum parameters of quasi-circular orbit from the 2.5PN accurate expressions\cite{kidder,tichy}. 
Evolution of the binary was dealt with ML BSSN \cite{Brown2008} thorn. The GW signal was computed from $l = 2, m = 2$ mode (only) of Weyl Newman-Penrose scalar \(\Psi_4\) by fixed frequency-integration \cite{Reisswig2011}, where its extraction radius is \(110M\). The resulting simulation has 21 GW cycles with \(q = 1.2\) and \(\chi_{1,2}=0.85\). While \(q, \chi_{1,2}\) and relative angles between BH spins and orbital angular momentum of Case 4 are same as those of Case 3, Case 4 has much higher initial orbital eccentricity (0.03) and initial GW frequency (0.0074) compared to Case 3.

\subsubsection{NR waveforms in LAL wave frame}
In general, the waveform modelling for compact binaries involves two coordinate frames: {\it source frame} and {\it wave frame}. The source frame is 
convenient to define the properties of the GW sources. The $\vek z$-axis of this coordinate 
system is along the orbital angular momentum $\vek k$ at certain reference time 
while $\vek x$-axis is chosen to be along the unit separation vector $\vek n$, pointing from second 
body to the first one, at the same reference time. The $\vek y$-axis has the usual definition $\vek y=\vek z \times \vek x$. The wave frame depends on the position of observer, i.e., the detector. The $\vek z$-axis of this frame points along the line-of-sight while the $\vek x$ and $\vek y$ axes are orthogonal to it spanning the 
plane of the sky. In practice, we compute waveforms in wave frame using {\scshape lalsuite}. Therefore, we call this frame the LAL wave frame. 
There exists another frame - NR frame - in which the NR simulations are performed. This frame has its origin at the center of mass of the source (as in the source frame), but its coordinate chart is chosen to enhance the stability of our numerical solutions to Einstein equations and enable long-duration evolutions.

\begin{figure*}
\begin{center}
$\begin{array}{cc}
\includegraphics[width=7.0in]{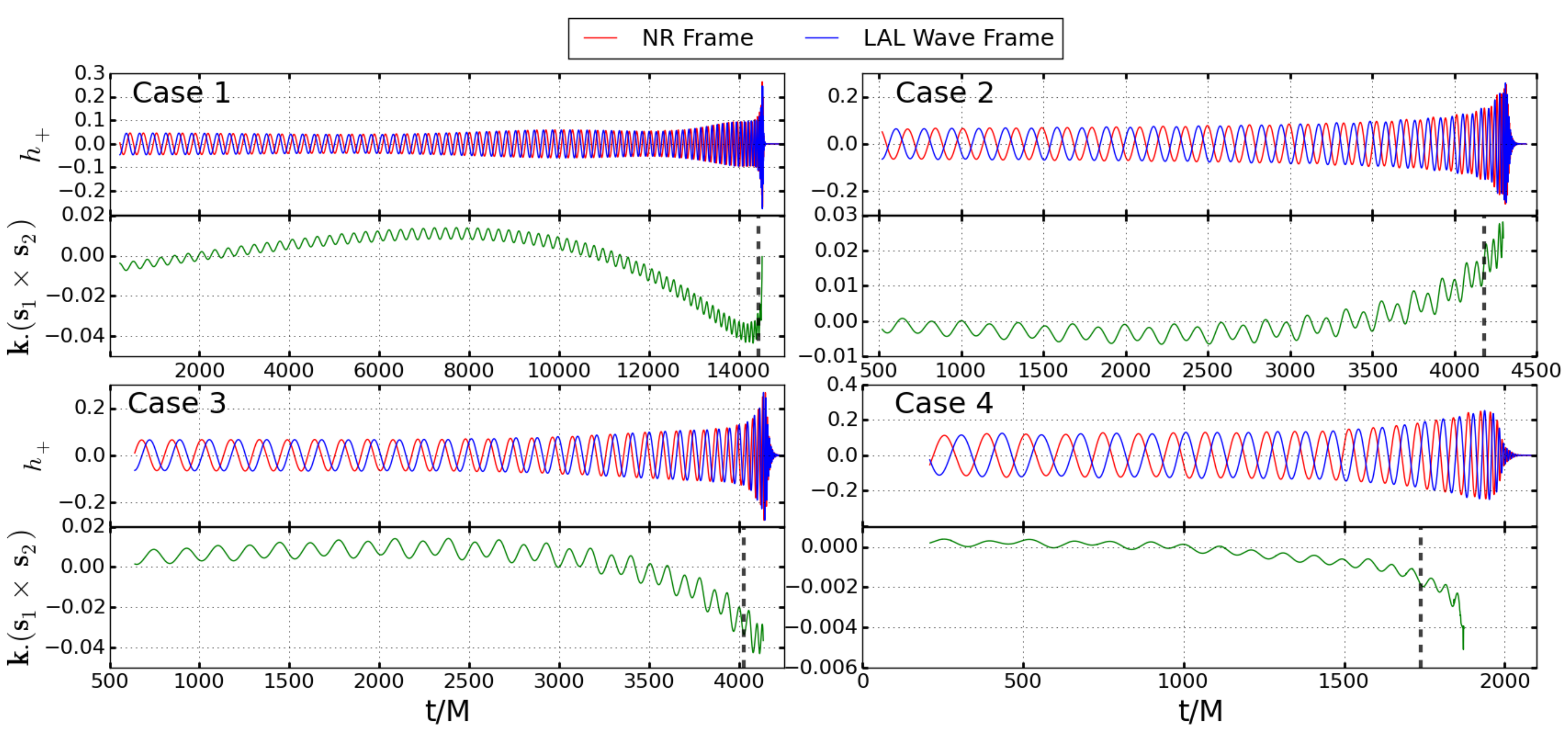}
\end{array}$
\end{center}
\caption{The figure shows the 4 cases of the NR waveforms with total mass of $20M_{\odot}$ with $\iota = 0^{\circ}$ and $\Phi_{\rm ref}=0^{\circ}$.
We have included higher GW harmonics up to $l=5$ in these waveforms except in Case 4 which has only $(2,2)$ mode. The waveforms are shown in the NR (source) frame as well as the LAL wave frame. The quantity $\vek k \cdot (\vek s_1 \times \vek s_2)$ is plotted beneath each of the waveforms. The vertical dashed line represents the time when the orbital separation reaches 6M.}
\label{Fig_NR_waveforms}
\end{figure*}

In order to compare various waveform models in the LAL wave frame, one must transform waveform multipoles from the NR frame to the LAL wave frame.
We do this 
transformation using the method introduced in Ref. \cite{NRInfrastructure}
(we refer the readers
to the same for exhaustive details on this transformation procedure). Even in the LAL wave frame, we have the freedom to specify different values of reference phase $\Phi_{\rm ref}$ (angle between line of ascending node and $\vek n$) and inclination angle $\iota$ (angle between $\vek L$ and the line-of-sight direction). 
In this paper, we fix the reference phase $\Phi_{\rm ref} = 0^{\circ}$ whereas the inclination angle $\iota$ takes values in $\{0^{\circ}, 22.5^{\circ}, 45^{\circ}, 67.5^{\circ}, 90^{\circ}, 112.5^{\circ}, 135^{\circ}, 157.5^{\circ}, 180^{\circ} \}$. Moreover, we include the sub-dominant modes in the waveform up to $l=5$.
In Fig.~\ref{Fig_NR_waveforms}, we plot these waveforms in the NR frame as well as in LAL wave frame. We also plot the spin-orbit resonance condition, i.e., $\vek k \cdot (\vek s_1 \times \vek s_2)$ for these binary configurations. We see that the resonant condition holds approximately even at the late inspiral phase before merger.

Note that the above described waveforms are fully precessing and accurate enough to model GWs from comparable mass precessing
binaries experiencing spin-orbit resonances. However, such waveforms are computationally too expensive to be used to construct a template
bank. Therefore, computationally less expensive, though less accurate, waveforms are used to search for GWs from precessing binaries.
These include non-precessing models and it has been shown that they perform well in detecting most of the binaries including precessing ones.
In fact, the non-precessing IMRPhenomB model \cite{Ajith_prl} 
claims to detect a significant fraction of precessing binaries in the comparable mass regime.
These observations and claims motivated us to investigate the performance of currently available state-of-the-art waveform models in detecting and
characterizing GWs from resonant binaries. This is the aim of this paper. \\

\section{Waveform models employed in the detection and parameter estimation pipelines}
\label{Sec_III}
In the present section we briefly describe the waveform models used in this paper
for detection and estimation of errors in the parameters of resonant binaries, namely,  IMRPhenomD, SEOBNRv4 and IMRPhenomP. 

\subsection{IMRPhenomD templates}
For low-latency GW searches it is convenient to have efficient though approximate search templates that are faster to compute.
That is why several frequency domain IMR phenomenological (phenom) templates have been developed in the past few years.
These phenom models are build upon certain hybrid waveforms which are constructed by stitching together the PN 
waveforms with NR waveforms. These hybrids are then fitted to a parameterized waveform
model containing a number of phenomenological coefficients which are finally mapped to the physical parameters.
For spinning, non-precessing binaries the first phenom model was IMRPhenomB, proposed by Ajith et al.~\cite{Ajith_prl}.
This is a 3 parameter waveform family that depends upon $M(=m_1+m_2)$, $\eta(=m_1\,m_2/M^2)$ and an effective spin parameter 
$\chi_{\rm eff} = (1+\delta m)\, (\vek \chi_1 \cdot \vek k)/2 + (1-\delta m)\, (\vek \chi_2 \cdot \vek k)/2$, where $\vek \chi_i=\vek S_i/m_i^2$. This model made use of
simple analytical ans\"{a}tze for the phase and amplitude of the PN-NR hybrids. Later on, these ans\"{a}tze were suitably 
modified to make smooth transitions between their inspiral, merger and ringdown forms, and improved the accuracy of the model \cite{Santamaria:2010yb}.
However, this improved model, also known as IMRPhenomC, is valid only to a limited region of parameter space as
it was calibrated only up to mass-ratio $q\leq4$ and BH spins $\chi_{i}\leq0.75$ ($0.85$ for equal mass systems).

Later on, a new phenom model - IMRPhenomD \cite{Khan:2015jqa, Husa:2015iqa}-  was developed which is basically an improvement to IMRPhenomC in terms of 
accuracy and range of validity. However, the IMRPhenomD model has several new features. 
First, it is calibrated to hybrid EOB+NR waveforms to constrain the model's free parameters.
Second, it exploits NR simulations with mass-ratio $q\in[1,18]$ and BH spins $\chi_i\leq0.85$ (0.99 for the
equal-mass case). Third, this model is parameterized by a reduced effective spin parameter
\begin{equation}
\label{eq_chi_cap}
 \hat{\chi} = \frac{\chi_{\rm PN}}{1-76\,\eta/113},
\end{equation}
where,
\begin{equation}
 \chi_{\rm PN} = \chi_{\rm eff} - \frac{38\,\eta}{113}\,(\vek \chi_1 \cdot \vek k + \vek \chi_2 \cdot \vek k),
\end{equation}
rather than $\chi_{\rm eff}$ which is a better parameter to use in IMR waveform models \cite{Kidder_Will, Purrer:2013xma}.
Fourth, the model is modular as it is free to use any inspiral model or a merger-ringdown model. This is important
because if, in future, one has access to more accurate inspiral waveforms or more accurate merger-ringdown model 
(calibrated to more accurate and longer NR simulations covering  a larger region of parameter space), then the existing 
models can be easily replaced by new ones without any adjustments.

\subsection{SEOBNRv4 templates}
\label{Sec_SEOBNRv2}
To tackle the two body dynamics of compact objects, a new approach - the EOB approach -
was developed by Buonanno and Damour in 1999 \cite{Buonanno99}. The main motivation 
behind this approach is to extend the analytical waveform models towards the last
stages of inspiral, merger and ringdown. The
EOB approach uses the results of PN theory, BH
perturbation theory and gravitational
self-force formalism. In EOB framework, the conservative dynamics
of two compact objects of masses $m_1$ and $m_2$ and spins  $\vek S_1$ and $\vek S_2$ is mapped 
into the dynamics of an effective particle of mass $\mu = m_1 m_2/(m_1 + m_2)$ and 
spin $\vek S_{*}$ moving in a deformed Kerr metric with mass $M = m_1 + m_2$ and spin $S_{\rm Kerr} = \vek S_1 + \vek S_2$.
Over a decade of improvements and developments, EOB model has now become the 
most accurate IMR waveform model for spinning and non-precessing binaries \cite{Taracchini:2012, Taracchini:2013rva}.
This is achieved by including in the EOB dynamics higher-order
(not formally known) PN terms and calibrating them to a large number of lengthy and more accurate NR
simulations.

In Ref.~\cite{Taracchini:2012}, Taracchini et al. proposed a 
prototype EOB model (SEOBNRv1) for non-precessing binaries which is calibrated to 5 non-spinning NR simulations with mass-ratio 
$q = \{1, 2, 3, 4, 6\}$ and 2 equal-mass equal-spin NR simulations. The SEOBNRv1 model is accurate for any mass-ratio but 
for individual BH spins in the range $-1\leq(\vek \chi_{\rm i} \cdot \vek k) \lesssim0.7$.
The improved version of this model - SEOBNRv2 - is calibrated to 38 new and longer NR waveforms (8 of them are non-spinning while 30 are spinning, non-precessing)
with  $q=\{1, 1.5, 2, 3, 4, 5, 6, 8\}$ and $-0.98\leq(\vek \chi_{\rm i} \cdot \vek  k) \leq0.98$ \cite{Taracchini:2013rva}. The SEOBNRv2 model is valid for any
mass-ratio and spin magnitude. More recently, Ref.~\cite{SEOBNRv4}
improved the accuracy of SEOBNRv2 by calibrating it with 141 new NR simulations that span larger mass-ratios and spins as compared to simulations which were calibrated with SEOBNRv2 model. This new model is known as `SEOBNRv4'. The fact that EOB models employ a set of
differential equations, it is computationally too expensive to use them as templates for detection and parameter estimation purposes. 
Therefore, we use a frequency-domain reduced order model - ${\rm SEOBNRv4\_ROM}$ - in our analysis. 
This reduced order model faithfully represents the original model with an accuracy that is better than the statistical uncertainty caused by the instrumental noise.

\subsection{IMRPhenomP templates}
Hannam et al. \cite{HSBH2014} proposed a frequency-domain
IMR waveform - IMRPhenomP - to model GW signal from precessing BBHs.
The key idea of this model is the fact that a precessing waveform can be approximated by appropriately rotating waveforms constructed in a co-precessing frame back to the inertial frame.  This is straightforward to do since waveform multipoles produced by a precessing BBH in a co-precessing frame are well modeled by multipoles produced by non-precessing BBHs in an inertial frame \cite{SHH2012,SHHA2011,OVHMS2011}.
Therefore, one can approximately model precessing waveforms by combining models for non-precessing waveforms and the rotation that tracks the precession of the orbital plane. The model is characterized by only 3 intrinsic
dimensionless physical parameters: mass-ratio $q$, an effective inspiral spin $\chi_{\rm eff}$ and an effective precession spin $\chi_{\rm p}$ \cite{Schmidt2015}, defined as 
\begin{equation}
\label{eq:chip}
\chi_{\rm p} = \frac{\max(A_1\, |\vek S_1 -(\vek S_1 \cdot \vek k) \vek k |, A_2\, |\vek S_2 -(\vek S_2 \cdot \vek k)\vek k|)}{A_1\, m_1^2}\,,
\end{equation}
where $A_1=2+ (3/2)m_2/m_1$ and $A_2=2+ (3/2)m_1/m_2$.
The ability of this model to compute generic waveforms for precessing BBH with only two spin parameters implies strong degeneracies that will make
it difficult to identify individual BH spins, in
particular, the spin of the smaller BH.
This model was initially constructed
by a transformation of IMRPhenomC.
The most recent version of the model - IMRPhenomPv2 -  employs IMRPhenomD approximant for the rotation to model precessing waveforms. For more details of IMRPhenomPv2 model, we refer the readers to Refs.~\cite{HSBH2014, phenompv2}.

In the next two sections, we present our results on the detection and characterization of spin-orbit resonant binaries using aLIGO detector. 

\section{Detection of Spin-orbit resonances}
\label{Sec_IV}

To check the performance of the aforementioned non-precessing and precessing template approximants, we define certain quantities which are commonly used in BBH searches by LIGO Scientific Collaboration. Let $s(t;\vek \lambda_s)$ be the expected GW signal from resonant binaries where $\vek \lambda_s$ represents a set of physical parameters of the binary system such as masses, spin's magnitude and orientation. Further, let $h(t; \vek \lambda_h)$ be the template waveform where $\vek \lambda_h$ represents a set of parameters upon which these models depend; here $\vek \lambda_h$ contains, for example, only the masses and spin magnitudes for non-precessing templates. We now define the normalized overlap between $s(t)$ and $h(t)$ as
\begin{equation}
{\cal O}(s, h) = \langle \hat{s}, \hat{h}\rangle = \frac{\langle s, h \rangle}{\sqrt{\langle s, s \rangle \langle h, h \rangle}}\,,
\end{equation}
where $\hat{s}$ and $\hat{h}$ stand for normalized $s(t)$ and $h(t)$, respectively. The angular bracket denotes the following noise-weighted inner product
\begin{equation}
\langle s, h \rangle = 4\, {\mathcal Re} \left[\int_{f_{\rm low}}^{f_{\rm high}} \frac{\tilde s(f)\, \tilde h^{*}(f)}{S_h(f)}\, df \right] \,,
\end{equation}
where $\tilde s(f)$ and $\tilde h(f)$ represent Fourier transforms of $s(t)$ and $h(t)$, respectively, and $S_h(f)$ is the one-sided power spectral density (PSD) of the noise of the detector. 
The maximization of ${\cal O}(s, h)$ over external parameters such as the time of arrival of the signal $t_0$ and associated phase $\Phi_0$ is known as {\it match}:
\begin{equation}
\label{eq_match}
{\cal M}(s, h) = \max_{t_0, \Phi_0} {\cal O}(s, h)\,.
\end{equation}
In the real searches, the signal parameters are not known a priory, and hence, a bank of templates is employed to search for GW. A template bank contains a discrete set of waveforms corresponding to different values of parameters $\vek \lambda_h$. An optimal template bank is one which minimizes the computational cost in a search without reducing the detectability of signals. Therefore, templates in a template bank are chosen (we call it template placement) using appropriate template placement algorithms. Among many algorithms, there are two, namely geometric \cite{geometric} and stochastic \cite{stochastic} placement algorithms, which are employed in real BBH searches. In these algorithms, the templates are placed in the parameter space such that it corresponds to an acceptable loss of SNR. In other words, the templates are placed such that the mismatch ($1- match$) between a template corresponding to any point in the parameter  space considered and at least one template in the bank is less than a certain prescribed value. In practice, we use a maximal acceptable mismatch of $0.03$.
The loss in SNR in GW searches using a template bank can be attributed to two factors. First, the placement of templates in the parameter space and second is the difference in waveform model and true GW signal. Therefore, the fraction of optimal SNR recovered by a template bank is estimated by maximizing ${\cal M}$ over all the template parameters. We call it {\it fitting factor} (FF):
\begin{equation}
FF = \max_{\vek \lambda_h} {\cal M}(s, h)\,.
\end{equation}
A template bank is effectual in detecting a GW signal if $FF\geq0.97$. 

In this paper, we use stochastic template placement algorithm \cite{stochastic,stochastic2_CHPB,stochastic_AFPNMW} 
to create non-precessing template banks using IMRPhenomD and SEOBNRv4 waveform approximants. 
We use $0.03$ as the maximum mismatch (i.e., ${\cal M}=0.97$) while aLIGOZeroDetHighPower PSD has been used in the match calculation using Eq.~(\ref{eq_match}). The template bank parameters are as follows: total mass $M\in [6, 100]$ and aligned spins $(\vek \chi_1 \cdot \vek k), (\vek \chi_2 \cdot \vek k ) \in [-0.99, 0.99]$. In our study, we use 10,000 precessing injections of each kind ($\Delta \phi=0^{\circ}$, $\Delta \phi=180^{\circ}$ and generic) using SEOBNRv3 approximants. All the three types of injections have same values of $M$, $q$, $\chi_1$, $\chi_2$, $\theta_1$, $\phi_1$, $\delta$, $\alpha$, $\iota$ and $\psi$ in their respective ranges as mentioned in Table \ref{Tab_injections}, except for $\theta_2$ and $\phi_2$. For $\Delta \phi=0^{\circ}$ injections we set $\phi_2=\phi_1$ and obtained solutions for $\theta_2$ by solving Eq.~(\ref{eq:SOR_cond}). Similarly for $\Delta \phi=180^{\circ}$ injections we set $\phi_2=\phi_1-\pi$ and solve Eq.~(\ref{eq:SOR_cond}) for $\theta_2$. However, for generic injections we choose values for $\theta_2$ and $\phi_2$ which are uniform in $[0, \pi]$ and $[0, 2\pi]$, respectively. In other words, our generic injections have isotropic spin orientations and some of them may arbitrarily be close to one of the resonant configurations. 
Additionally, we use {\it four} NR waveforms which model GW from resonant binaries as our injections.

As mentioned above, one of the factors that causes the loss in SNR in binary searches is the mismatch between model and signal waveforms.  To attribute the loss in SNR only due to this factor, we further improve the FFs by performing a continuous search over template parameters.
We do this as follows: we first compute the FF and the corresponding best matched parameters for IMPhenomD and SEOBNRv4 template banks while using PyCBC \cite{pycbc_Canton_2014}. We then use {\it particle swarm optimization} (PSO) algorithm  \cite{pso} for further maximization of match given by Eq.~(\ref{eq_match}). This algorithm requires a range over which the best matched parameters will be searched and match will be maximized. We assume this range to be centered around the best matched parameters obtained from our FF calculations using PyCBC. The ranges for the parameters are defined as follows: For each injection with $M_c<30 M_{\odot}$, we set the range for $M_c$ to be $\pm 10 \% $ of the chirp mass value obtained from our template bank analysis. However, for injections having $M_c \ge 30 M_{\odot}$, we set this margin to be $\pm 30 \%$. For all injections, we vary $\eta$ between $[0.10, 0.25]$ which is based on the results we obtained from PyCBC. Finally, for all the injections, we give full range of $[-0.99, 0.99]$ to all the spin components $\chi_{1x}$, $\chi_{1y}$, $\chi_{1z}$, $\chi_{2x}$, $\chi_{2y}$, $\chi_{2z}$ such that the dimensionless spin magnitudes $\chi_1$ and $\chi_2$ are less than equal to $0.99$. In order to make sure that we obtain the true maximum match in our analyses, we perform the PSO runs multiple times and select the best match template by selecting the maximum match over all these multiple PSO runs. We verify that this procedure indeed allows us to obtain the best match values. As it become harder and harder to obtain the maximum of an underlying function as the number of parameters and complexity increases, we ran the PSO different number of times for different template approximants.  For example, we conduct the PSO search $8$ times while recovering injections with non-precessing templates and with IMRPhenomPv2, we use the best out of $28$ PSO trials for low masses $(m_1  + m_2 \le 30 M_{\odot})$ and best out of $16$ PSO trials for high masses $(m_1  + m_2 > 30 M_{\odot})$. Since the PSO runs improve the FF values as compared to the template bank runs, all the results presented in this paper are obtained from PSO runs.

For non-precessing template approximants, IMRPhenomD and SEOBNRv4, the match is maximized over $m_1, m_2, \chi_{1z}$ and $\chi_{2z}$ to compute the FF and need not be maximized over $D_L$, $\delta$, $\alpha$, $\psi$ and $\iota$. This is because the dependence of non-precessing templates on these parameters results in an overall phase and amplitude in the waveform model which is taken care of while maximizing overlap ${\cal O}$ over $\Phi_0$ and using normalized templates (see, e.g., Ref~\cite{Harryetal} for more detail). On the other hand, for precessing template approximant, IMRPhenomPv2, the match is not only maximized over $m_1, m_2, \chi_{1x}, \chi_{1y}, \chi_{1z}, \chi_{2x}, \chi_{2y}$ and $\chi_{2z}$ but should also over $D_L$, $\delta$, $\alpha$, $\psi$ and $\iota$. This is because the orientation of a precessing binary with respect to the detector changes with time and as a result the precessing waveform can not be written in a simple form where the dependence of angular parameters $\delta$, $\alpha$, $\psi$ and $\iota$ can be factored out. The maximization of match over these angular parameters is necessary otherwise one loses the information of sub-dominant modes present in the precessing waveform model in detector frame\footnote{Though the precessing approximant, IMRPhenomPv2,  contains only $(2,2)$ modes in the co-precessing frame, some power from the $(2,2)$ mode gets leaked into $(2,\pm1)$ modes in the inertial (detector) frame. Therefore, IMRPhenomPv2 approximant has partial $(2,\pm1)$ modes in detector frame.}. Very recently, Harry et al. \cite{Harryetal} derived a new statistic which maximizes the overlap also over the sky-location-dependent parameters, i.e., $\delta$, $\alpha$, $\psi$ along with $t_0$, $\Phi_0$ and $D_L$. Ref.~\cite{Harryetal} defined a sky-maxed SNR,
\begin{equation}
\rho_{\rm SM} = \sqrt{2\, \lambda}\,,
\end{equation}
such that, 
\begin{widetext}
\begin{equation}
 \lambda = \max_{D_L, t_0, \Phi_0,\delta, \alpha, \psi}(\lambda)  
 = \frac{1}{4}\left( \frac{
      |\hat{\rho}_{+}|^2 - 2 \hat{\gamma} I_{+\times} + |\hat{\rho}_\times|^2
      + \sqrt{(|\hat{\rho}_+|^2 - |\hat{\rho}_\times|^2)^2 + 4 (I_{+\times} |\hat{\rho}_+|^2 - \hat{\gamma})
              (I_{+\times} |\hat{\rho}_\times|^2 - \hat{\gamma}) }}
  {1 - I_{+\times}^2} \right),
\end{equation}
\end{widetext}
where,
\begin{eqnarray}
 \hat{\rho}_{+,\times} &=& \langle s | \hat{h}_{+,\times} \rangle \,, \\
 \hat{\gamma} &=& {\mathcal Re} \left[ \hat{\rho}_+ \hat{\rho}^*_{\times} \right] \,, \\
I_{+ \times} &=& {\mathcal Re} [\langle \hat{h}_+ | \hat{h}_{\times} \rangle ]  \,,
\end{eqnarray}
and $\hat{h}_+$ and $\hat{h}_{\times}$ are the two normalized GW polarizations of the precessing waveform model. The sky-maxed SNR $\rho_{\rm SM}$ is again maximized over the remaining parameters, namely, $m_1, m_2, S_{1x}, S_{1y}, S_{1z}, S_{2x}, S_{2y}$, $S_{2z}$ and $\iota$ to compute the sky-maxed FF ($=\rho_{\rm SM}/\sqrt{<s,s>}$) for precessing templates.  This is our so-called FF for precessing templates whereas the parameter values at which it is maximized correspond to the best match parameters. 
The FF computed in this way is, therefore, the maximum fraction of SNR a waveform model (precessing or non-precessing) can recover and any deviation from unity is purely due the differences in the template and signal models. 

The FF value associated with each precessing signal can be sometime misleading and one can not say whether the signal will be detectable by the detector or not. The aLIGO like detectors have sensitivity that depends upon the direction and orientation of the binary systems.  Signals that are not partially aligned to the detector may not have enough SNR to be detected, regardless of their FF value. It has been shown that many highly precessing binaries precess in such a way that they land to a point where the detector has very little sensitivity \cite{Brown2012}. 
Therefore, we'll use what is called {\it signal recovery fraction} (SRF), $\alpha$, to quantify the performance of a template approximant which is defined as \cite{Harry_Nitz_Brown_2014} 

\begin{equation}
\alpha = \frac{\Sigma_i^{\rm N} \, {\rm FF}^3 \, \sigma_i^3}{\Sigma_i^{\rm N} \, \sigma_i^3}\,,
\end{equation}
where, $\sigma_i = \sqrt{<s_i, s_i>}$ is the optimal SNR of the $i^{\rm th}$ binary and ${\rm N}$ is the number of binaries in a population.  The signal recovery fraction  $\alpha$, thus, represents the fraction of sources in a binary population that can be detected using the underlying approximant as detection templates. In the next section, we will calculate this fraction for $\Delta \phi = 0^{\circ}$, $\Delta \phi = 180^{\circ}$, and generic precessing binary populations using IMRPhenomD, SEOBNRv4 and IMRPhenomPv2 as search templates.  

Below we will also show how {\it signal recovery fraction} depends upon the various binary parameters and its orientation in the sky. We do this as follows while considering any two parameters (say, $X$ and $Y$) at a time. We split the FFs (and the associated $\sigma_i$s) into a series of bins corresponding to ranges in both the parameters and for each bin we compute average values of ${\rm FF}^3 \, \sigma^3$ and $\sigma^3$ within that bin. Therefore, we define
\begin{equation}
\alpha(X, Y) = \frac{\overline{{\rm FF(X, Y)}^3 \, \sigma(X, Y)^3}}{\overline{\sigma(X, Y)^3}}\,,
\end{equation}
where $\overline{Z}$ represents the average of $Z$ in a bin. In the plots below we divided the ranges in parameters X and Y into 20 bins each.

In the next section we investigate the performance of IMRPhenomD, SEOBNRv4 and IMRPhenomPv2 template approximants in detecting GW signals from resonant binaries modeled by SEOBNRv3 approximant and four NR simulations. 

\begin{figure*}[ht]
\begin{center}
\hspace*{-0.75cm}
\includegraphics[width=7.5in]{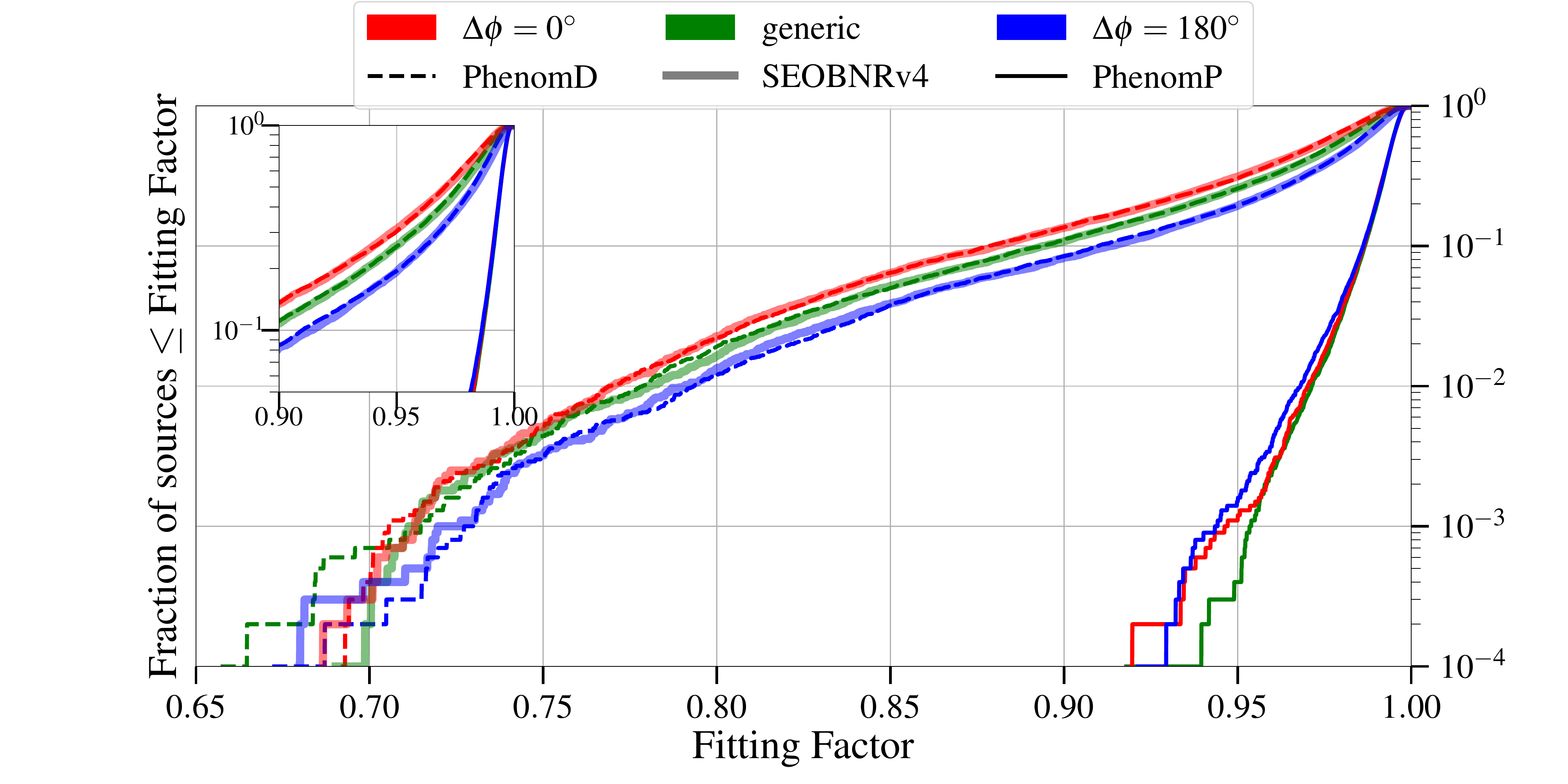}
\end{center}
 \caption{Comparison of performances of IMRPhenomD, SEOBNRv4 and IMRPhenomPv2 templates in recovering SEOBNRv3 injections with arbitrary orbital inclination. Cumulative histogram of FFs against the fraction of sources is plotted in log-scale. The FFs for $\Delta \phi = 0^{\circ}$ resonance family are depicted in red; $\Delta \phi = 180^{\circ}$ resonance family are depicted in blue and generic binaries are depicted in green. The dashed line represents IMRPhenomD templates, the translucent solid line SEOBNRv4 templates while dark solid line represents IMRPhenomPv2 templates.}
\label{template_comparison_fig_v3}
\end{figure*}

\subsection{Results: SEOBNRv3 as injections}
\label{sec:SEOBNRv3results}
In this section, we use SEOBNRv3 approximant to model GWs from resonant binaries, {\it i.e.,} binaries with $\Delta \phi=0^{\circ}$ and $180^{\circ}$. For comparison, we also present results for generic precessing binaries -- injections which may or may not satisfy resonant condition(s). In Fig.~\ref{template_comparison_fig_v3}, we compare the performances of all three template approximants in recovering all three types of injections. The figure shows cumulative histograms of FFs against the fraction of sources, which are shown in log-scale on the y-axis. We find that non-precessing approximants roughly recover $\sim 70\%$ of injections (all three kind; $\Delta \phi=0^{\circ}$, $\Delta \phi=180^{\circ}$ and generic) with FF$>0.97$ (or $90\%$ of injections with FF$>0.9$). The precessing approximant IMRPhenomPv2, on the other hand, stands out with flying colors and recovers $\sim 99\%$ of the injections with sky-maxed FF$>0.97$. 

\begin{figure*}
\begin{center}

\centering
\includegraphics[width=7in]{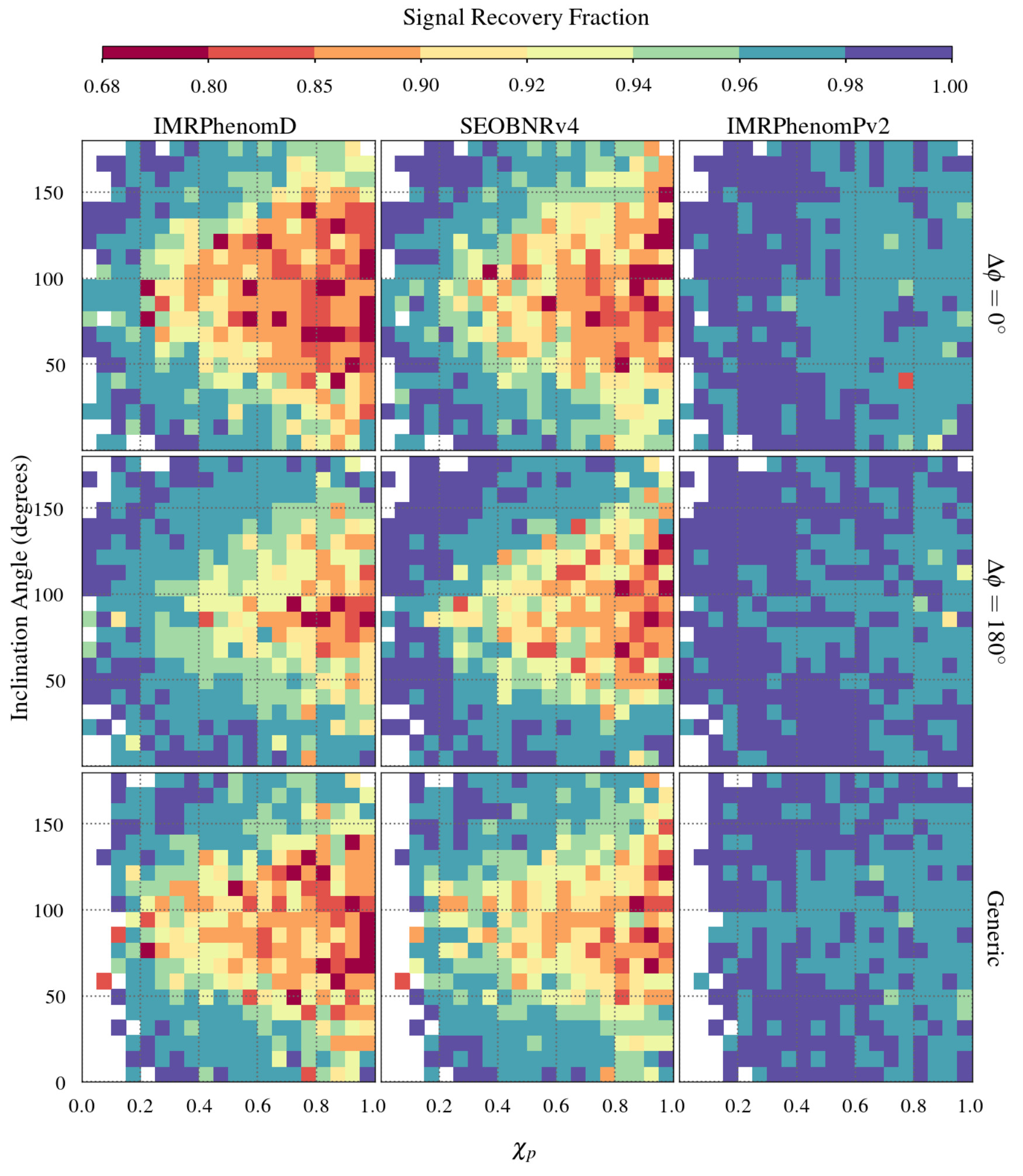}
\end{center}
 \caption{Depicting color coded SRF as a function of $\iota$ and $\chi_p$. The first, second and third columns are for IMRPhenomD, SEOBNRv4 and IMRPhenomPv2 templates, respectively. The first, second and third rows are for $\Delta \phi = 0^{\circ}$ injections, $\Delta \phi = 180^{\circ}$ injections and generic injections, respectively.}
\label{v3_chip_inclination}
\end{figure*}

Note that the SEOBNRv3 injections have arbitrary inclination angle $\iota$ uniformly distributed in $[0^{\circ}, 180^{\circ}]$ and such loss in SNR (or low FF) comes about when the binary is inclined away from ``face-on" or ``face-off" orientations with respect to the detector, and consequently its orbital precession is coupled strongly to the resulting waveform in detector frame. Our signal model includes $(l, m)=(2, \pm2), (2, \pm1)$ modes and when the binary is neither face-on ($\iota=0^{\circ}$) nor face-off ($\iota=180^{\circ}$), the sub-dominant modes $(l, m)=(2, \pm 1)$ become important and lead to a substantial mismatch with template that contains only the dominant $(l, m)=(2,\pm 2)$ mode in the inertial frame. The other factor that causes low FFs is the effect of spin precession in GW signals, as modelled by SEOBNRv3. As mentioned earlier, the template models IMRPhenomD and SEOBNRv4 are non-precessing and fail to capture the precessional features of SEOBNRv3 model. However, IMRPhenomPv2 have both (i) the sub-dominant GW modes $(l,m) = (2, \pm1)$ and (ii) precession effects incorporated in its modeling but still fails to recover the remaining $1\%$ of the injections. This can be attributed to the fact that IMRPhenomPv2 models spin and precession effects in an effective way using only two spin parameters, namely $\chi_{\rm eff}$ and $\chi_{\rm p}$, whereas SEOBNRv3 is a full precessing waveform model described by all six spin parameters.  In Fig.~\ref{v3_chip_inclination}, we plot SRF for all $10000$ injections as a function of inclination angle ($\iota$) and in-plane spin parameter ($\chi_{\rm p}$). Note that larger value of $\chi_{\rm p}$ means large precessional effects in GW signal. As expected, we receive relatively lower SRF for injections having $45^{\circ} \lesssim \iota \lesssim 135^{\circ}$ and $\chi_{\rm p}>0.5$ for all the three approximants.

\begin{figure*}
\begin{center}

\centering
\includegraphics[width=7in]{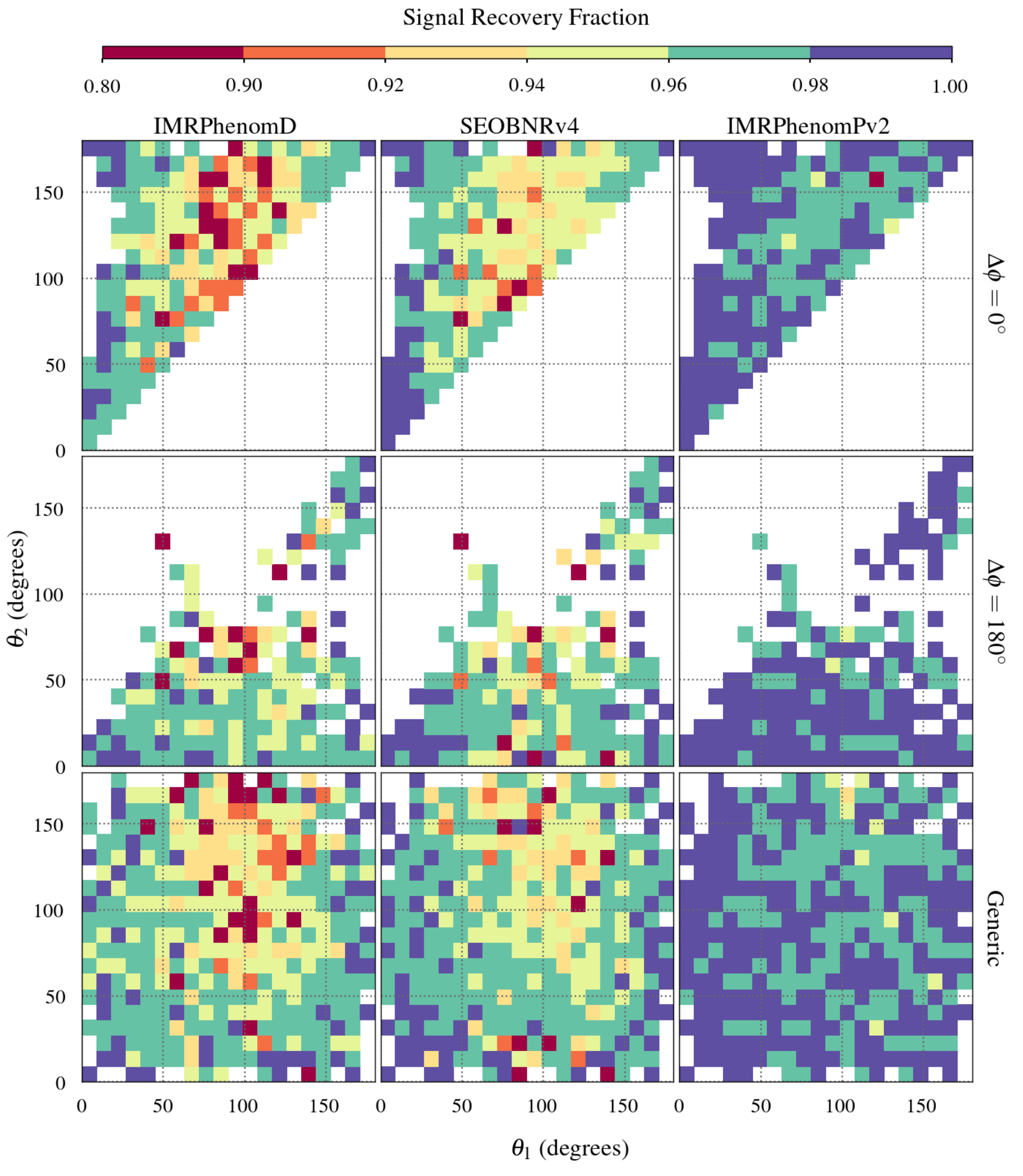}
\end{center}
 \caption{Depicting color coded SRF as a function of $\theta_1$ and $\theta_2$ for signals having $0^{\circ} \leq \iota  \leq 45^{\circ}$ and $135^{\circ} \leq \iota  \leq 180^{\circ}$. The first, second and third columns are for IMRPhenomD, SEOBNRv4 and IMRPhenomPv2 templates, respectively. The first, second and third rows are for $\Delta \phi = 0^{\circ}$ injections, $\Delta \phi = 180^{\circ}$ injections and generic injections, respectively.}
\label{v3_delta_plots_faceon}
\end{figure*}

\begin{figure*}
\begin{center}

\centering
\includegraphics[width=7in]{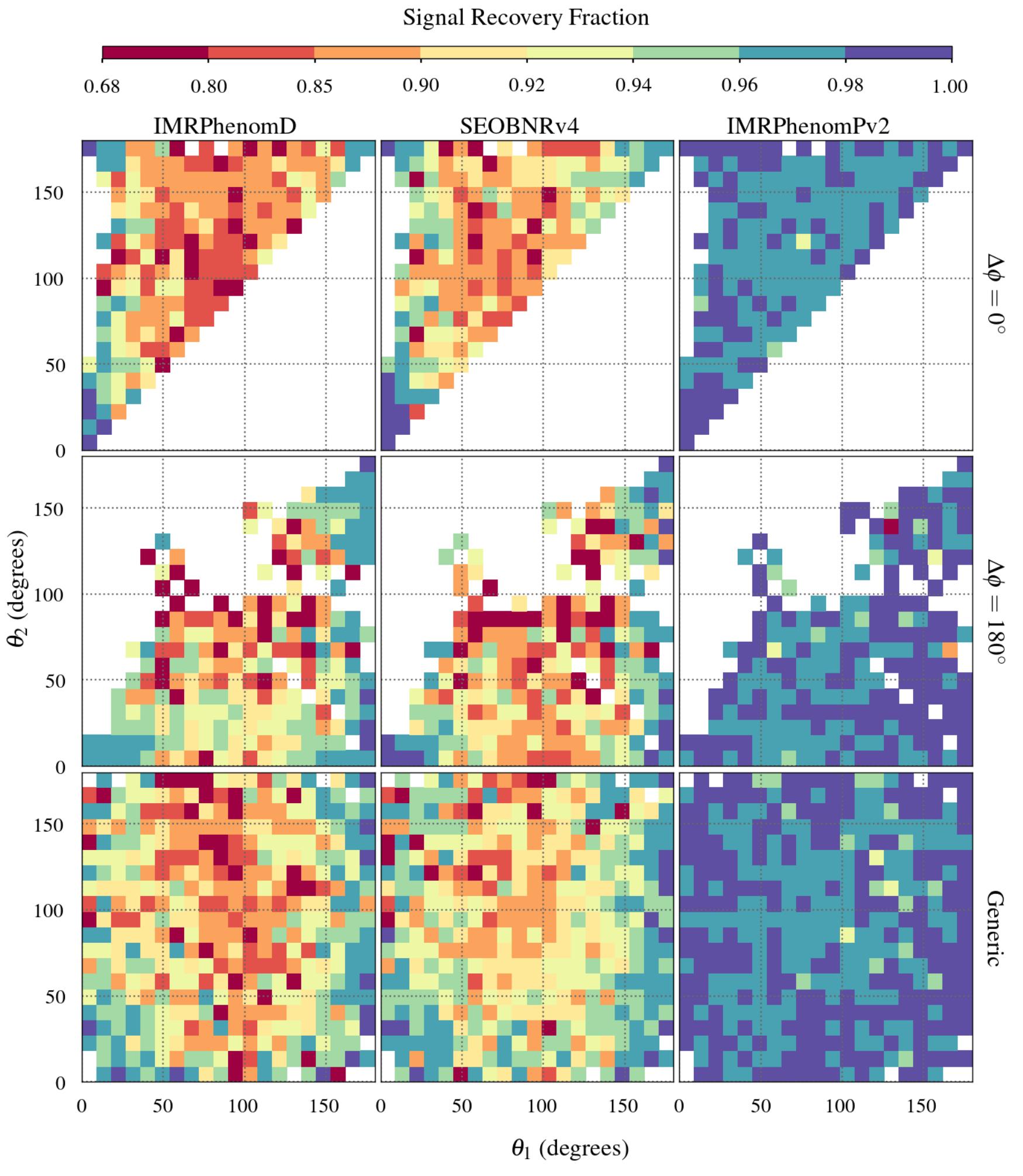}
\end{center}
 \caption{Depicting color coded SRF as a function of $\theta_1$ and $\theta_2$ for signals having $45^{\circ} \leq \iota  \leq 135^{\circ}$. The First, second and third column is for IMRPhenomD, SEOBNRv4 and IMRPhenomPv2 templates, respectively. The first, second and third rows are for $\Delta \phi = 0^{\circ}$ injections, $\Delta \phi = 180^{\circ}$ injections and generic injections, respectively.}
\label{v3_delta_plots_edgeon}
\end{figure*}

\begin{figure*}
\begin{center}

\centering
\includegraphics[width=7in]{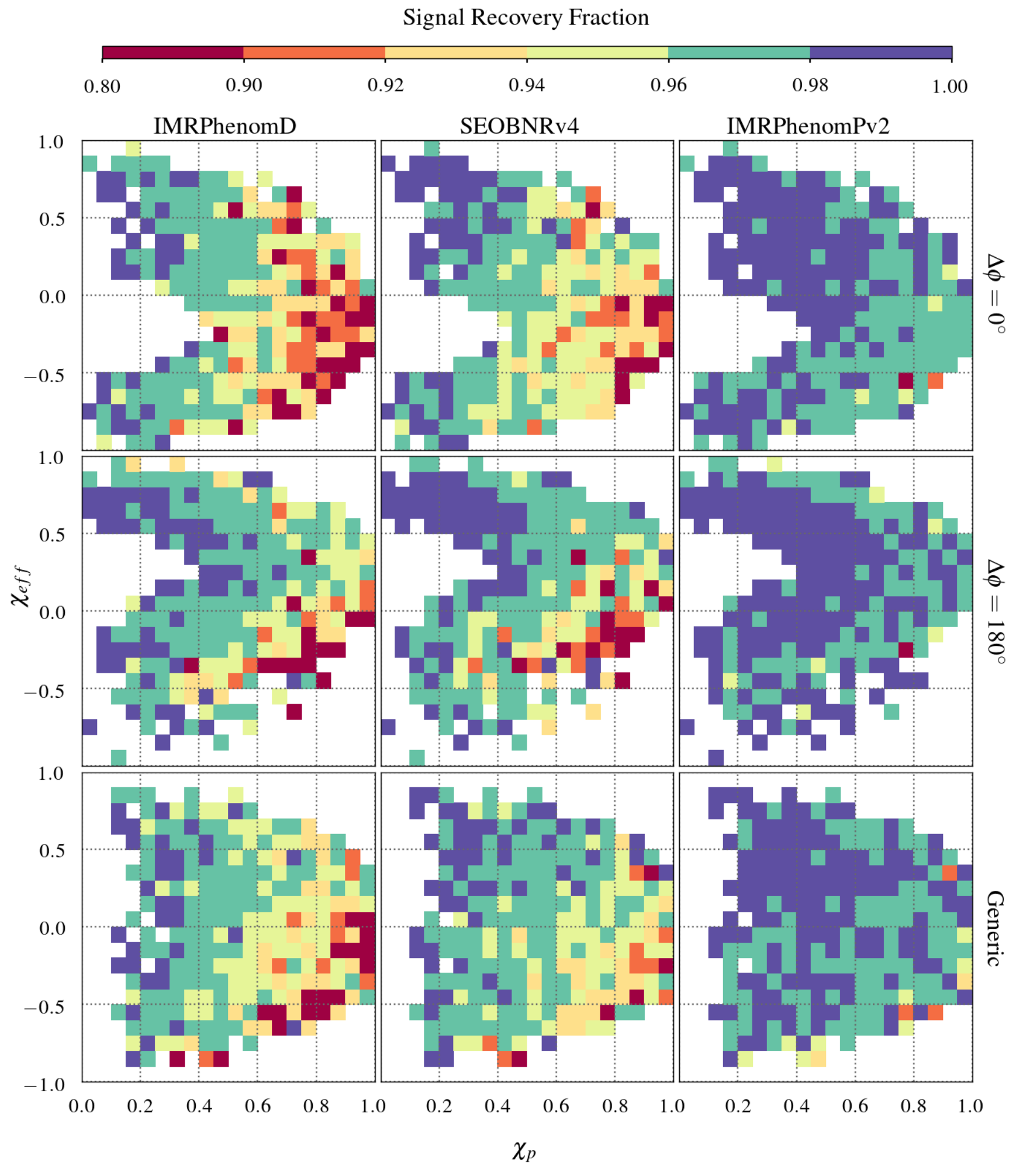}
\end{center}
 \caption{Depicting color coded SRF as a function of aligned spin parameter $\chi_{\rm eff}$ and in-plane spin parameter $\chi_{\rm p}$ for signals having $0^{\circ} \leq \iota  \leq 45^{\circ}$ and $135^{\circ} \leq \iota  \leq 180^{\circ}$. The First, second and third column is for IMRPhenomD, SEOBNRv4 and IMRPhenomPv2 templates, respectively. The first, second and third row is for $\Delta \phi = 0^{\circ}$ injections, $\Delta \phi = 180^{\circ}$ injections and generic injections, respectively.}
\label{v3_delta_plots_chi_face_on}
\end{figure*}
\begin{figure*}
\begin{center}

\centering
\includegraphics[width=7in]{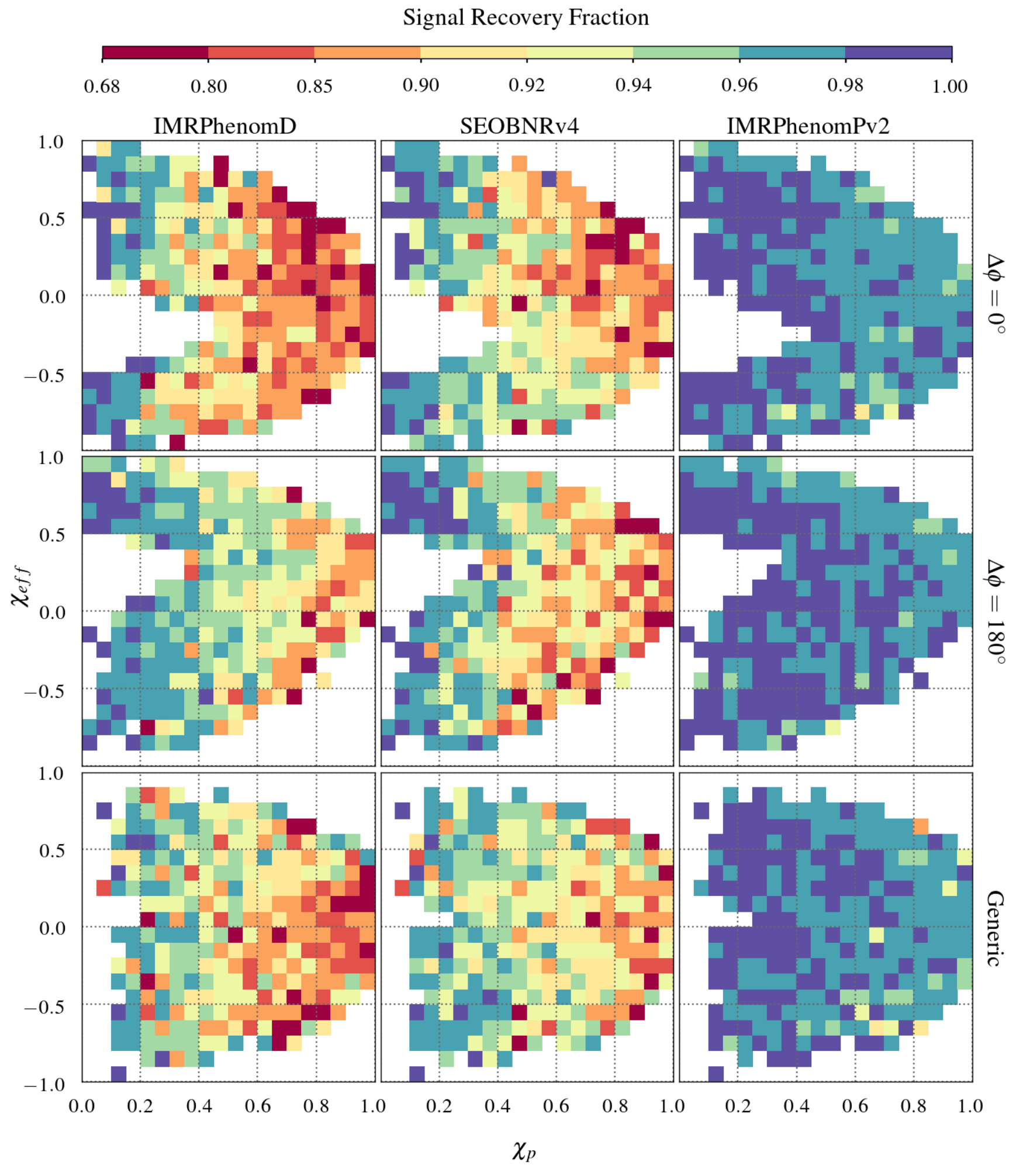}
\end{center}
 \caption{Depicting color coded SRF as a function of aligned spin parameter $\chi_{\rm eff}$ and in-plane spin parameter $\chi_{\rm p}$ for signals having $45^{\circ} \leq \iota  \leq 135^{\circ}$. The First, second and third column are for IMRPhenomD, SEOBNRv4 and IMRPhenomPv2 templates, respectively. The first, second and third row is for $\Delta \phi = 0^{\circ}$ injections, $\Delta \phi = 180^{\circ}$ injections and generic injections, respectively.}
\label{v3_delta_plots_chi_edge_on}
\end{figure*}

Interestingly, it turns out that resonant binaries with $\Delta \phi = 180^{\circ}$ are more likely to be detectable as compared to $\Delta \phi = 0^{\circ}$ resonant binaries as well as the generic precessing binaries. This is because FF and SRF values for $\Delta \phi = 180^{\circ}$ resonant binaries are slightly higher than that for generic precessing binaries which are in turn higher than for $\Delta \phi = 0^{\circ}$ resonant binaries as can be seen in Figs.~\ref{template_comparison_fig_v3} and \ref{v3_chip_inclination}. Below we attempt to explain this trend in FF/SRF values. 

The three types of injections differ from each other only in their $\theta_2$ and $\Delta \phi$ values while all other binary parameters are the same. This is because for given values of binary parameters ($\theta_1, m_1, m_2, \chi_1, \chi_2$ and $x$), the two resonant families cover different parts of $\theta_1-\theta_2$ plane. The $\Delta \phi = 0^{\circ}$ resonants tend to have $\theta_1 < \theta_2$ while  $\Delta \phi = 180^{\circ}$ resonants tend to have $\theta_1 > \theta_2$ (see, e.g., Figs.~3 \& 4 in Ref.~{\cite{JS}} and Figs.~1 \& 2 in Ref.~\cite{KSB10}). 
On the other hand, there is no particular trend in $\theta_1-\theta_2$ values for the generic precessing binaries. These difference in spin configurations of three types of injections lead to difference in their precession dynamics. In Fig.~\ref{v3_delta_plots_faceon} and \ref{v3_delta_plots_edgeon} we plot SRFs as a function of $\theta_1$ and $\theta_2$ for all three types of injections and three template approximants for weakly precession-coupled sources with $\{0^{\circ} \leq \iota  \leq 45^{\circ}, 135^{\circ} \leq \iota  \leq 180^{\circ}\}$ and for strongly precession-coupled sources $45^{\circ} \leq \iota  \leq 135^{\circ}$, respectively, in order to separate out the effects of inclination angle on SRFs. 

The difference in $\theta_1-\theta_2$ phase space for the three kinds of injections result in different distributions for their $\chi_{\rm eff}$ and $\chi_p$. Therefore, to see the effect of different spin distributions on the computed SRFs, in Figs.~\ref{v3_delta_plots_chi_face_on} and \ref{v3_delta_plots_chi_edge_on} we plot SRF as a function of aligned spin parameter $\chi_{\rm eff}$ and in-plane spin parameter $\chi_p$ of injections. In  $\Delta \phi = 0^{\circ}$ resonant configurations the in-plane spin components add constructively giving orbital angular momentum $\vek L$ greater misalignment with respect to the total angular momentum $\vek J$, resulting in greater precession modulation of the orbital plane. On the other hand for  $\Delta \phi = 180^{\circ}$ resonants it is the opposite: the in-plane spin components add destructively allowing $\vek L$ to be more aligned with $\vek J$ and hence less precession modulation of the orbital plane. This difference in precession for two resonant families is also shown in Fig.~3 in Ref.~\cite{Gerosa2014}. It is merely an artifact of `max' in the definition of $\chi_{\rm p}$ in Eq.~(\ref{eq:chip}) that the distribution of $\chi_{\rm p}$ for $\Delta \phi = 0^{\circ}$ and $\Delta \phi = 180^{\circ}$ resonants are similar but inherently $\Delta \phi = 0^{\circ}$ resonance are more precessing than $\Delta \phi = 180^{\circ}$ ones and hence have lower SRFs.   
Moreover, given the trends in $\theta_1$ and $\theta_2$ for resonant binaries (i.e., $\theta_1<\theta_2$ for $\Delta \phi = 0^{\circ}$ and $\theta_1>\theta_2$ for $\Delta \phi = 180^{\circ}$ resonances), one also expects more negative $\chi_{\rm eff}$ values for $\Delta \phi = 0^{\circ}$ resonances than for $\Delta \phi = 180^{\circ}$ resonances for unequal mass binaries. It has been known that the negative effective spin binaries possess more precession  modulation in their signal as compared to binaries having relatively positive effective spins. Therefore, what we gather from above discussion is that $\Delta \phi = 0^{\circ}$ resonances possess higher precession and more negative $\chi_{\rm eff}$ values than $\Delta \phi = 180^{\circ}$ resonances. This means it will be harder for non-precessing templates to search for GW signal from such binaries. Figures~\ref{v3_delta_plots_chi_face_on} and \ref{v3_delta_plots_chi_edge_on} indeed depict that high in-plane spins and negative effective spins of injections cause low SRFs. 
It, thus, explains why $\Delta \phi  = 180^{\circ}$  resonant binaries have higher SRFs as compared to $\Delta \phi=0^{\circ}$ resonant and generic binaries. Higher SRFs for $\Delta \phi=180^{\circ}$ and lower SRFs for $\Delta \phi=0^{\circ}$ resonant binaries as compared to generic ones indicate that we will have slight observation bias towards $\Delta \phi  = 180^{\circ}$ resonant binaries and away from $\Delta \phi=0^{\circ}$ resonant binaries if we employ these non-precessing approximants as our search templates.

In Table~\ref{table:SRF_values}, we present the SRF for all injection-template pairs and see that $\Delta \phi = 180^{\circ}$ injections indeed have highest SRF for all three template approximants as compared to $\Delta \phi=0^{\circ}$ and generic injections. The overall SRF values are highest for IMRPhenomPv2 followed by SEOBNRv4 and IMRPhenomD respectively. The expected SRF trend for three types of injections, i.e., ${\rm SRF}_{\Delta \phi=0^{\circ}}<{\rm SRF}_{\rm generic}<{\rm SRF}_{\Delta \phi=180^{\circ}}$ does not hold good for IMRPhenomD while it does for SEOBNRv4 approximant. This is because in the case of IMRPhenomD there are a few generic injections ($\sim6$) that have very low FF values ($\sim \in[0.66-0.7]$) as compared to $\Delta \phi=0^{\circ}$ injections (see the tail of dashed green line in Fig.~\ref{template_comparison_fig_v3}).  We verified that if we remove those low FF injections and recompute SRF, we get the desired trend. 

\renewcommand{\arraystretch}{1.5}
\begin{table}[t]
\caption{Signal recovery fraction (SRF) values for the three approximants IMRPhenomD, SEOBNRv4, and IMRPhenomPv2 in recovering the three types of SEOBNRv3 injections. The numbers in the parenthesis represent values at the fourth and fifth decimal places.
}

\centering
\begin{tabular}{|P{3cm} | P{1.7cm} | P{1.7cm} |P{1.7cm} |} 
 \hline
 Approximant & $\Delta \phi=0^{\circ}$ & $\Delta \phi=180^{\circ}$ & Generic \\ [1ex] 
 \hline
 IMRPhenomD & 0.931 & 0.952 & 0.929 \\ 
 SEOBNRv4 & 0.942 & 0.943 & 0.953 \\
 IMRPhenomPv2 & 0.978(13) & 0.980 & 0.978(42) \\ [1ex] 
 \hline
\end{tabular}
\label{table:SRF_values}
\end{table}

Therefore, what we gather from our above study is that the non-precessing waveform models are not accurate enough in detecting BBHs which possess relatively high precession even if one restrict the mass ratio in the range [1,3]. This implies that these waveform models need improvements which shall be done by including full two-spin precession effects as well as non-quadrupole modes in the model. The IMRPhenomPv2 has these features but it will require us to implement search strategies such as {\it sky-maxed-SNR} Ref.~\cite{Harryetal} that can handle the effect of non-quadrupole modes in the precessing GW signals.

\subsection{Results: NR waveform as injections}
\label{NR_inj}

\begin{figure*}[!ht]
\includegraphics[width=7.0in]{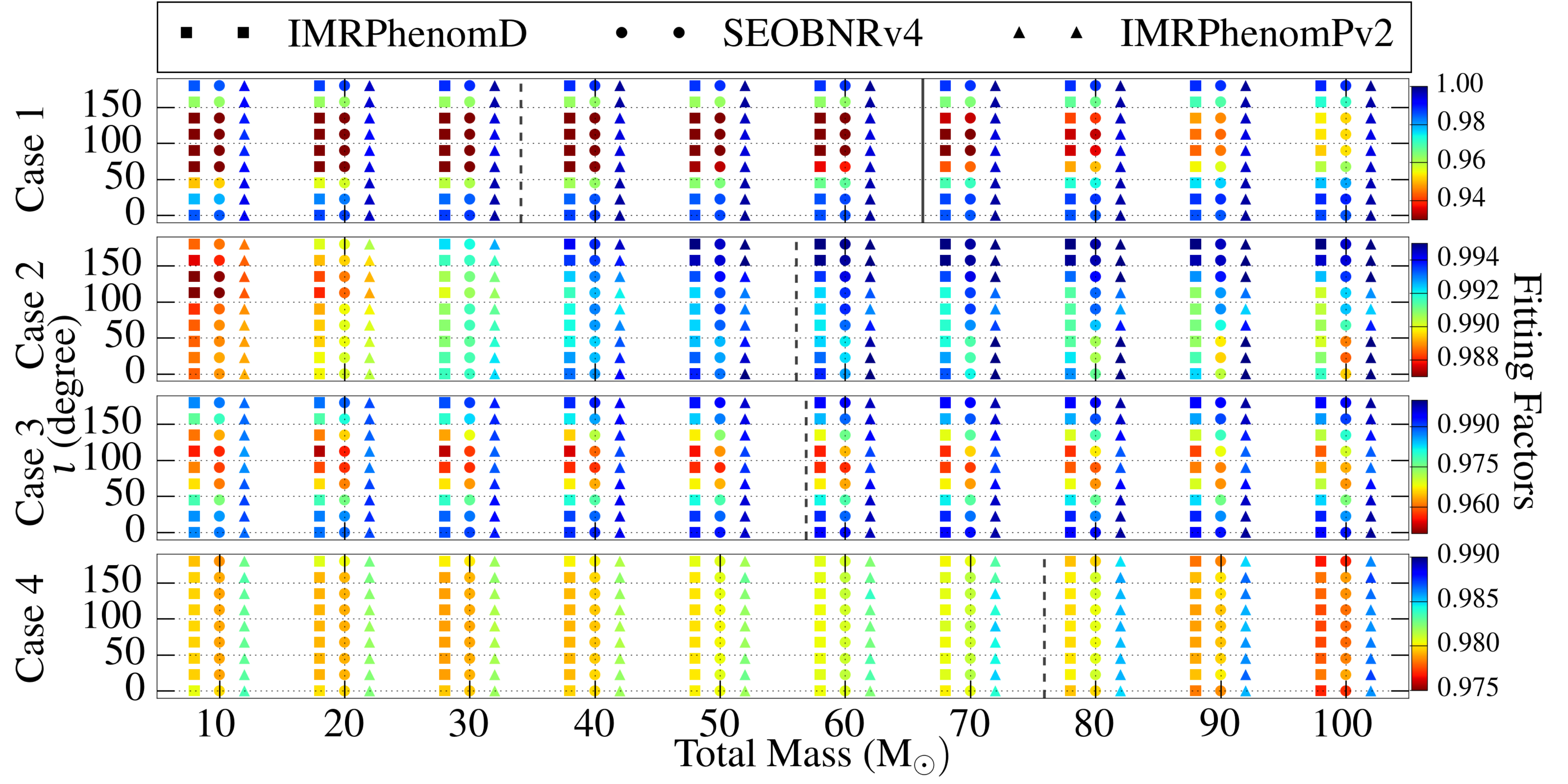}
\caption{Comparison of the performances of three template approximants, IMRPhenomD, SEOBNRv4, and IMRPhenomPv2, in recovering NR injections. $x-$ and $y-$axis represent the total mass (in $M_{\odot}$) and inclination angle (in degree) while FFs are given in colorbar. Different markers are for different approximants.  The solid vertical line  on the top panel for Case 1 represents the mass for which the lowest frequency of the NR waveform is equal to $10$ Hz. All the masses on the right hand side to this line have lowest frequency less than $10$ Hz and hence we loose a few GW cycles while calculating FF for these NR waveforms as we keep $f_{\rm low}$ to be $10$ Hz. Such a line for Cases 2, 3 and 4 exists for total mass greater than $100M_{\odot}$ and hence can not be seen in the plot. Similarly, the dashed vertical lines in all the four panels represent the total mass for which the lowest frequency of NR waveform is equal to $20$ Hz.}
\label{FF_NR}
\end{figure*}

In this section, we use our {\it four} NR waveforms that model GWs from fully precessing spin-orbit resonant binaries. Three of them are in $\Delta \phi= 0^{\circ}$ resonance (i.e., Cases 1, 3 and 4) while other one is in $\Delta \phi= 180^{\circ}$ resonance (i.e., Case 2). We scale these waveforms with total mass $M\in\{10, 20, 30, 40, 50, 60, 70, 80, 90, 100\} M_{\odot}$ and rotate them to inclination angles $\iota\in\{0^{\circ}, 22.5^{\circ}, 45^{\circ}, 67.5^{\circ}, 90^{\circ}, 112.5^{\circ}, 135^{\circ}, 157.5^{\circ}, 180^{\circ}\}$. We randomly choose a set of values for $\delta$, $\alpha$ and $\psi$ to assign the sky location and verified that the sky-maxed FF (in case of IMRPhenomPv2 approximant) is insensitive to this choice. We compute FF for each of these NR waveforms (total $4\times10\times9=360$ waveforms) using IMRPhenomD, SEOBNRv4 and IMRPhenomPv2 approximants. In this exercise, we do not generate any template bank but compute the FF by maximizing match over the template parameters using PSO algorithm as discussed in Sec.~\ref{Sec_IV}. 
Note that in this subsection, we use $f_{\rm low}$ different than $20$ Hz for FF calculations. For each of $360$ waveforms, the $f_{\rm low}$ is set equal to the lowest possible frequency possessed by the NR waveform. However, if the lowest allowed frequency for a given NR waveform is less than $10$ Hz, we set $f_{\rm low}$ to be $10$ Hz (e.g., masses $70 M_{\odot} - 100 M_{\odot}$ in Case 1 waveform).  

In Fig.~\ref{FF_NR}, we compare the performances of all three template approximants in recovering resonant NR injections as a function of inclination angle and total mass. As expected, for any given Case, the FF is lowest for $\iota =90^{\circ}$ (edge-on) and maximum for $\iota=\{0^{\circ}, 180^{\circ}\}$ (face-on/off). This is because for $\iota$ value other than $0^{\circ}$ and $180^{\circ}$ the contributions from sub-dominant modes become important which lead to additional mismatch between the NR and our dominant-mode template waveforms. Such trend between FF and $\iota$ is not visible for Case 4 as for this simulation, we could extract only the $(l, m)=(2, 2)$ mode. We find that for a given template approximant and inclination angle $\iota$, FF does not vary much with total mass for all of the NR waveforms. This is expected as the number of cycles in the NR waveforms are constant in the aLIGO frequency band we considered while computing the FF. However, if we see in the top panel of Fig.~\ref{FF_NR}, the FF increases with total mass for masses on the right hand side of the solid vertical line. This line represents the total mass above which the lowest frequency of NR waveform is less than $10$ Hz. As we compute FFs with $f_{\rm low}=10$ Hz for these masses, we are using fewer and fewer cycles in the NR waveform with increasing total mass. The fact that there are comparatively less cycles in the waveform for these higher masses, it leads to reduced precessional effects causing less cumulative phase mismatch which results in higher FF.  

We note that the IMRPhenomPv2 approximant performs slightly better in recovering all the resonant NR injections as compared to IMRPhenomD and SEOBNRv4, as expected. For example, the minimum FF (over all the masses and inclination angles) for IMRPhenomPv2 is $0.986$ whereas the same is $0.90$ and $0.899$ for IMRPhonomD and SEOBNRv4, respectively.   
We finally note that our FF results with NR injections are consistent with results obtained in previous subsection where we use SEOBNRv3 injections, as expected. 

In the next section, we study the systematic errors in estimating the intrinsic parameters of resonant binaries.

\begin{figure*}
\begin{center}

\centering
\includegraphics[width=7in]{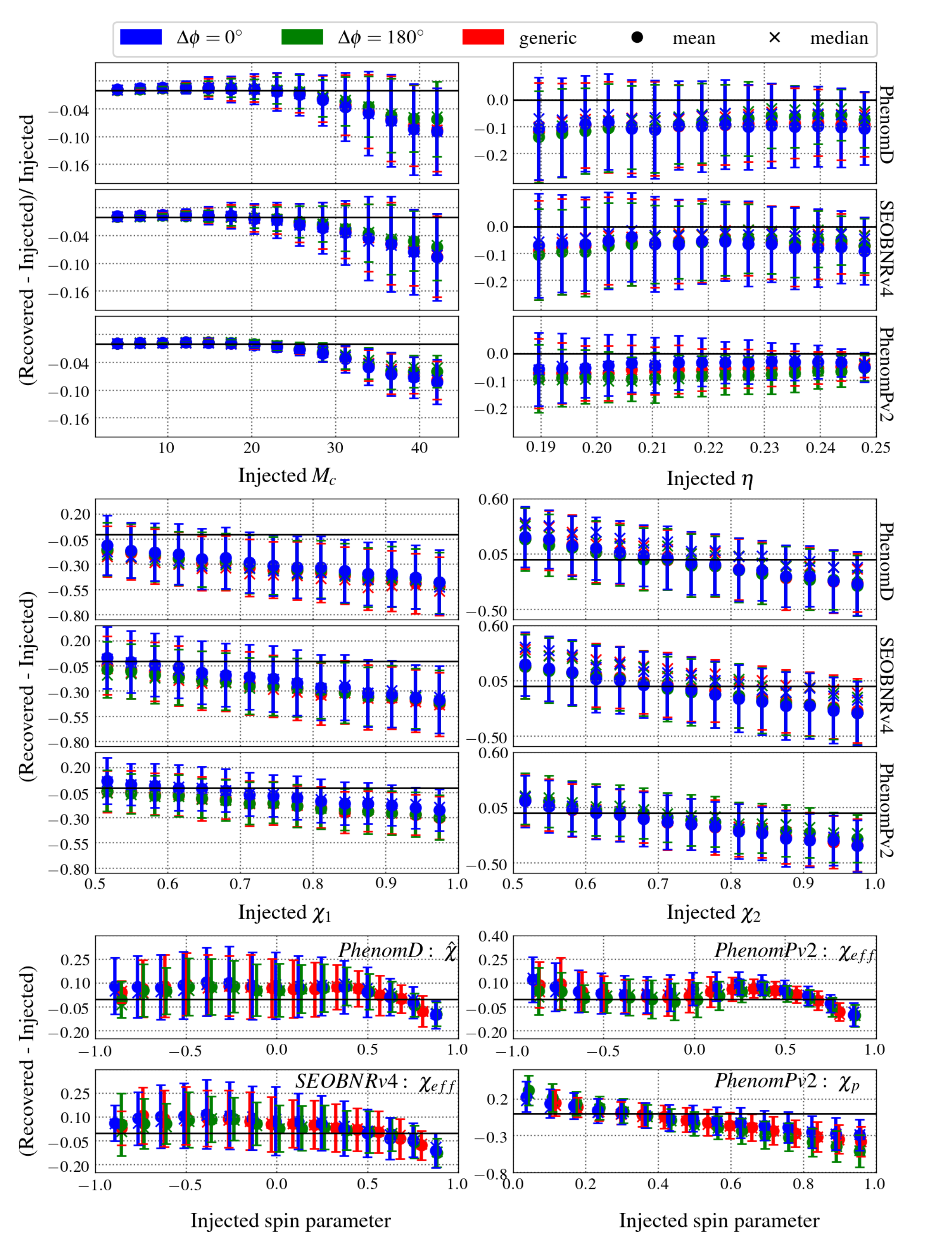}
\end{center}
 \caption{Systematic error in $M_c$, $\eta$,$\chi_1, \chi_2$, $\hat{\chi}$, $\chi_{\rm eff}$ and $\chi_{\rm p}$ for SEOBNRv3 injections. The quantities on the $x$-axis represent injected parameter values and are divided into equal bins. The mean (circle), standard deviations (error bars are symmetric with respect to mean) and median (cross) for systematic biases corresponding to each bin are plotted on the y-axis.}
\label{v3_delta_plots_baises}
\end{figure*}
\section{Systematic errors in the recovery of parameters for spin-orbit resonances}
\label{Sec_V}

For detection of GW signals frequency domain ROMs of time-domain approximants are used since these are computationally faster to generate. However, to determine the parameters of the source from its GW signal, a Bayesian parameter estimation analysis is employed, which sensitively depends on the accuracy of the waveform models that are being used. Current frequency domain waveforms are approximation to the true signals, they contain inaccuracies which show up as systematic errors in the parameters of the recovered waveform. In this section we present the systematic biases that we encounter while recovering source parameters from GW signals from resonant binaries.

Instead of looking at recovered component masses, $m_1$ and $m_2$, we use chirp mass $M_c$ and symmetric mass ratio $\eta$.  Moreover, we look for relevant spin parameters for each of the template waveforms, discussed in Sec.~\ref{Sec_III}, along with the spin of individual BHs. This is because all of our template approximants use either aligned and anti-aligned spins or some effective spin parameter to model spin and precession effects. For example, IMRPhenomPv2 apart from effective spin uses planar spin parameter $\chi_{\rm p}$ which represents components of the spins in the orbital plane to model precession. Note that we do not perform a detailed Bayesian parameter estimation study using various approximants as this is computationally too expensive and beyond the scope this paper. We will present the results for the Bayesian analysis in a future publication. 
 
We discuss results for specific kind of injection sets (SEOBNRv3 and NR) in the next subsections.

\subsection{Results: SEOBNRv3 as injections}
\label{PE_SEOBNRv3}

\begin{figure*}
\begin{center}
\includegraphics[width=7.0in]{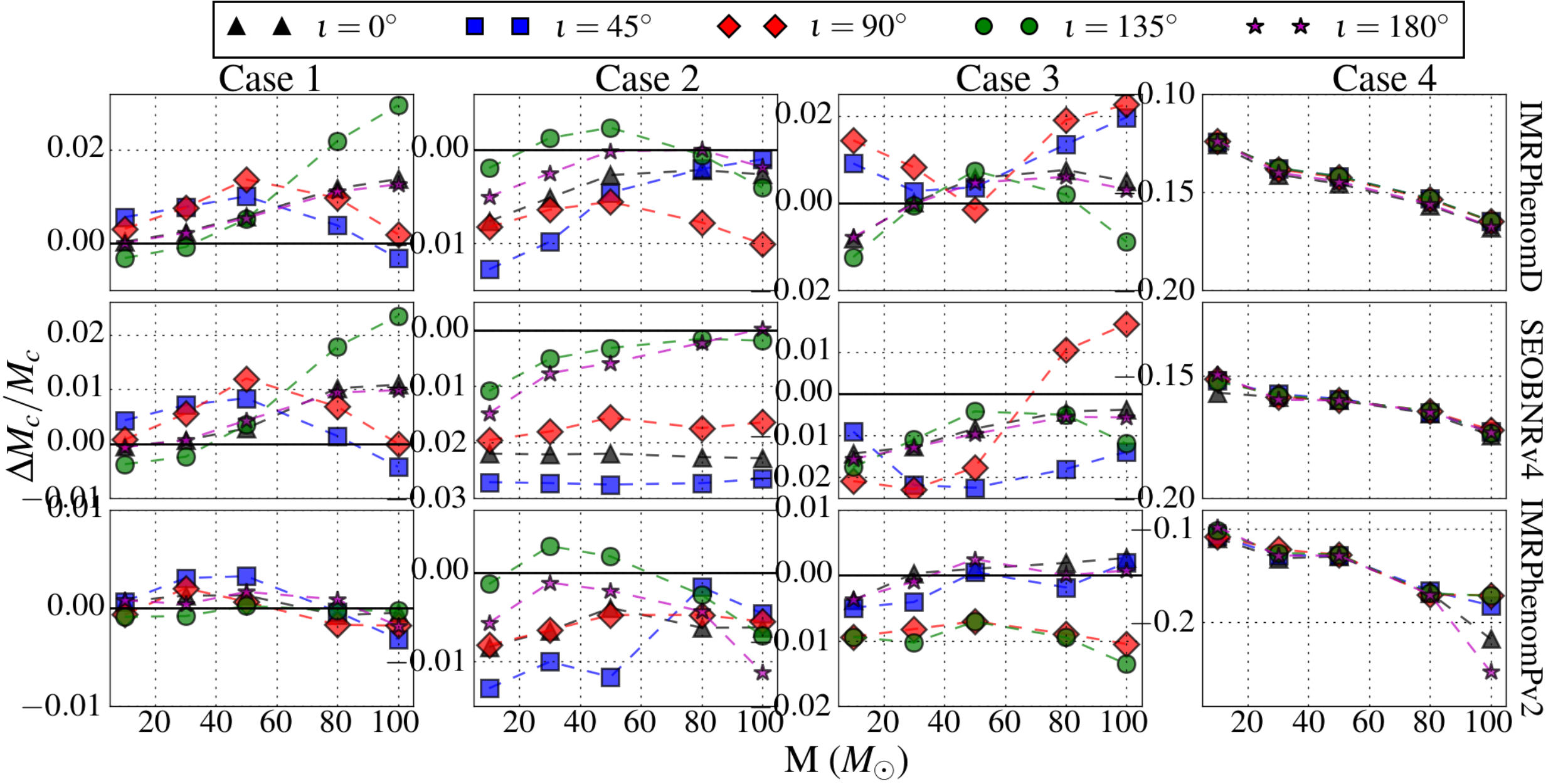} 
\end{center}
 \caption{Systematic biases in the estimation of chirp mass $M_c$  for the all the four NR injections while using IMRPhenomD, SEOBNRv4 and IMRPhenomPv2 templates as a function of total mass and inclination angle $\iota$.} 
\label{mchirp_NR}
\end{figure*}
\begin{figure*}
\begin{center}
\includegraphics[width=7.0in]{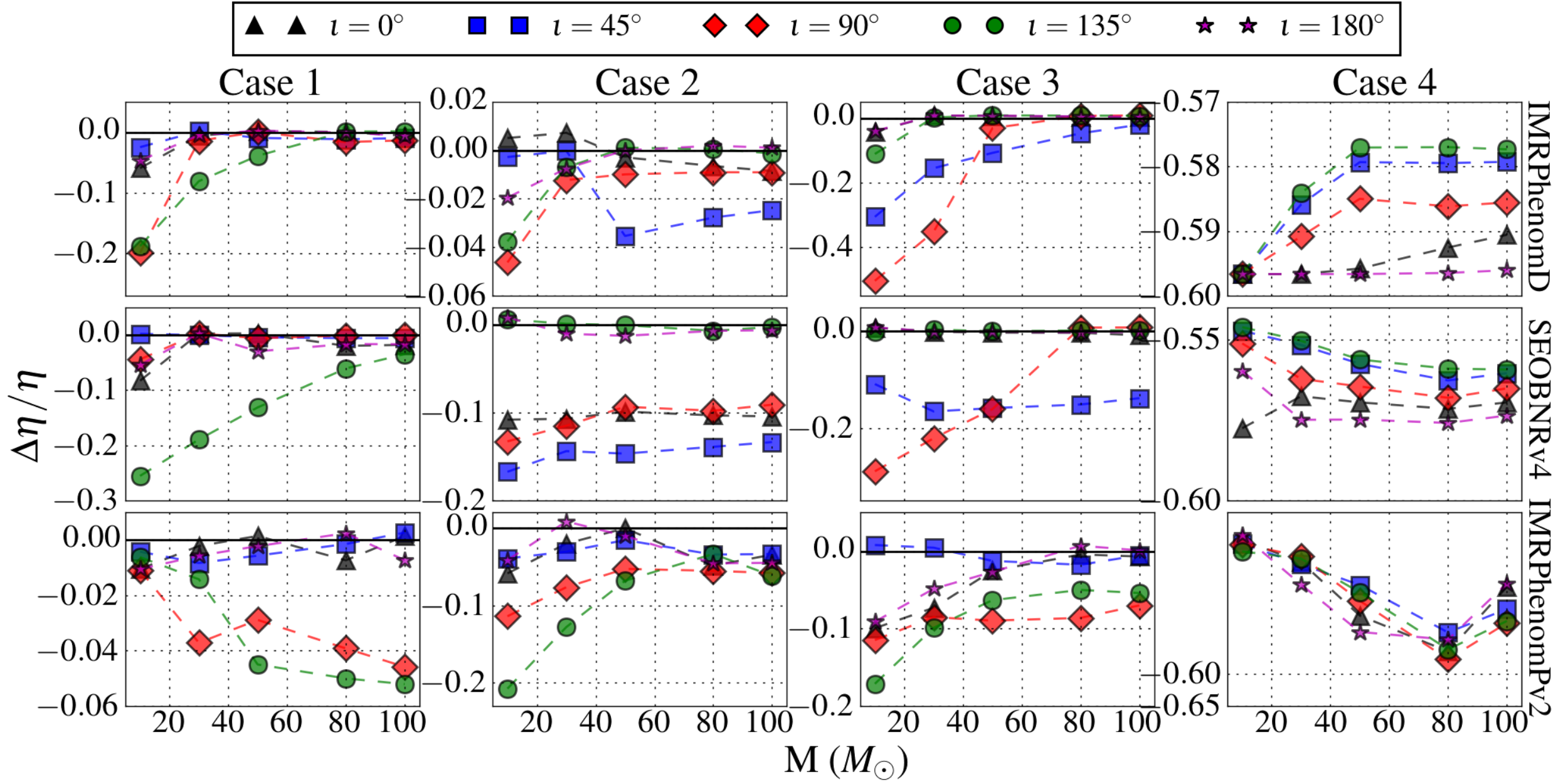} 
\end{center}
 \caption{Systematic biases in the estimation of symmetric mass ratio $\eta$ for the all the four NR injections while using IMRPhenomD, SEOBNRv4 and IMRPhenomPv2 templates as a function of total mass and inclination angle $\iota$.}
\label{Eta_NR}
\end{figure*}
\begin{figure*}
\begin{center}
\includegraphics[width=7in]{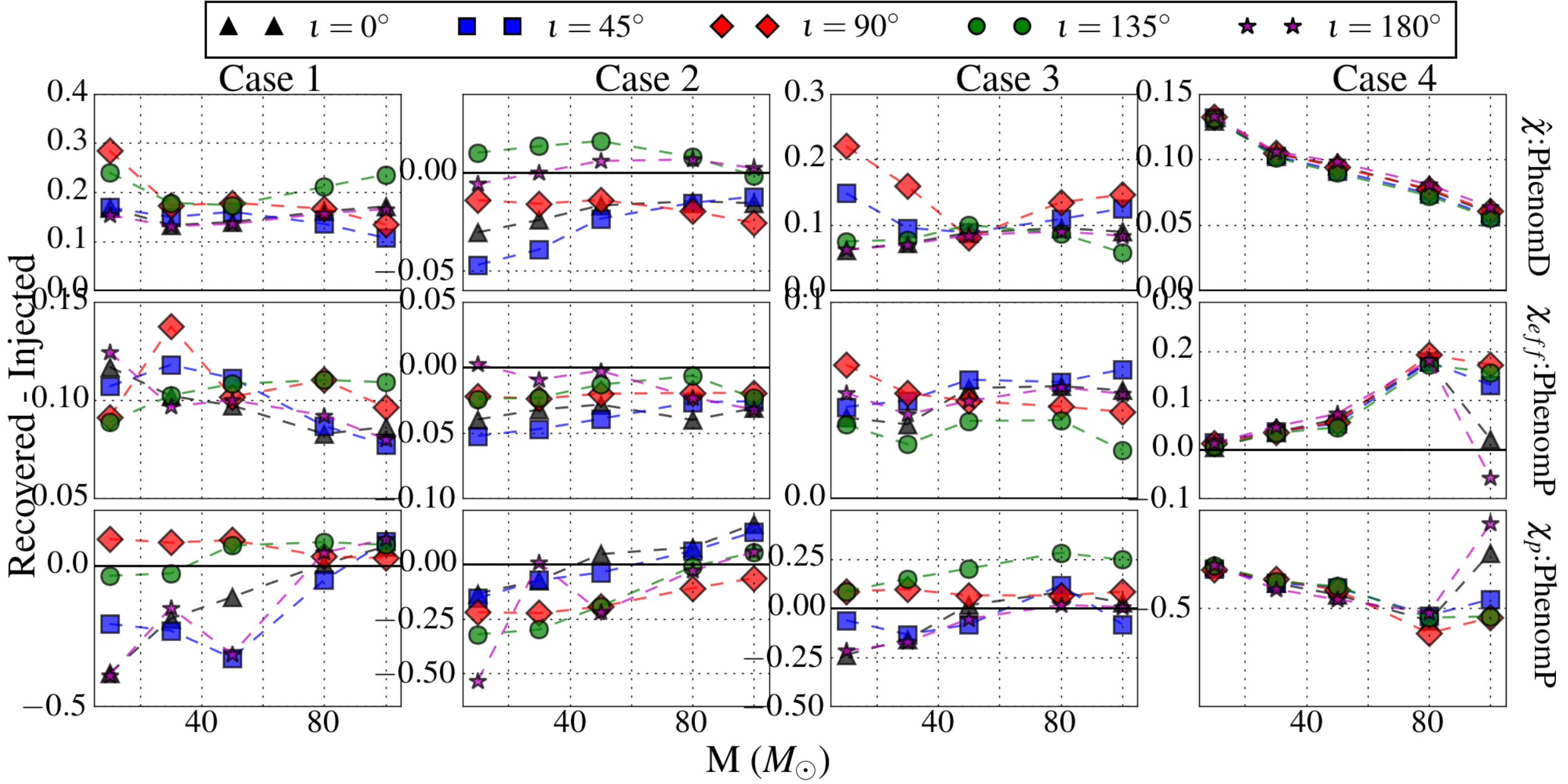} 
\end{center}
 \caption{Systematic biases in the estimation of spin parameters for the all the four NR injections while using IMRPhenomD and IMRPhenomPv2 templates as a function of total mass and inclination angle $\iota$. The spin components evolve with time and here in this figure the injected and recovered spin parameters are computed at $f_0$ (see Table~\ref{tab:Injection_NR}).}
\label{phenom_spin_NR}
\end{figure*}
\begin{figure*}
\begin{center}
\includegraphics[width=7in]{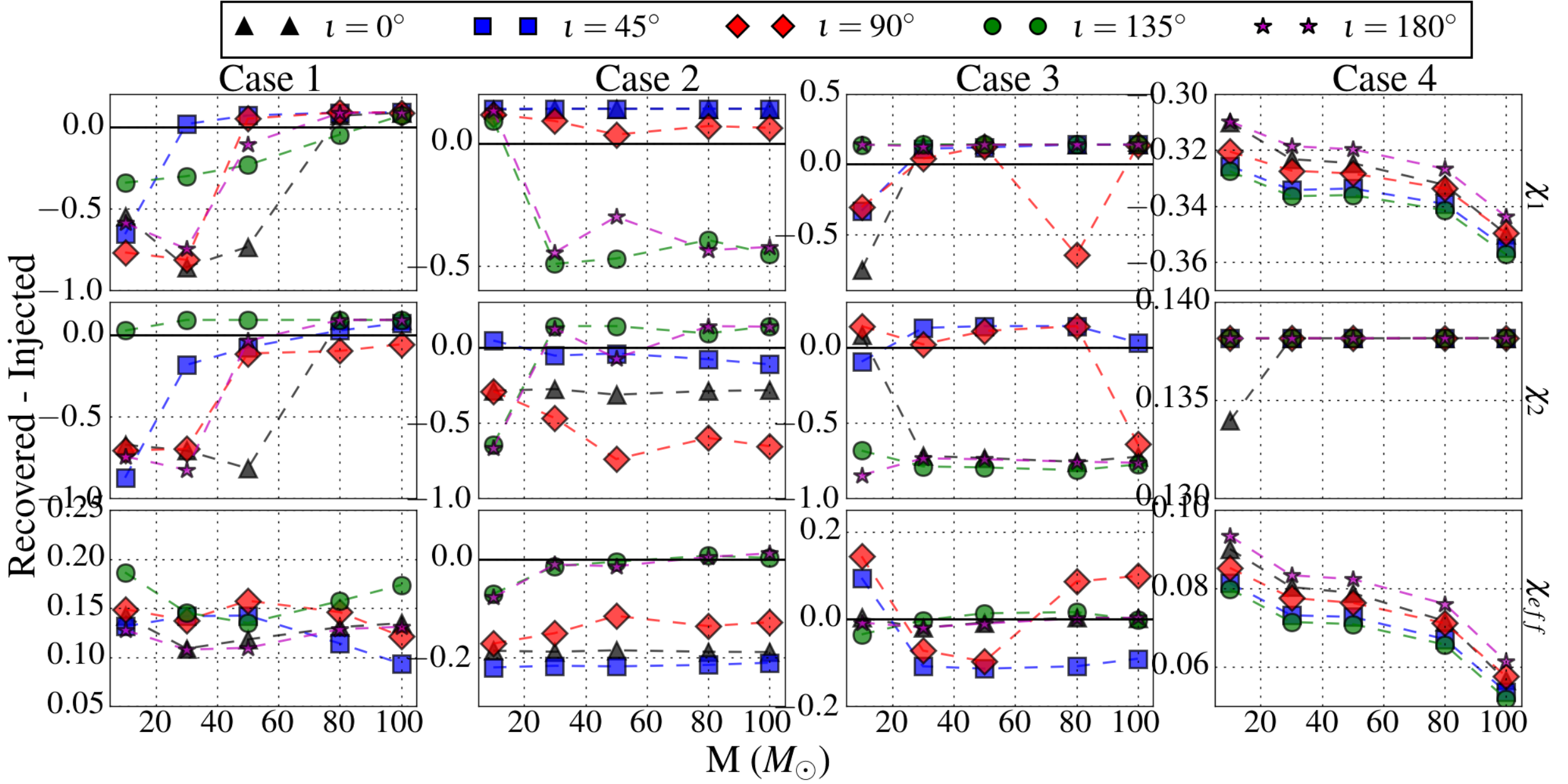} 
\end{center}
 \caption{Systematic biases in the estimation of spin parameters for the all the four NR injections while using SEOBNRv4 templates as a function of total mass and inclination angle $\iota$. The spin components evolve with time and here in this figure the injected and recovered spin parameters are computed at $f_0$ (see Table~\ref{tab:Injection_NR}).}
\label{EOB_spin_NR}
\end{figure*}
In this section, we present systematic errors in the estimation of binary parameters of resonant binaries as well as generic precessing binaries whose GWs are modeled using SEOBNRv3 approximant. Figure~\ref{v3_delta_plots_baises} shows biases in $M_{\rm c}$, $\eta$, $\chi_1$, $\chi_2$, $\hat{\chi}$, $\chi_{\rm eff}$ and $\chi_{\rm p}$. 
For the plots in this subsection, the quantities on the $x$-axis like $M_c$, $\eta$ are divided into equal bins; the mean (circle), standard deviations (error bars are symmetric with respect to mean) and median (cross) corresponding to each bin are plotted. This is so as to get a broad idea of the distributions of the systematic errors in the recovered parameters.
The chirp mass $M_{\rm c}$ is very well estimated with biases $<2\%$ for low mass binaries ($M_{\rm c}<25M_{\odot}$) whereas they can be as large as $10\%$ for high mass injections ($M_{\rm c}>25M_{\odot}$). This is expected as for high mass binaries, where merger and ringdown phases dominate, chirp mass is not a good parameter to model template waveform. Instead, it is the total mass which can be estimated with low errors. The $\Delta \phi = 180^{\circ}$ resonant binary injections have slightly smaller errors in $M_{\rm c}$ as compared to the other two types of injections. For example, the error in $M_{\rm c}$ for $\Delta \phi = 180^{\circ}$ injections lie in the range $\in [0\%, 6\%]$ while it is $\in [0\%, 8\%]$ for both $\Delta \phi = 0^{\circ}$ and generic precessing injections. In the top right panel of Fig.~\ref{v3_delta_plots_baises}, we present systematic errors in symmetric mass ratio $\eta$ and we see that $\eta$ is always underestimated. For high $\eta_{\rm injected}$, it is inevitable as $\eta$ is bounded in the range $[0, 0.25]$ and there is more room for the template to match with signal at lower $\eta$ side. Unlike in the case of chirp mass, biases in $\eta$ are not reduced for $\Delta \phi=180^{\circ}$ injections but increase or decrease depending upon the value of $\eta_{\rm injected}$. For the three types of injections the overall errors in $\eta$ lie in the range $\in [4\%, 15\%]$. In general, IMRPhenomPv2 approximant estimates mass parameters way better than IMRPhenomD and SEOBNRv4.  

In the middle panel of Fig.~\ref{v3_delta_plots_baises}, we plot absolute error in the spin magnitude of the individual BHs. As $\eta$, the spin magnitude of the heavier BH $\chi_1$ is mostly underestimated. The larger the injected spin magnitude, the more is the absolute error in $\chi_1$ which can be as high as $0.55$ for all three template approximants and all three types of injections. The spin magnitude of lighter BH $\chi_2$, on the other hand, is overestimated when injected $\chi_2$ is small and is underestimated when injected $\chi_2$ is high. The precessing IMRPhenomPv2 approximant gives smaller errors in the estimation of $\chi_1$ and $\chi_2$ as compared to the non-precessing approximants. The lower panel of Fig.~\ref{v3_delta_plots_baises} presents absolute error in the (respective) aligned spin and in-plane spin parameters of the template approximants. The effective spin parameters $\chi_{\rm eff}$ and $\hat{\chi}$ are always recovered with relatively positive values unless the injected $\chi_{\rm eff}/\hat{\chi}$ is close to $+1$. This means that a misaligned spin binary will be recovered with a relatively aligned spin binary and if the binary has very high spin magnitude ($\sim1$) and both the spins are aligned to the orbital angular momentum, we will be bias to recover it with either ``smaller magnitude aligned spin" or ``higher spin misaligned spin" binary. The absolute error in the measurement of $\chi_{\rm eff}/\hat{\chi}$ can be as high as $0.13$ for all the template approximants and all three types of injections. The in-plane spin parameter $\chi_p$ is mostly underestimated  implying that a precessing GW signal may be recovered as a relatively less precessing signal if we use IMRPhenomPv2 for parameter estimation of GW signals.  
 
In summary, what we gather from the above results is that our waveform models need to be made more accurate even in this restricted range of parameter space in order to make accurate measurements of the parameters of resonant binaries as well as generic precessing binaries.

\subsection{Results: NR waveforms as injections}
\label{PE_NR}
We compute error in the recovered parameters of the NR injections discussed
in Sec.~\ref{Sec_NR_inj} while using IMRPhenomD, SEOBNRv4 and IMRPhenomPv2 as our recovery templates. In all the figures in this subsection, we plot the error in the parameters like masses and spins for NR waveforms having total mass $M\in\{10, 30, 50, 80, 100\} M_{\odot}$ and inclination angles $\iota\in\{0^{\circ}, 45^{\circ}, 90^{\circ}, 135^{\circ}, 180^{\circ}\}$ for visual clarity in the figures. The four columns are for four NR spin-orbit resonance configurations whereas different markers correspond to different $\iota$ value. 

Figure~\ref{mchirp_NR} shows the error in chirp mass $M_c$. We can see that for first three NR waveforms (i.e., Case 1, 2 and 3), all the three approximants recover $M_c$ within $3\%$ or less. This is true for all masses in the range $[10, 100]$ $M_{\odot}$. However, in Case 4 the errors in $M_{\rm c}$ are relatively higher and it can be as large as $20\%$. The relatively large error in $M_c$ for Case 4 can be attributed to the non-quasi-circularity in the simulation (e.g., see discussion in Sec.~\ref{Sec_NR_inj}).  All the three (quasi-circular) template approximants yield FFs $>0.97$ for this (non-circular) NR simulation (see the last row in Fig.~\ref{FF_NR}) at the cost of larger error in chirp mass. We see that the error in $M_c$ varies with inclination angle for Case 1, 2 and 3 while it is insensitive to $\iota$ for Case 4, as expected.

Figure~\ref{Eta_NR} shows the error in symmetric mass ratio $\eta$. We can see that for first three NR waveforms (i.e., Case 1, 2 and 3), all the three approximants recover $\eta$ within $50\%$ or better. This is true for all masses in the range $[10, 100]$ $M_{\odot}$. The recovered $\eta$ is mostly less than injected value similar to SEOBNRv3 injections. On the other hand in Case 4 the errors $\eta$ are relatively large and it can be as large as $\sim 60\%$. As the recovered $\eta $ is always lower than the injected one for Case 4, it  indicates that a comparable mass eccentric binary emits GW signal similar to asymmetric mass circular binary. 

In Figs.~\ref{phenom_spin_NR} and ~\ref{EOB_spin_NR}, we present absolute error in various spin parameters with which our template approximants are characterized. As we can see in these figures, the absolute error in aligned-spin parameters ($\chi_{\rm eff}, \hat{\chi}$) can be as high as $1$ depending upon the inclination angle. However, the in-plane spin parameter $\chi_p$ is estimated relatively better with absolute error $<0.6$. 
Fig.~\ref{EOB_spin_NR} shows the absolute error in the individual spin magnitudes $\chi_{1}$ and $\chi_{2}$. We see for Cases 1, 2 and 3, the spin magnitudes are mostly underestimated. 

From Figs. \ref{v3_chip_inclination} and \ref{FF_NR}, we notice that the SRF/FF values are mostly symmetric about $\iota = 90^{\circ}$. This is expected as the SNR of the GW signal from a binary is not much affected under the transformation $\iota\rightarrow180^{\circ}-\iota$. As a result, we also found the biases in the recovery of parameters for SEOBNRv3 injections (discussed in the previous subsection) as well symmetric about $\iota = 90^{\circ}$. However, this trend is not prominent for NR injections in Figs.~\ref{mchirp_NR}, \ref{Eta_NR}, \ref{phenom_spin_NR}, and \ref{EOB_spin_NR}.  This is just because we have only a handful of NR injections: there are only five inclination cases for a given mass ratio and spins. We expect the NR injections to exhibit this trend as well if we would have more NR simulations covering a bit larger parameter space.
In Summary, we find that the errors in the parameters of NR injections (except for Case 4) are consistent with that of SEOBNRv3 injections. This is expected as Ref.~\cite{BTB2017} demonstrated the reliability of the model for nearly equal-mass binaries.

\section{Conclusions}
\label{sec:conclusions}
In this paper, we investigate the performance of currently available state-of-the-art waveforms in detecting and characterizing spin-orbit resonant binaries. Explicitly, we employ IMRPhemonD, SEOBNRv4 and IMRPhenomPv2 approximants as our templates and SEOBNRv3 as well as NR waveforms to simulate GW signals from resonant binaries. We find that the non-precessing template approximants IMRPhemonD and SEOBNRv4 recover more than $70\%$ of injections with FF$>0.97$ (or $90\%$ injections with FF$>0.9$). Such loss in SNR is mainly due to non-availability of higher harmonics in the template waveform models as well as missing physics regarding the precession in GW signals. For the precessing approximant IMRPhenomPv2, we computed `sky-maxed' FF by maximizing the match over $\delta$, $\alpha$, $\psi$ and $\iota$ as well and found that it performs impressively better than its non-precessing counterparts recovering $99\%$ of the injections with FF$>0.97$. Interestingly, $\Delta \phi =180^{\circ}$ resonant binaries yield higher FFs while $\Delta \phi =0^{\circ}$ resonant binaries obtain lower FFs as compared to generic precessing binaries. This is essentially because our template waveform models do not account for high precession and negative effective spins even in the comparable mass regime ($q\in[1,3]$). This means that aLIGO will miss a few more $\Delta \phi = 0^{\circ}$ resonant binaries compared to their $\Delta \phi =180^{\circ}$ counterparts if we use current non-precessing waveform approximants as search templates. We recommend to use better statistic such as sky-maxed FF and precessing waveform models such as IMRPhenomPv2 if we do not want to miss any of them. 

For low mass binaries the systematic error in chirp mass is $<2\%$  while it can go as high as $10\%$ for high mass binaries. The symmetric mass ratio is always recovered with value lower than the injected one and maximum error it can have is $15\%$. The aligned spin parameters $\chi_{\rm eff}/\hat{\chi}$ are mostly recovered with relatively positive values indicating that a misaligned spin binary can be recovered as aligned spin binary. The absolute error in the $\chi_{\rm eff}/\hat{\chi}$ measurement can be as high as $\sim 0.13$. The in-plane spin parameter $\chi_p$ is mostly underestimated implying that a precessing GW signal can be measured as a relatively less precessing signal if we use IMRPhenomPv2 for parameter estimation. The absolute error in the measurement of $\chi_p$ can be as high as $0.6$. The results from our investigation with NR waveforms are consistent with that of SEOBNRv3 results as expected. In summary, the current waveform models are just as good for aLIGO to detect most of the resonant binaries from both the families as they are at detecting generic precessing binaries. However, these models still need improvements to accurately estimate the parameters of such binaries. 

In future, we plan to perform a full Bayesian parameter estimation study to investigate the statistical errors in recovering physical parameters of resonant binaries. It will be worthwhile to expand the range of parameters like mass ratio ($q>3$) and spin magnitude ($\chi_{1,2}\in[0, 1]$) to see how well our template approximants perform in the detection and parameter estimation of generic precessing binaries. Such a study will point us towards of the region of parameter space where our template models need to be improved and the region where more NR simulations should be performed.

\section*{Acknowledgments}
We thank Archisman Ghosh for helpful discussion, and Richard O'Shaughnessy for providing valuable comments on the manuscript. CA thanks Inter-University Centre for Astronomy and Astrophysics, Pune for the hospitality, where most of the work is done. AG acknowledges support from SERB-NPDF grant (PDF/2015/000263), NSF grants AST-1716394 and AST-1708146, and the Charles E. Kaufman Foundation of The Pittsburgh Foundation. BG acknowledges the support of University Grants Commission (UGC), India. SM acknowledges support from the Department of Science \& Technology (DST), India provided under the Swarna Jayanti Fellowships scheme.
The work done by CSUF authors was supported in part by NSF PHY-1307489 and PHY-1606522 and the computations we did were done on ORCA, a cluster supported in part by NSF PHY-1429873, the Research Corporation, and Cal State Fullerton. We gratefully acknowledge support for this research at CITA from
NSERC of Canada, the Ontario Early Researcher Awards
Program, the Canada Research Chairs Program, and
the Canadian Institute for Advanced Research; at Caltech
from the Sherman Fairchild Foundation and NSF
grants PHY-1404569 and AST-1333520; and at Cornell from
the Sherman Fairchild Foundation and NSF grants PHY1306125
and AST-1333129. This research is benefited from a grant awarded to IUCAA by the  Navajbai Ratan Tata Trust (NRTT). We are grateful for computational resources provided by the Leonard E Parker Center for Gravitation, Cosmology and Astrophysics at University of Wisconsin-Milwaukee, the Max Planck Institute for Gravitational Physics (Albert Einstein Institute), Hannover, Inter-University Centre for Astronomy and Astrophysics, Pune and LIGO-Caltech. This paper has LIGO document number LIGO-P1500266.

\bibliography{SORMS}

\begin{thebibliography}{103}%
\makeatletter
\providecommand \@ifxundefined [1]{%
 \@ifx{#1\undefined}
}%
\providecommand \@ifnum [1]{%
 \ifnum #1\expandafter \@firstoftwo
 \else \expandafter \@secondoftwo
 \fi
}%
\providecommand \@ifx [1]{%
 \ifx #1\expandafter \@firstoftwo
 \else \expandafter \@secondoftwo
 \fi
}%
\providecommand \natexlab [1]{#1}%
\providecommand \enquote  [1]{``#1''}%
\providecommand \bibnamefont  [1]{#1}%
\providecommand \bibfnamefont [1]{#1}%
\providecommand \citenamefont [1]{#1}%
\providecommand \href@noop [0]{\@secondoftwo}%
\providecommand \href [0]{\begingroup \@sanitize@url \@href}%
\providecommand \@href[1]{\@@startlink{#1}\@@href}%
\providecommand \@@href[1]{\endgroup#1\@@endlink}%
\providecommand \@sanitize@url [0]{\catcode `\\12\catcode `\$12\catcode
  `\&12\catcode `\#12\catcode `\^12\catcode `\_12\catcode `\%12\relax}%
\providecommand \@@startlink[1]{}%
\providecommand \@@endlink[0]{}%
\providecommand \url  [0]{\begingroup\@sanitize@url \@url }%
\providecommand \@url [1]{\endgroup\@href {#1}{\urlprefix }}%
\providecommand \urlprefix  [0]{URL }%
\providecommand \Eprint [0]{\href }%
\providecommand \doibase [0]{http://dx.doi.org/}%
\providecommand \selectlanguage [0]{\@gobble}%
\providecommand \bibinfo  [0]{\@secondoftwo}%
\providecommand \bibfield  [0]{\@secondoftwo}%
\providecommand \translation [1]{[#1]}%
\providecommand \BibitemOpen [0]{}%
\providecommand \bibitemStop [0]{}%
\providecommand \bibitemNoStop [0]{.\EOS\space}%
\providecommand \EOS [0]{\spacefactor3000\relax}%
\providecommand \BibitemShut  [1]{\csname bibitem#1\endcsname}%
\let\auto@bib@innerbib\@empty
\bibitem [{\citenamefont {{Gerosa}}\ \emph {et~al.}(2013)\citenamefont
  {{Gerosa}}, \citenamefont {{Kesden}}, \citenamefont {{Berti}}, \citenamefont
  {{O'Shaughnessy}},\ and\ \citenamefont {{Sperhake}}}]{Gerosa13}%
  \BibitemOpen
  \bibfield  {author} {\bibinfo {author} {\bibfnamefont {D.}~\bibnamefont
  {{Gerosa}}}, \bibinfo {author} {\bibfnamefont {M.}~\bibnamefont {{Kesden}}},
  \bibinfo {author} {\bibfnamefont {E.}~\bibnamefont {{Berti}}}, \bibinfo
  {author} {\bibfnamefont {R.}~\bibnamefont {{O'Shaughnessy}}}, \ and\ \bibinfo
  {author} {\bibfnamefont {U.}~\bibnamefont {{Sperhake}}},\ }\href {\doibase
  10.1103/PhysRevD.87.104028} {\bibfield  {journal} {\bibinfo  {journal}
  {Physical Review D}\ }\textbf {\bibinfo {volume} {87}},\ \bibinfo {eid}
  {104028} (\bibinfo {year} {2013})},\ \Eprint {http://arxiv.org/abs/1302.4442}
  {arXiv:1302.4442 [gr-qc]} \BibitemShut {NoStop}%
\bibitem [{\citenamefont {{Abbott}}\ \emph
  {et~al.}(2016{\natexlab{a}})\citenamefont {{Abbott}}, \citenamefont
  {{Abbott}}, \citenamefont {{Abbott}}, \citenamefont {{Abernathy}},
  \citenamefont {{Acernese}}, \citenamefont {{Ackley}}, \citenamefont
  {{Adams}}, \citenamefont {{Adams}}, \citenamefont {{Addesso}}, \citenamefont
  {{Adhikari}},\ and\ \citenamefont {et~al.}}]{detection}%
  \BibitemOpen
  \bibfield  {author} {\bibinfo {author} {\bibfnamefont {B.~P.}\ \bibnamefont
  {{Abbott}}}, \bibinfo {author} {\bibfnamefont {R.}~\bibnamefont {{Abbott}}},
  \bibinfo {author} {\bibfnamefont {T.~D.}\ \bibnamefont {{Abbott}}}, \bibinfo
  {author} {\bibfnamefont {M.~R.}\ \bibnamefont {{Abernathy}}}, \bibinfo
  {author} {\bibfnamefont {F.}~\bibnamefont {{Acernese}}}, \bibinfo {author}
  {\bibfnamefont {K.}~\bibnamefont {{Ackley}}}, \bibinfo {author}
  {\bibfnamefont {C.}~\bibnamefont {{Adams}}}, \bibinfo {author} {\bibfnamefont
  {T.}~\bibnamefont {{Adams}}}, \bibinfo {author} {\bibfnamefont
  {P.}~\bibnamefont {{Addesso}}}, \bibinfo {author} {\bibfnamefont {R.~X.}\
  \bibnamefont {{Adhikari}}}, \ and\ \bibinfo {author} {\bibnamefont
  {et~al.}},\ }\href {\doibase 10.1103/PhysRevLett.116.061102} {\bibfield
  {journal} {\bibinfo  {journal} {Physical Review Letters}\ }\textbf {\bibinfo
  {volume} {116}},\ \bibinfo {eid} {061102} (\bibinfo {year}
  {2016}{\natexlab{a}})},\ \Eprint {http://arxiv.org/abs/1602.03837}
  {arXiv:1602.03837 [gr-qc]} \BibitemShut {NoStop}%
\bibitem [{\citenamefont {{Abbott}}\ \emph
  {et~al.}(2016{\natexlab{b}})\citenamefont {{Abbott}}, \citenamefont
  {{Abbott}}, \citenamefont {{Abbott}}, \citenamefont {{Abernathy}},
  \citenamefont {{Acernese}}, \citenamefont {{Ackley}}, \citenamefont
  {{Adams}}, \citenamefont {{Adams}}, \citenamefont {{Addesso}}, \citenamefont
  {{Adhikari}},\ and\ \citenamefont {et~al.}}]{GW151226}%
  \BibitemOpen
  \bibfield  {author} {\bibinfo {author} {\bibfnamefont {B.~P.}\ \bibnamefont
  {{Abbott}}}, \bibinfo {author} {\bibfnamefont {R.}~\bibnamefont {{Abbott}}},
  \bibinfo {author} {\bibfnamefont {T.~D.}\ \bibnamefont {{Abbott}}}, \bibinfo
  {author} {\bibfnamefont {M.~R.}\ \bibnamefont {{Abernathy}}}, \bibinfo
  {author} {\bibfnamefont {F.}~\bibnamefont {{Acernese}}}, \bibinfo {author}
  {\bibfnamefont {K.}~\bibnamefont {{Ackley}}}, \bibinfo {author}
  {\bibfnamefont {C.}~\bibnamefont {{Adams}}}, \bibinfo {author} {\bibfnamefont
  {T.}~\bibnamefont {{Adams}}}, \bibinfo {author} {\bibfnamefont
  {P.}~\bibnamefont {{Addesso}}}, \bibinfo {author} {\bibfnamefont {R.~X.}\
  \bibnamefont {{Adhikari}}}, \ and\ \bibinfo {author} {\bibnamefont
  {et~al.}},\ }\href {\doibase 10.1103/PhysRevLett.116.241103} {\bibfield
  {journal} {\bibinfo  {journal} {Physical Review Letters}\ }\textbf {\bibinfo
  {volume} {116}},\ \bibinfo {eid} {241103} (\bibinfo {year}
  {2016}{\natexlab{b}})},\ \Eprint {http://arxiv.org/abs/1606.04855}
  {arXiv:1606.04855 [gr-qc]} \BibitemShut {NoStop}%
\bibitem [{\citenamefont {{Abbott}}\ \emph {et~al.}(2017)\citenamefont
  {{Abbott}}, \citenamefont {{Abbott}}, \citenamefont {{Abbott}}, \citenamefont
  {{Acernese}}, \citenamefont {{Ackley}}, \citenamefont {{Adams}},
  \citenamefont {{Adams}}, \citenamefont {{Addesso}}, \citenamefont
  {{Adhikari}}, \citenamefont {{Adya}},\ and\ \citenamefont
  {et~al.}}]{GW170104}%
  \BibitemOpen
  \bibfield  {author} {\bibinfo {author} {\bibfnamefont {B.~P.}\ \bibnamefont
  {{Abbott}}}, \bibinfo {author} {\bibfnamefont {R.}~\bibnamefont {{Abbott}}},
  \bibinfo {author} {\bibfnamefont {T.~D.}\ \bibnamefont {{Abbott}}}, \bibinfo
  {author} {\bibfnamefont {F.}~\bibnamefont {{Acernese}}}, \bibinfo {author}
  {\bibfnamefont {K.}~\bibnamefont {{Ackley}}}, \bibinfo {author}
  {\bibfnamefont {C.}~\bibnamefont {{Adams}}}, \bibinfo {author} {\bibfnamefont
  {T.}~\bibnamefont {{Adams}}}, \bibinfo {author} {\bibfnamefont
  {P.}~\bibnamefont {{Addesso}}}, \bibinfo {author} {\bibfnamefont {R.~X.}\
  \bibnamefont {{Adhikari}}}, \bibinfo {author} {\bibfnamefont {V.~B.}\
  \bibnamefont {{Adya}}}, \ and\ \bibinfo {author} {\bibnamefont {et~al.}},\
  }\href {\doibase 10.1103/PhysRevLett.118.221101} {\bibfield  {journal}
  {\bibinfo  {journal} {Physical Review Letters}\ }\textbf {\bibinfo {volume}
  {118}},\ \bibinfo {eid} {221101} (\bibinfo {year} {2017})},\ \Eprint
  {http://arxiv.org/abs/1706.01812} {arXiv:1706.01812 [gr-qc]} \BibitemShut
  {NoStop}%
\bibitem [{\citenamefont {Abbott}\ \emph {et~al.}(2017)\citenamefont {Abbott}
  \emph {et~al.}}]{GW170608}%
  \BibitemOpen
  \bibfield  {author} {\bibinfo {author} {\bibfnamefont {B.~P.}\ \bibnamefont
  {Abbott}} \emph {et~al.} (\bibinfo {collaboration} {Virgo, LIGO
  Scientific}),\ }\href {\doibase 10.3847/2041-8213/aa9f0c} {\bibfield
  {journal} {\bibinfo  {journal} {Astrophys. J.}\ }\textbf {\bibinfo {volume}
  {851}},\ \bibinfo {pages} {L35} (\bibinfo {year} {2017})},\ \Eprint
  {http://arxiv.org/abs/1711.05578} {arXiv:1711.05578 [astro-ph.HE]}
  \BibitemShut {NoStop}%
\bibitem [{\citenamefont {{Abbott}}\ \emph
  {et~al.}(2017{\natexlab{a}})\citenamefont {{Abbott}}, \citenamefont
  {{Abbott}}, \citenamefont {{Abbott}}, \citenamefont {{Acernese}},
  \citenamefont {{Ackley}}, \citenamefont {{Adams}}, \citenamefont {{Adams}},
  \citenamefont {{Addesso}}, \citenamefont {{Adhikari}}, \citenamefont
  {{Adya}},\ and\ \citenamefont {et~al.}}]{GW170814}%
  \BibitemOpen
  \bibfield  {author} {\bibinfo {author} {\bibfnamefont {B.~P.}\ \bibnamefont
  {{Abbott}}}, \bibinfo {author} {\bibfnamefont {R.}~\bibnamefont {{Abbott}}},
  \bibinfo {author} {\bibfnamefont {T.~D.}\ \bibnamefont {{Abbott}}}, \bibinfo
  {author} {\bibfnamefont {F.}~\bibnamefont {{Acernese}}}, \bibinfo {author}
  {\bibfnamefont {K.}~\bibnamefont {{Ackley}}}, \bibinfo {author}
  {\bibfnamefont {C.}~\bibnamefont {{Adams}}}, \bibinfo {author} {\bibfnamefont
  {T.}~\bibnamefont {{Adams}}}, \bibinfo {author} {\bibfnamefont
  {P.}~\bibnamefont {{Addesso}}}, \bibinfo {author} {\bibfnamefont {R.~X.}\
  \bibnamefont {{Adhikari}}}, \bibinfo {author} {\bibfnamefont {V.~B.}\
  \bibnamefont {{Adya}}}, \ and\ \bibinfo {author} {\bibnamefont {et~al.}},\
  }\href {\doibase 10.1103/PhysRevLett.119.141101} {\bibfield  {journal}
  {\bibinfo  {journal} {Physical Review Letters}\ }\textbf {\bibinfo {volume}
  {119}},\ \bibinfo {eid} {141101} (\bibinfo {year} {2017}{\natexlab{a}})},\
  \Eprint {http://arxiv.org/abs/1709.09660} {arXiv:1709.09660 [gr-qc]}
  \BibitemShut {NoStop}%
\bibitem [{\citenamefont {{Acernese}}\ \emph {et~al.}(2015)\citenamefont
  {{Acernese}}, \citenamefont {{Agathos}}, \citenamefont {{Agatsuma}},
  \citenamefont {{Aisa}}, \citenamefont {{Allemandou}}, \citenamefont
  {{Allocca}}, \citenamefont {{Amarni}}, \citenamefont {{Astone}},
  \citenamefont {{Balestri}}, \citenamefont {{Ballardin}},\ and\ \citenamefont
  {et~al.}}]{virgo2}%
  \BibitemOpen
  \bibfield  {author} {\bibinfo {author} {\bibfnamefont {F.}~\bibnamefont
  {{Acernese}}}, \bibinfo {author} {\bibfnamefont {M.}~\bibnamefont
  {{Agathos}}}, \bibinfo {author} {\bibfnamefont {K.}~\bibnamefont
  {{Agatsuma}}}, \bibinfo {author} {\bibfnamefont {D.}~\bibnamefont {{Aisa}}},
  \bibinfo {author} {\bibfnamefont {N.}~\bibnamefont {{Allemandou}}}, \bibinfo
  {author} {\bibfnamefont {A.}~\bibnamefont {{Allocca}}}, \bibinfo {author}
  {\bibfnamefont {J.}~\bibnamefont {{Amarni}}}, \bibinfo {author}
  {\bibfnamefont {P.}~\bibnamefont {{Astone}}}, \bibinfo {author}
  {\bibfnamefont {G.}~\bibnamefont {{Balestri}}}, \bibinfo {author}
  {\bibfnamefont {G.}~\bibnamefont {{Ballardin}}}, \ and\ \bibinfo {author}
  {\bibnamefont {et~al.}},\ }\href {\doibase 10.1088/0264-9381/32/2/024001}
  {\bibfield  {journal} {\bibinfo  {journal} {Classical and Quantum Gravity}\
  }\textbf {\bibinfo {volume} {32}},\ \bibinfo {eid} {024001} (\bibinfo {year}
  {2015})},\ \Eprint {http://arxiv.org/abs/1408.3978} {arXiv:1408.3978 [gr-qc]}
  \BibitemShut {NoStop}%
\bibitem [{\citenamefont {{Abbott}}\ \emph
  {et~al.}(2017{\natexlab{b}})\citenamefont {{Abbott}}, \citenamefont
  {{Abbott}}, \citenamefont {{Abbott}}, \citenamefont {{Acernese}},
  \citenamefont {{Ackley}}, \citenamefont {{Adams}}, \citenamefont {{Adams}},
  \citenamefont {{Addesso}}, \citenamefont {{Adhikari}}, \citenamefont
  {{Adya}},\ and\ \citenamefont {et~al.}}]{GW170817}%
  \BibitemOpen
  \bibfield  {author} {\bibinfo {author} {\bibfnamefont {B.~P.}\ \bibnamefont
  {{Abbott}}}, \bibinfo {author} {\bibfnamefont {R.}~\bibnamefont {{Abbott}}},
  \bibinfo {author} {\bibfnamefont {T.~D.}\ \bibnamefont {{Abbott}}}, \bibinfo
  {author} {\bibfnamefont {F.}~\bibnamefont {{Acernese}}}, \bibinfo {author}
  {\bibfnamefont {K.}~\bibnamefont {{Ackley}}}, \bibinfo {author}
  {\bibfnamefont {C.}~\bibnamefont {{Adams}}}, \bibinfo {author} {\bibfnamefont
  {T.}~\bibnamefont {{Adams}}}, \bibinfo {author} {\bibfnamefont
  {P.}~\bibnamefont {{Addesso}}}, \bibinfo {author} {\bibfnamefont {R.~X.}\
  \bibnamefont {{Adhikari}}}, \bibinfo {author} {\bibfnamefont {V.~B.}\
  \bibnamefont {{Adya}}}, \ and\ \bibinfo {author} {\bibnamefont {et~al.}},\
  }\href {\doibase 10.1103/PhysRevLett.119.161101} {\bibfield  {journal}
  {\bibinfo  {journal} {Physical Review Letters}\ }\textbf {\bibinfo {volume}
  {119}},\ \bibinfo {eid} {161101} (\bibinfo {year} {2017}{\natexlab{b}})},\
  \Eprint {http://arxiv.org/abs/1710.05832} {arXiv:1710.05832 [gr-qc]}
  \BibitemShut {NoStop}%
\bibitem [{\citenamefont {Somiya}(2012)}]{KS11}%
  \BibitemOpen
  \bibfield  {author} {\bibinfo {author} {\bibfnamefont {K.}~\bibnamefont
  {Somiya}},\ }\href {http://stacks.iop.org/0264-9381/29/i=12/a=124007}
  {\bibfield  {journal} {\bibinfo  {journal} {Classical and Quantum Gravity}\
  }\textbf {\bibinfo {volume} {29}},\ \bibinfo {pages} {124007} (\bibinfo
  {year} {2012})}\BibitemShut {NoStop}%
\bibitem [{\citenamefont {Unnikrishnan}(2013)}]{LIGO_I_UNNI}%
  \BibitemOpen
  \bibfield  {author} {\bibinfo {author} {\bibfnamefont {C.~S.}\ \bibnamefont
  {Unnikrishnan}},\ }\href {\doibase 10.1142/S0218271813410101} {\bibfield
  {journal} {\bibinfo  {journal} {International Journal of Modern Physics D}\
  }\textbf {\bibinfo {volume} {22}},\ \bibinfo {pages} {1341010} (\bibinfo
  {year} {2013})}\BibitemShut {NoStop}%
\bibitem [{\citenamefont {Sathyaprakash}\ and\ \citenamefont
  {Dhurandhar}(1991)}]{SathyaDhurandhar1991}%
  \BibitemOpen
  \bibfield  {author} {\bibinfo {author} {\bibfnamefont {B.~S.}\ \bibnamefont
  {Sathyaprakash}}\ and\ \bibinfo {author} {\bibfnamefont {S.~V.}\ \bibnamefont
  {Dhurandhar}},\ }\href {\doibase 10.1103/PhysRevD.44.3819} {\bibfield
  {journal} {\bibinfo  {journal} {Phys. Rev. D}\ }\textbf {\bibinfo {volume}
  {44}},\ \bibinfo {pages} {3819} (\bibinfo {year} {1991})}\BibitemShut
  {NoStop}%
\bibitem [{\citenamefont {Dhurandhar}\ and\ \citenamefont
  {Sathyaprakash}(1994)}]{SathyaDhurandhar1994}%
  \BibitemOpen
  \bibfield  {author} {\bibinfo {author} {\bibfnamefont {S.~V.}\ \bibnamefont
  {Dhurandhar}}\ and\ \bibinfo {author} {\bibfnamefont {B.~S.}\ \bibnamefont
  {Sathyaprakash}},\ }\href {\doibase 10.1103/PhysRevD.49.1707} {\bibfield
  {journal} {\bibinfo  {journal} {Phys. Rev.}\ }\textbf {\bibinfo {volume}
  {D49}},\ \bibinfo {pages} {1707} (\bibinfo {year} {1994})}\BibitemShut
  {NoStop}%
\bibitem [{\citenamefont {Shibata}\ and\ \citenamefont {Taniguchi}(2011)}]{NR}%
  \BibitemOpen
  \bibfield  {author} {\bibinfo {author} {\bibfnamefont {M.}~\bibnamefont
  {Shibata}}\ and\ \bibinfo {author} {\bibfnamefont {K.}~\bibnamefont
  {Taniguchi}},\ }\href {\doibase 10.1007/lrr-2011-6} {\bibfield  {journal}
  {\bibinfo  {journal} {Living Reviews in Relativity}\ }\textbf {\bibinfo
  {volume} {14}} (\bibinfo {year} {2011}),\ 10.1007/lrr-2011-6}\BibitemShut
  {NoStop}%
\bibitem [{\citenamefont {Ajith}\ \emph {et~al.}(2007)\citenamefont {Ajith},
  \citenamefont {Babak}, \citenamefont {Chen}, \citenamefont {Hewitson},
  \citenamefont {Krishnan}, \citenamefont {Whelan}, \citenamefont
  {BrÃ¼gmann}, \citenamefont {Diener}, \citenamefont {Gonzalez},
  \citenamefont {Hannam}, \citenamefont {Husa}, \citenamefont {Koppitz},
  \citenamefont {Pollney}, \citenamefont {Rezzolla}, \citenamefont
  {SantamarÃ­a}, \citenamefont {Sintes}, \citenamefont {Sperhake},\ and\
  \citenamefont {Thornburg}}]{Ajith2007}%
  \BibitemOpen
  \bibfield  {author} {\bibinfo {author} {\bibfnamefont {P.}~\bibnamefont
  {Ajith}}, \bibinfo {author} {\bibfnamefont {S.}~\bibnamefont {Babak}},
  \bibinfo {author} {\bibfnamefont {Y.}~\bibnamefont {Chen}}, \bibinfo {author}
  {\bibfnamefont {M.}~\bibnamefont {Hewitson}}, \bibinfo {author}
  {\bibfnamefont {B.}~\bibnamefont {Krishnan}}, \bibinfo {author}
  {\bibfnamefont {J.~T.}\ \bibnamefont {Whelan}}, \bibinfo {author}
  {\bibfnamefont {B.}~\bibnamefont {BrÃ¼gmann}}, \bibinfo {author}
  {\bibfnamefont {P.}~\bibnamefont {Diener}}, \bibinfo {author} {\bibfnamefont
  {J.}~\bibnamefont {Gonzalez}}, \bibinfo {author} {\bibfnamefont
  {M.}~\bibnamefont {Hannam}}, \bibinfo {author} {\bibfnamefont
  {S.}~\bibnamefont {Husa}}, \bibinfo {author} {\bibfnamefont {M.}~\bibnamefont
  {Koppitz}}, \bibinfo {author} {\bibfnamefont {D.}~\bibnamefont {Pollney}},
  \bibinfo {author} {\bibfnamefont {L.}~\bibnamefont {Rezzolla}}, \bibinfo
  {author} {\bibfnamefont {L.}~\bibnamefont {SantamarÃ­a}}, \bibinfo {author}
  {\bibfnamefont {A.~M.}\ \bibnamefont {Sintes}}, \bibinfo {author}
  {\bibfnamefont {U.}~\bibnamefont {Sperhake}}, \ and\ \bibinfo {author}
  {\bibfnamefont {J.}~\bibnamefont {Thornburg}},\ }\href
  {http://stacks.iop.org/0264-9381/24/i=19/a=S31} {\bibfield  {journal}
  {\bibinfo  {journal} {Classical and Quantum Gravity}\ }\textbf {\bibinfo
  {volume} {24}},\ \bibinfo {pages} {S689} (\bibinfo {year}
  {2007})}\BibitemShut {NoStop}%
\bibitem [{\citenamefont {{The LIGO Scientific Collaboration}}\ \emph
  {et~al.}(2016{\natexlab{a}})\citenamefont {{The LIGO Scientific
  Collaboration}}, \citenamefont {{the Virgo Collaboration}}, \citenamefont
  {{Abbott}}, \citenamefont {{Abbott}}, \citenamefont {{Abbott}}, \citenamefont
  {{Abernathy}}, \citenamefont {{Acernese}}, \citenamefont {{Ackley}},
  \citenamefont {{Adams}}, \citenamefont {{Adams}},\ and\ \citenamefont
  {et~al.}}]{O1BBH}%
  \BibitemOpen
  \bibfield  {author} {\bibinfo {author} {\bibnamefont {{The LIGO Scientific
  Collaboration}}}, \bibinfo {author} {\bibnamefont {{the Virgo
  Collaboration}}}, \bibinfo {author} {\bibfnamefont {B.~P.}\ \bibnamefont
  {{Abbott}}}, \bibinfo {author} {\bibfnamefont {R.}~\bibnamefont {{Abbott}}},
  \bibinfo {author} {\bibfnamefont {T.~D.}\ \bibnamefont {{Abbott}}}, \bibinfo
  {author} {\bibfnamefont {M.~R.}\ \bibnamefont {{Abernathy}}}, \bibinfo
  {author} {\bibfnamefont {F.}~\bibnamefont {{Acernese}}}, \bibinfo {author}
  {\bibfnamefont {K.}~\bibnamefont {{Ackley}}}, \bibinfo {author}
  {\bibfnamefont {C.}~\bibnamefont {{Adams}}}, \bibinfo {author} {\bibfnamefont
  {T.}~\bibnamefont {{Adams}}}, \ and\ \bibinfo {author} {\bibnamefont
  {et~al.}},\ }\href@noop {} {\bibfield  {journal} {\bibinfo  {journal} {ArXiv
  e-prints}\ } (\bibinfo {year} {2016}{\natexlab{a}})},\ \Eprint
  {http://arxiv.org/abs/1606.04856} {arXiv:1606.04856 [gr-qc]} \BibitemShut
  {NoStop}%
\bibitem [{\citenamefont {{Schnittman}}(2004)}]{JS}%
  \BibitemOpen
  \bibfield  {author} {\bibinfo {author} {\bibfnamefont {J.~D.}\ \bibnamefont
  {{Schnittman}}},\ }\href {\doibase 10.1103/PhysRevD.70.124020} {\bibfield
  {journal} {\bibinfo  {journal} {Physical Review D}\ }\textbf {\bibinfo
  {volume} {70}},\ \bibinfo {eid} {124020} (\bibinfo {year} {2004})},\ \Eprint
  {http://arxiv.org/abs/astro-ph/0409174} {astro-ph/0409174} \BibitemShut
  {NoStop}%
\bibitem [{\citenamefont {{Kesden}}\ \emph {et~al.}(2010)\citenamefont
  {{Kesden}}, \citenamefont {{Sperhake}},\ and\ \citenamefont
  {{Berti}}}]{KSB09}%
  \BibitemOpen
  \bibfield  {author} {\bibinfo {author} {\bibfnamefont {M.}~\bibnamefont
  {{Kesden}}}, \bibinfo {author} {\bibfnamefont {U.}~\bibnamefont
  {{Sperhake}}}, \ and\ \bibinfo {author} {\bibfnamefont {E.}~\bibnamefont
  {{Berti}}},\ }\href {\doibase 10.1088/0004-637X/715/2/1006} {\bibfield
  {journal} {\bibinfo  {journal} {The Astrophysical Journal}\ }\textbf
  {\bibinfo {volume} {715}},\ \bibinfo {pages} {1006} (\bibinfo {year}
  {2010})},\ \Eprint {http://arxiv.org/abs/1003.4993} {arXiv:1003.4993
  [astro-ph.CO]} \BibitemShut {NoStop}%
\bibitem [{\citenamefont {{Berti}}\ \emph {et~al.}(2012)\citenamefont
  {{Berti}}, \citenamefont {{Kesden}},\ and\ \citenamefont
  {{Sperhake}}}]{BKS12}%
  \BibitemOpen
  \bibfield  {author} {\bibinfo {author} {\bibfnamefont {E.}~\bibnamefont
  {{Berti}}}, \bibinfo {author} {\bibfnamefont {M.}~\bibnamefont {{Kesden}}}, \
  and\ \bibinfo {author} {\bibfnamefont {U.}~\bibnamefont {{Sperhake}}},\
  }\href {\doibase 10.1103/PhysRevD.85.124049} {\bibfield  {journal} {\bibinfo
  {journal} {Physical Review D}\ }\textbf {\bibinfo {volume} {85}},\ \bibinfo
  {eid} {124049} (\bibinfo {year} {2012})},\ \Eprint
  {http://arxiv.org/abs/1203.2920} {arXiv:1203.2920 [astro-ph.HE]} \BibitemShut
  {NoStop}%
\bibitem [{\citenamefont {Dal~Canton}\ \emph {et~al.}(2015)\citenamefont
  {Dal~Canton}, \citenamefont {Lundgren},\ and\ \citenamefont
  {Nielsen}}]{Tito}%
  \BibitemOpen
  \bibfield  {author} {\bibinfo {author} {\bibfnamefont {T.}~\bibnamefont
  {Dal~Canton}}, \bibinfo {author} {\bibfnamefont {A.~P.}\ \bibnamefont
  {Lundgren}}, \ and\ \bibinfo {author} {\bibfnamefont {A.~B.}\ \bibnamefont
  {Nielsen}},\ }\href {\doibase 10.1103/PhysRevD.91.062010} {\bibfield
  {journal} {\bibinfo  {journal} {Phys. Rev. D}\ }\textbf {\bibinfo {volume}
  {91}},\ \bibinfo {pages} {062010} (\bibinfo {year} {2015})}\BibitemShut
  {NoStop}%
\bibitem [{\citenamefont {Harry}\ \emph {et~al.}(2014)\citenamefont {Harry},
  \citenamefont {Nitz}, \citenamefont {Brown}, \citenamefont {Lundgren},
  \citenamefont {Ochsner},\ and\ \citenamefont {Keppel}}]{Ian}%
  \BibitemOpen
  \bibfield  {author} {\bibinfo {author} {\bibfnamefont {I.~W.}\ \bibnamefont
  {Harry}}, \bibinfo {author} {\bibfnamefont {A.~H.}\ \bibnamefont {Nitz}},
  \bibinfo {author} {\bibfnamefont {D.~A.}\ \bibnamefont {Brown}}, \bibinfo
  {author} {\bibfnamefont {A.~P.}\ \bibnamefont {Lundgren}}, \bibinfo {author}
  {\bibfnamefont {E.}~\bibnamefont {Ochsner}}, \ and\ \bibinfo {author}
  {\bibfnamefont {D.}~\bibnamefont {Keppel}},\ }\href {\doibase
  10.1103/PhysRevD.89.024010} {\bibfield  {journal} {\bibinfo  {journal} {Phys.
  Rev. D}\ }\textbf {\bibinfo {volume} {89}},\ \bibinfo {pages} {024010}
  (\bibinfo {year} {2014})}\BibitemShut {NoStop}%
\bibitem [{\citenamefont {Ajith}\ \emph {et~al.}(2011)\citenamefont {Ajith},
  \citenamefont {Hannam}, \citenamefont {Husa}, \citenamefont {Chen},
  \citenamefont {Br\"ugmann}, \citenamefont {Dorband}, \citenamefont
  {M\"uller}, \citenamefont {Ohme}, \citenamefont {Pollney}, \citenamefont
  {Reisswig}, \citenamefont {Santamar\'{\i}a},\ and\ \citenamefont
  {Seiler}}]{Ajith_prl}%
  \BibitemOpen
  \bibfield  {author} {\bibinfo {author} {\bibfnamefont {P.}~\bibnamefont
  {Ajith}}, \bibinfo {author} {\bibfnamefont {M.}~\bibnamefont {Hannam}},
  \bibinfo {author} {\bibfnamefont {S.}~\bibnamefont {Husa}}, \bibinfo {author}
  {\bibfnamefont {Y.}~\bibnamefont {Chen}}, \bibinfo {author} {\bibfnamefont
  {B.}~\bibnamefont {Br\"ugmann}}, \bibinfo {author} {\bibfnamefont
  {N.}~\bibnamefont {Dorband}}, \bibinfo {author} {\bibfnamefont
  {D.}~\bibnamefont {M\"uller}}, \bibinfo {author} {\bibfnamefont
  {F.}~\bibnamefont {Ohme}}, \bibinfo {author} {\bibfnamefont {D.}~\bibnamefont
  {Pollney}}, \bibinfo {author} {\bibfnamefont {C.}~\bibnamefont {Reisswig}},
  \bibinfo {author} {\bibfnamefont {L.}~\bibnamefont {Santamar\'{\i}a}}, \ and\
  \bibinfo {author} {\bibfnamefont {J.}~\bibnamefont {Seiler}},\ }\href
  {\doibase 10.1103/PhysRevLett.106.241101} {\bibfield  {journal} {\bibinfo
  {journal} {Phys. Rev. Lett.}\ }\textbf {\bibinfo {volume} {106}},\ \bibinfo
  {pages} {241101} (\bibinfo {year} {2011})}\BibitemShut {NoStop}%
\bibitem [{\citenamefont {Santamar\'{\i}a}\ \emph {et~al.}(2010)\citenamefont
  {Santamar\'{\i}a}, \citenamefont {Ohme}, \citenamefont {Ajith}, \citenamefont
  {Br\"ugmann}, \citenamefont {Dorband}, \citenamefont {Hannam}, \citenamefont
  {Husa}, \citenamefont {M\"osta}, \citenamefont {Pollney}, \citenamefont
  {Reisswig}, \citenamefont {Robinson}, \citenamefont {Seiler},\ and\
  \citenamefont {Krishnan}}]{Santamaria:2010yb}%
  \BibitemOpen
  \bibfield  {author} {\bibinfo {author} {\bibfnamefont {L.}~\bibnamefont
  {Santamar\'{\i}a}}, \bibinfo {author} {\bibfnamefont {F.}~\bibnamefont
  {Ohme}}, \bibinfo {author} {\bibfnamefont {P.}~\bibnamefont {Ajith}},
  \bibinfo {author} {\bibfnamefont {B.}~\bibnamefont {Br\"ugmann}}, \bibinfo
  {author} {\bibfnamefont {N.}~\bibnamefont {Dorband}}, \bibinfo {author}
  {\bibfnamefont {M.}~\bibnamefont {Hannam}}, \bibinfo {author} {\bibfnamefont
  {S.}~\bibnamefont {Husa}}, \bibinfo {author} {\bibfnamefont {P.}~\bibnamefont
  {M\"osta}}, \bibinfo {author} {\bibfnamefont {D.}~\bibnamefont {Pollney}},
  \bibinfo {author} {\bibfnamefont {C.}~\bibnamefont {Reisswig}}, \bibinfo
  {author} {\bibfnamefont {E.~L.}\ \bibnamefont {Robinson}}, \bibinfo {author}
  {\bibfnamefont {J.}~\bibnamefont {Seiler}}, \ and\ \bibinfo {author}
  {\bibfnamefont {B.}~\bibnamefont {Krishnan}},\ }\href {\doibase
  10.1103/PhysRevD.82.064016} {\bibfield  {journal} {\bibinfo  {journal} {Phys.
  Rev. D}\ }\textbf {\bibinfo {volume} {82}},\ \bibinfo {pages} {064016}
  (\bibinfo {year} {2010})}\BibitemShut {NoStop}%
\bibitem [{\citenamefont {Khan}\ \emph {et~al.}(2016)\citenamefont {Khan},
  \citenamefont {Husa}, \citenamefont {Hannam}, \citenamefont {Ohme},
  \citenamefont {P\"urrer}, \citenamefont {Forteza},\ and\ \citenamefont
  {Boh\'e}}]{Khan:2015jqa}%
  \BibitemOpen
  \bibfield  {author} {\bibinfo {author} {\bibfnamefont {S.}~\bibnamefont
  {Khan}}, \bibinfo {author} {\bibfnamefont {S.}~\bibnamefont {Husa}}, \bibinfo
  {author} {\bibfnamefont {M.}~\bibnamefont {Hannam}}, \bibinfo {author}
  {\bibfnamefont {F.}~\bibnamefont {Ohme}}, \bibinfo {author} {\bibfnamefont
  {M.}~\bibnamefont {P\"urrer}}, \bibinfo {author} {\bibfnamefont {X.~J.}\
  \bibnamefont {Forteza}}, \ and\ \bibinfo {author} {\bibfnamefont
  {A.}~\bibnamefont {Boh\'e}},\ }\href {\doibase 10.1103/PhysRevD.93.044007}
  {\bibfield  {journal} {\bibinfo  {journal} {Phys. Rev. D}\ }\textbf {\bibinfo
  {volume} {93}},\ \bibinfo {pages} {044007} (\bibinfo {year}
  {2016})}\BibitemShut {NoStop}%
\bibitem [{\citenamefont {Husa}\ \emph {et~al.}(2016)\citenamefont {Husa},
  \citenamefont {Khan}, \citenamefont {Hannam}, \citenamefont {P\"urrer},
  \citenamefont {Ohme}, \citenamefont {Forteza},\ and\ \citenamefont
  {Boh\'e}}]{Husa:2015iqa}%
  \BibitemOpen
  \bibfield  {author} {\bibinfo {author} {\bibfnamefont {S.}~\bibnamefont
  {Husa}}, \bibinfo {author} {\bibfnamefont {S.}~\bibnamefont {Khan}}, \bibinfo
  {author} {\bibfnamefont {M.}~\bibnamefont {Hannam}}, \bibinfo {author}
  {\bibfnamefont {M.}~\bibnamefont {P\"urrer}}, \bibinfo {author}
  {\bibfnamefont {F.}~\bibnamefont {Ohme}}, \bibinfo {author} {\bibfnamefont
  {X.~J.}\ \bibnamefont {Forteza}}, \ and\ \bibinfo {author} {\bibfnamefont
  {A.}~\bibnamefont {Boh\'e}},\ }\href {\doibase 10.1103/PhysRevD.93.044006}
  {\bibfield  {journal} {\bibinfo  {journal} {Phys. Rev. D}\ }\textbf {\bibinfo
  {volume} {93}},\ \bibinfo {pages} {044006} (\bibinfo {year}
  {2016})}\BibitemShut {NoStop}%
\bibitem [{\citenamefont {Hannam}\ \emph
  {et~al.}(2014{\natexlab{a}})\citenamefont {Hannam}, \citenamefont {Schmidt},
  \citenamefont {Boh\'e}, \citenamefont {Haegel}, \citenamefont {Husa},
  \citenamefont {Ohme}, \citenamefont {Pratten},\ and\ \citenamefont
  {P\"urrer}}]{Hannam2014}%
  \BibitemOpen
  \bibfield  {author} {\bibinfo {author} {\bibfnamefont {M.}~\bibnamefont
  {Hannam}}, \bibinfo {author} {\bibfnamefont {P.}~\bibnamefont {Schmidt}},
  \bibinfo {author} {\bibfnamefont {A.}~\bibnamefont {Boh\'e}}, \bibinfo
  {author} {\bibfnamefont {L.}~\bibnamefont {Haegel}}, \bibinfo {author}
  {\bibfnamefont {S.}~\bibnamefont {Husa}}, \bibinfo {author} {\bibfnamefont
  {F.}~\bibnamefont {Ohme}}, \bibinfo {author} {\bibfnamefont {G.}~\bibnamefont
  {Pratten}}, \ and\ \bibinfo {author} {\bibfnamefont {M.}~\bibnamefont
  {P\"urrer}},\ }\href {\doibase 10.1103/PhysRevLett.113.151101} {\bibfield
  {journal} {\bibinfo  {journal} {Phys. Rev. Lett.}\ }\textbf {\bibinfo
  {volume} {113}},\ \bibinfo {pages} {151101} (\bibinfo {year}
  {2014}{\natexlab{a}})}\BibitemShut {NoStop}%
\bibitem [{\citenamefont {Taracchini}\ \emph {et~al.}(2014)\citenamefont
  {Taracchini}, \citenamefont {Buonanno}, \citenamefont {Pan}, \citenamefont
  {Hinderer}, \citenamefont {Boyle}, \citenamefont {Hemberger}, \citenamefont
  {Kidder}, \citenamefont {Lovelace}, \citenamefont {Mrou\'e}, \citenamefont
  {Pfeiffer}, \citenamefont {Scheel}, \citenamefont {Szil\'agyi}, \citenamefont
  {Taylor},\ and\ \citenamefont {Zenginoglu}}]{Taracchini:2013rva}%
  \BibitemOpen
  \bibfield  {author} {\bibinfo {author} {\bibfnamefont {A.}~\bibnamefont
  {Taracchini}}, \bibinfo {author} {\bibfnamefont {A.}~\bibnamefont
  {Buonanno}}, \bibinfo {author} {\bibfnamefont {Y.}~\bibnamefont {Pan}},
  \bibinfo {author} {\bibfnamefont {T.}~\bibnamefont {Hinderer}}, \bibinfo
  {author} {\bibfnamefont {M.}~\bibnamefont {Boyle}}, \bibinfo {author}
  {\bibfnamefont {D.~A.}\ \bibnamefont {Hemberger}}, \bibinfo {author}
  {\bibfnamefont {L.~E.}\ \bibnamefont {Kidder}}, \bibinfo {author}
  {\bibfnamefont {G.}~\bibnamefont {Lovelace}}, \bibinfo {author}
  {\bibfnamefont {A.~H.}\ \bibnamefont {Mrou\'e}}, \bibinfo {author}
  {\bibfnamefont {H.~P.}\ \bibnamefont {Pfeiffer}}, \bibinfo {author}
  {\bibfnamefont {M.~A.}\ \bibnamefont {Scheel}}, \bibinfo {author}
  {\bibfnamefont {B.}~\bibnamefont {Szil\'agyi}}, \bibinfo {author}
  {\bibfnamefont {N.~W.}\ \bibnamefont {Taylor}}, \ and\ \bibinfo {author}
  {\bibfnamefont {A.}~\bibnamefont {Zenginoglu}},\ }\href {\doibase
  10.1103/PhysRevD.89.061502} {\bibfield  {journal} {\bibinfo  {journal} {Phys.
  Rev. D}\ }\textbf {\bibinfo {volume} {89}},\ \bibinfo {pages} {061502}
  (\bibinfo {year} {2014})}\BibitemShut {NoStop}%
\bibitem [{\citenamefont {{Boh{\'e}}}\ \emph {et~al.}(2016)\citenamefont
  {{Boh{\'e}}}, \citenamefont {{Shao}}, \citenamefont {{Taracchini}},
  \citenamefont {{Buonanno}}, \citenamefont {{Babak}}, \citenamefont {{Harry}},
  \citenamefont {{Hinder}}, \citenamefont {{Ossokine}}, \citenamefont
  {{P{\"u}rrer}}, \citenamefont {{Raymond}}, \citenamefont {{Chu}},
  \citenamefont {{Fong}}, \citenamefont {{Kumar}}, \citenamefont {{Pfeiffer}},
  \citenamefont {{Boyle}}, \citenamefont {{Hemberger}}, \citenamefont
  {{Kidder}}, \citenamefont {{Lovelace}}, \citenamefont {{Scheel}},\ and\
  \citenamefont {{Szil{\'a}gyi}}}]{SEOBNRv4}%
  \BibitemOpen
  \bibfield  {author} {\bibinfo {author} {\bibfnamefont {A.}~\bibnamefont
  {{Boh{\'e}}}}, \bibinfo {author} {\bibfnamefont {L.}~\bibnamefont {{Shao}}},
  \bibinfo {author} {\bibfnamefont {A.}~\bibnamefont {{Taracchini}}}, \bibinfo
  {author} {\bibfnamefont {A.}~\bibnamefont {{Buonanno}}}, \bibinfo {author}
  {\bibfnamefont {S.}~\bibnamefont {{Babak}}}, \bibinfo {author} {\bibfnamefont
  {I.~W.}\ \bibnamefont {{Harry}}}, \bibinfo {author} {\bibfnamefont
  {I.}~\bibnamefont {{Hinder}}}, \bibinfo {author} {\bibfnamefont
  {S.}~\bibnamefont {{Ossokine}}}, \bibinfo {author} {\bibfnamefont
  {M.}~\bibnamefont {{P{\"u}rrer}}}, \bibinfo {author} {\bibfnamefont
  {V.}~\bibnamefont {{Raymond}}}, \bibinfo {author} {\bibfnamefont
  {T.}~\bibnamefont {{Chu}}}, \bibinfo {author} {\bibfnamefont
  {H.}~\bibnamefont {{Fong}}}, \bibinfo {author} {\bibfnamefont
  {P.}~\bibnamefont {{Kumar}}}, \bibinfo {author} {\bibfnamefont {H.~P.}\
  \bibnamefont {{Pfeiffer}}}, \bibinfo {author} {\bibfnamefont
  {M.}~\bibnamefont {{Boyle}}}, \bibinfo {author} {\bibfnamefont {D.~A.}\
  \bibnamefont {{Hemberger}}}, \bibinfo {author} {\bibfnamefont {L.~E.}\
  \bibnamefont {{Kidder}}}, \bibinfo {author} {\bibfnamefont {G.}~\bibnamefont
  {{Lovelace}}}, \bibinfo {author} {\bibfnamefont {M.~A.}\ \bibnamefont
  {{Scheel}}}, \ and\ \bibinfo {author} {\bibfnamefont {B.}~\bibnamefont
  {{Szil{\'a}gyi}}},\ }\href@noop {} {\bibfield  {journal} {\bibinfo  {journal}
  {ArXiv e-prints}\ } (\bibinfo {year} {2016})},\ \Eprint
  {http://arxiv.org/abs/1611.03703} {arXiv:1611.03703 [gr-qc]} \BibitemShut
  {NoStop}%
\bibitem [{\citenamefont {Pan}\ \emph {et~al.}(2014{\natexlab{a}})\citenamefont
  {Pan}, \citenamefont {Buonanno}, \citenamefont {Taracchini}, \citenamefont
  {Kidder}, \citenamefont {Mrou\'e}, \citenamefont {Pfeiffer}, \citenamefont
  {Scheel},\ and\ \citenamefont {Szil\'agyi}}]{Pan2013}%
  \BibitemOpen
  \bibfield  {author} {\bibinfo {author} {\bibfnamefont {Y.}~\bibnamefont
  {Pan}}, \bibinfo {author} {\bibfnamefont {A.}~\bibnamefont {Buonanno}},
  \bibinfo {author} {\bibfnamefont {A.}~\bibnamefont {Taracchini}}, \bibinfo
  {author} {\bibfnamefont {L.~E.}\ \bibnamefont {Kidder}}, \bibinfo {author}
  {\bibfnamefont {A.~H.}\ \bibnamefont {Mrou\'e}}, \bibinfo {author}
  {\bibfnamefont {H.~P.}\ \bibnamefont {Pfeiffer}}, \bibinfo {author}
  {\bibfnamefont {M.~A.}\ \bibnamefont {Scheel}}, \ and\ \bibinfo {author}
  {\bibfnamefont {B.}~\bibnamefont {Szil\'agyi}},\ }\href {\doibase
  10.1103/PhysRevD.89.084006} {\bibfield  {journal} {\bibinfo  {journal} {Phys.
  Rev. D}\ }\textbf {\bibinfo {volume} {89}},\ \bibinfo {pages} {084006}
  (\bibinfo {year} {2014}{\natexlab{a}})}\BibitemShut {NoStop}%
\bibitem [{\citenamefont {Abbott}\ \emph
  {et~al.}(2016{\natexlab{a}})\citenamefont {Abbott} \emph {et~al.}}]{O1data}%
  \BibitemOpen
  \bibfield  {author} {\bibinfo {author} {\bibfnamefont {B.~P.}\ \bibnamefont
  {Abbott}} \emph {et~al.} (\bibinfo {collaboration} {Virgo, LIGO
  Scientific}),\ }\href@noop {} {\bibfield  {journal} {\bibinfo  {journal}
  {Phys. Rev. D}\ }\textbf {\bibinfo {volume} {93}},\ \bibinfo {pages} {122003}
  (\bibinfo {year} {2016}{\natexlab{a}})},\ \Eprint
  {http://arxiv.org/abs/1602.03839} {arXiv:1602.03839 [gr-qc]} \BibitemShut
  {NoStop}%
\bibitem [{\citenamefont {Abbott}\ \emph
  {et~al.}(2016{\natexlab{b}})\citenamefont {Abbott} \emph
  {et~al.}}]{GW150914props}%
  \BibitemOpen
  \bibfield  {author} {\bibinfo {author} {\bibfnamefont {B.~P.}\ \bibnamefont
  {Abbott}} \emph {et~al.} (\bibinfo {collaboration} {Virgo, LIGO
  Scientific}),\ }\href@noop {} {\bibfield  {journal} {\bibinfo  {journal}
  {Phys. Rev. Lett.}\ }\textbf {\bibinfo {volume} {116}},\ \bibinfo {pages}
  {241102} (\bibinfo {year} {2016}{\natexlab{b}})},\ \Eprint
  {http://arxiv.org/abs/1602.03840} {arXiv:1602.03840 [gr-qc]} \BibitemShut
  {NoStop}%
\bibitem [{\citenamefont {{The LIGO Scientific Collaboration}}\ \emph
  {et~al.}(2016{\natexlab{b}})\citenamefont {{The LIGO Scientific
  Collaboration}}, \citenamefont {{the Virgo Collaboration}}, \citenamefont
  {{Abbott}}, \citenamefont {{Abbott}}, \citenamefont {{Abbott}}, \citenamefont
  {{Abernathy}}, \citenamefont {{Acernese}}, \citenamefont {{Ackley}},
  \citenamefont {{Adams}}, \citenamefont {{Adams}},\ and\ \citenamefont
  {et~al.}}]{PE_precessing}%
  \BibitemOpen
  \bibfield  {author} {\bibinfo {author} {\bibnamefont {{The LIGO Scientific
  Collaboration}}}, \bibinfo {author} {\bibnamefont {{the Virgo
  Collaboration}}}, \bibinfo {author} {\bibfnamefont {B.~P.}\ \bibnamefont
  {{Abbott}}}, \bibinfo {author} {\bibfnamefont {R.}~\bibnamefont {{Abbott}}},
  \bibinfo {author} {\bibfnamefont {T.~D.}\ \bibnamefont {{Abbott}}}, \bibinfo
  {author} {\bibfnamefont {M.~R.}\ \bibnamefont {{Abernathy}}}, \bibinfo
  {author} {\bibfnamefont {F.}~\bibnamefont {{Acernese}}}, \bibinfo {author}
  {\bibfnamefont {K.}~\bibnamefont {{Ackley}}}, \bibinfo {author}
  {\bibfnamefont {C.}~\bibnamefont {{Adams}}}, \bibinfo {author} {\bibfnamefont
  {T.}~\bibnamefont {{Adams}}}, \ and\ \bibinfo {author} {\bibnamefont
  {et~al.}},\ }\href@noop {} {\bibfield  {journal} {\bibinfo  {journal} {ArXiv
  e-prints}\ } (\bibinfo {year} {2016}{\natexlab{b}})},\ \Eprint
  {http://arxiv.org/abs/1606.01210} {arXiv:1606.01210 [gr-qc]} \BibitemShut
  {NoStop}%
\bibitem [{\citenamefont {{The LIGO Scientific Collaboration}}\ \emph
  {et~al.}(2016{\natexlab{c}})\citenamefont {{The LIGO Scientific
  Collaboration}}, \citenamefont {{the Virgo Collaboration}}, \citenamefont
  {{Abbott}}, \citenamefont {{Abbott}}, \citenamefont {{Abbott}}, \citenamefont
  {{Abernathy}}, \citenamefont {{Acernese}}, \citenamefont {{Ackley}},
  \citenamefont {{Adams}}, \citenamefont {{Adams}},\ and\ \citenamefont
  {et~al.}}]{SEOBNRv3_PE}%
  \BibitemOpen
  \bibfield  {author} {\bibinfo {author} {\bibnamefont {{The LIGO Scientific
  Collaboration}}}, \bibinfo {author} {\bibnamefont {{the Virgo
  Collaboration}}}, \bibinfo {author} {\bibfnamefont {B.~P.}\ \bibnamefont
  {{Abbott}}}, \bibinfo {author} {\bibfnamefont {R.}~\bibnamefont {{Abbott}}},
  \bibinfo {author} {\bibfnamefont {T.~D.}\ \bibnamefont {{Abbott}}}, \bibinfo
  {author} {\bibfnamefont {M.~R.}\ \bibnamefont {{Abernathy}}}, \bibinfo
  {author} {\bibfnamefont {F.}~\bibnamefont {{Acernese}}}, \bibinfo {author}
  {\bibfnamefont {K.}~\bibnamefont {{Ackley}}}, \bibinfo {author}
  {\bibfnamefont {C.}~\bibnamefont {{Adams}}}, \bibinfo {author} {\bibfnamefont
  {T.}~\bibnamefont {{Adams}}}, \ and\ \bibinfo {author} {\bibnamefont
  {et~al.}} (\bibinfo {collaboration} {LIGO Scientific Collaboration and Virgo
  Collaboration}),\ }\href {\doibase 10.1103/PhysRevX.6.041014} {\bibfield
  {journal} {\bibinfo  {journal} {Phys. Rev. X}\ }\textbf {\bibinfo {volume}
  {6}},\ \bibinfo {pages} {041014} (\bibinfo {year}
  {2016}{\natexlab{c}})}\BibitemShut {NoStop}%
\bibitem [{sxs()}]{sxs}%
  \BibitemOpen
  \href@noop {} {}\bibinfo {howpublished}
  {\url{https://www.black-holes.org/waveforms/}}\BibitemShut {NoStop}%
\bibitem [{\citenamefont {Harry}\ \emph {et~al.}(2016)\citenamefont {Harry},
  \citenamefont {Privitera}, \citenamefont {Boh\'e},\ and\ \citenamefont
  {Buonanno}}]{Harryetal}%
  \BibitemOpen
  \bibfield  {author} {\bibinfo {author} {\bibfnamefont {I.}~\bibnamefont
  {Harry}}, \bibinfo {author} {\bibfnamefont {S.}~\bibnamefont {Privitera}},
  \bibinfo {author} {\bibfnamefont {A.}~\bibnamefont {Boh\'e}}, \ and\ \bibinfo
  {author} {\bibfnamefont {A.}~\bibnamefont {Buonanno}},\ }\href {\doibase
  10.1103/PhysRevD.94.024012} {\bibfield  {journal} {\bibinfo  {journal} {Phys.
  Rev. D}\ }\textbf {\bibinfo {volume} {94}},\ \bibinfo {pages} {024012}
  (\bibinfo {year} {2016})}\BibitemShut {NoStop}%
\bibitem [{\citenamefont {Babak}\ \emph
  {et~al.}(2017{\natexlab{a}})\citenamefont {Babak}, \citenamefont
  {Taracchini},\ and\ \citenamefont {Buonanno}}]{Babaketal}%
  \BibitemOpen
  \bibfield  {author} {\bibinfo {author} {\bibfnamefont {S.}~\bibnamefont
  {Babak}}, \bibinfo {author} {\bibfnamefont {A.}~\bibnamefont {Taracchini}}, \
  and\ \bibinfo {author} {\bibfnamefont {A.}~\bibnamefont {Buonanno}},\ }\href
  {\doibase 10.1103/PhysRevD.95.024010} {\bibfield  {journal} {\bibinfo
  {journal} {Phys. Rev. D}\ }\textbf {\bibinfo {volume} {95}},\ \bibinfo
  {pages} {024010} (\bibinfo {year} {2017}{\natexlab{a}})}\BibitemShut
  {NoStop}%
\bibitem [{\citenamefont {Kumar}\ \emph {et~al.}(2016)\citenamefont {Kumar},
  \citenamefont {Chu}, \citenamefont {Fong}, \citenamefont {Pfeiffer},
  \citenamefont {Boyle}, \citenamefont {Hemberger}, \citenamefont {Kidder},
  \citenamefont {Scheel},\ and\ \citenamefont {Szilagyi}}]{Prayush2016}%
  \BibitemOpen
  \bibfield  {author} {\bibinfo {author} {\bibfnamefont {P.}~\bibnamefont
  {Kumar}}, \bibinfo {author} {\bibfnamefont {T.}~\bibnamefont {Chu}}, \bibinfo
  {author} {\bibfnamefont {H.}~\bibnamefont {Fong}}, \bibinfo {author}
  {\bibfnamefont {H.~P.}\ \bibnamefont {Pfeiffer}}, \bibinfo {author}
  {\bibfnamefont {M.}~\bibnamefont {Boyle}}, \bibinfo {author} {\bibfnamefont
  {D.~A.}\ \bibnamefont {Hemberger}}, \bibinfo {author} {\bibfnamefont {L.~E.}\
  \bibnamefont {Kidder}}, \bibinfo {author} {\bibfnamefont {M.~A.}\
  \bibnamefont {Scheel}}, \ and\ \bibinfo {author} {\bibfnamefont
  {B.}~\bibnamefont {Szilagyi}},\ }\href {\doibase 10.1103/PhysRevD.93.104050}
  {\bibfield  {journal} {\bibinfo  {journal} {Phys. Rev. D}\ }\textbf {\bibinfo
  {volume} {93}},\ \bibinfo {pages} {104050} (\bibinfo {year}
  {2016})}\BibitemShut {NoStop}%
\bibitem [{\citenamefont {{Varma}}\ and\ \citenamefont
  {{Ajith}}(2016)}]{Verma}%
  \BibitemOpen
  \bibfield  {author} {\bibinfo {author} {\bibfnamefont {V.}~\bibnamefont
  {{Varma}}}\ and\ \bibinfo {author} {\bibfnamefont {P.}~\bibnamefont
  {{Ajith}}},\ }\href@noop {} {\bibfield  {journal} {\bibinfo  {journal} {ArXiv
  e-prints}\ } (\bibinfo {year} {2016})},\ \Eprint
  {http://arxiv.org/abs/1612.05608} {arXiv:1612.05608 [gr-qc]} \BibitemShut
  {NoStop}%
\bibitem [{\citenamefont {{The LIGO Scientific Collaboration}}\ \emph
  {et~al.}(2016{\natexlab{d}})\citenamefont {{The LIGO Scientific
  Collaboration}}, \citenamefont {{the Virgo Collaboration}}, \citenamefont
  {{Abbott}}, \citenamefont {{Abbott}}, \citenamefont {{Abbott}}, \citenamefont
  {{Abernathy}}, \citenamefont {{Acernese}}, \citenamefont {{Ackley}},
  \citenamefont {{Adams}}, \citenamefont {{Adams}},\ and\ \citenamefont
  {et~al.}}]{NRsystematics}%
  \BibitemOpen
  \bibfield  {author} {\bibinfo {author} {\bibnamefont {{The LIGO Scientific
  Collaboration}}}, \bibinfo {author} {\bibnamefont {{the Virgo
  Collaboration}}}, \bibinfo {author} {\bibfnamefont {B.~P.}\ \bibnamefont
  {{Abbott}}}, \bibinfo {author} {\bibfnamefont {R.}~\bibnamefont {{Abbott}}},
  \bibinfo {author} {\bibfnamefont {T.~D.}\ \bibnamefont {{Abbott}}}, \bibinfo
  {author} {\bibfnamefont {M.~R.}\ \bibnamefont {{Abernathy}}}, \bibinfo
  {author} {\bibfnamefont {F.}~\bibnamefont {{Acernese}}}, \bibinfo {author}
  {\bibfnamefont {K.}~\bibnamefont {{Ackley}}}, \bibinfo {author}
  {\bibfnamefont {C.}~\bibnamefont {{Adams}}}, \bibinfo {author} {\bibfnamefont
  {T.}~\bibnamefont {{Adams}}}, \ and\ \bibinfo {author} {\bibnamefont
  {et~al.}},\ }\href@noop {} {\bibfield  {journal} {\bibinfo  {journal} {ArXiv
  e-prints}\ } (\bibinfo {year} {2016}{\natexlab{d}})},\ \Eprint
  {http://arxiv.org/abs/1611.07531} {arXiv:1611.07531 [gr-qc]} \BibitemShut
  {NoStop}%
\bibitem [{\citenamefont {Kesden}\ \emph {et~al.}(2010)\citenamefont {Kesden},
  \citenamefont {Sperhake},\ and\ \citenamefont {Berti}}]{KSB10}%
  \BibitemOpen
  \bibfield  {author} {\bibinfo {author} {\bibfnamefont {M.}~\bibnamefont
  {Kesden}}, \bibinfo {author} {\bibfnamefont {U.}~\bibnamefont {Sperhake}}, \
  and\ \bibinfo {author} {\bibfnamefont {E.}~\bibnamefont {Berti}},\ }\href
  {\doibase 10.1103/PhysRevD.81.084054} {\bibfield  {journal} {\bibinfo
  {journal} {Phys. Rev. D}\ }\textbf {\bibinfo {volume} {81}},\ \bibinfo
  {pages} {084054} (\bibinfo {year} {2010})}\BibitemShut {NoStop}%
\bibitem [{\citenamefont {Gupta}\ and\ \citenamefont
  {Gopakumar}(2014)}]{GG2014}%
  \BibitemOpen
  \bibfield  {author} {\bibinfo {author} {\bibfnamefont {A.}~\bibnamefont
  {Gupta}}\ and\ \bibinfo {author} {\bibfnamefont {A.}~\bibnamefont
  {Gopakumar}},\ }\href {http://stacks.iop.org/0264-9381/31/i=10/a=105017}
  {\bibfield  {journal} {\bibinfo  {journal} {Classical and Quantum Gravity}\
  }\textbf {\bibinfo {volume} {31}},\ \bibinfo {pages} {105017} (\bibinfo
  {year} {2014})}\BibitemShut {NoStop}%
\bibitem [{\citenamefont {Gerosa}\ \emph {et~al.}(2014)\citenamefont {Gerosa},
  \citenamefont {O'Shaughnessy}, \citenamefont {Kesden}, \citenamefont
  {Berti},\ and\ \citenamefont {Sperhake}}]{Gerosa2014}%
  \BibitemOpen
  \bibfield  {author} {\bibinfo {author} {\bibfnamefont {D.}~\bibnamefont
  {Gerosa}}, \bibinfo {author} {\bibfnamefont {R.}~\bibnamefont
  {O'Shaughnessy}}, \bibinfo {author} {\bibfnamefont {M.}~\bibnamefont
  {Kesden}}, \bibinfo {author} {\bibfnamefont {E.}~\bibnamefont {Berti}}, \
  and\ \bibinfo {author} {\bibfnamefont {U.}~\bibnamefont {Sperhake}},\ }\href
  {\doibase 10.1103/PhysRevD.89.124025} {\bibfield  {journal} {\bibinfo
  {journal} {Phys. Rev. D}\ }\textbf {\bibinfo {volume} {89}},\ \bibinfo
  {pages} {124025} (\bibinfo {year} {2014})}\BibitemShut {NoStop}%
\bibitem [{\citenamefont {Trifir\`o}\ \emph {et~al.}(2016)\citenamefont
  {Trifir\`o}, \citenamefont {O'Shaughnessy}, \citenamefont {Gerosa},
  \citenamefont {Berti}, \citenamefont {Kesden}, \citenamefont {Littenberg},\
  and\ \citenamefont {Sperhake}}]{Trifiro2016}%
  \BibitemOpen
  \bibfield  {author} {\bibinfo {author} {\bibfnamefont {D.}~\bibnamefont
  {Trifir\`o}}, \bibinfo {author} {\bibfnamefont {R.}~\bibnamefont
  {O'Shaughnessy}}, \bibinfo {author} {\bibfnamefont {D.}~\bibnamefont
  {Gerosa}}, \bibinfo {author} {\bibfnamefont {E.}~\bibnamefont {Berti}},
  \bibinfo {author} {\bibfnamefont {M.}~\bibnamefont {Kesden}}, \bibinfo
  {author} {\bibfnamefont {T.}~\bibnamefont {Littenberg}}, \ and\ \bibinfo
  {author} {\bibfnamefont {U.}~\bibnamefont {Sperhake}},\ }\href {\doibase
  10.1103/PhysRevD.93.044071} {\bibfield  {journal} {\bibinfo  {journal} {Phys.
  Rev. D}\ }\textbf {\bibinfo {volume} {93}},\ \bibinfo {pages} {044071}
  (\bibinfo {year} {2016})}\BibitemShut {NoStop}%
\bibitem [{lal()}]{lalsuite}%
  \BibitemOpen
  \href@noop {} {}\bibinfo {howpublished}
  {\url{https://www.lsc-group.phys.uwm.edu/daswg/projects/lalsuite.html}}\BibitemShut
  {NoStop}%
\bibitem [{\citenamefont {Kesden}\ \emph {et~al.}(2015)\citenamefont {Kesden},
  \citenamefont {Gerosa}, \citenamefont {O'Shaughnessy}, \citenamefont
  {Berti},\ and\ \citenamefont {Sperhake}}]{Kesdenetal2015}%
  \BibitemOpen
  \bibfield  {author} {\bibinfo {author} {\bibfnamefont {M.}~\bibnamefont
  {Kesden}}, \bibinfo {author} {\bibfnamefont {D.}~\bibnamefont {Gerosa}},
  \bibinfo {author} {\bibfnamefont {R.}~\bibnamefont {O'Shaughnessy}}, \bibinfo
  {author} {\bibfnamefont {E.}~\bibnamefont {Berti}}, \ and\ \bibinfo {author}
  {\bibfnamefont {U.}~\bibnamefont {Sperhake}},\ }\href {\doibase
  10.1103/PhysRevLett.114.081103} {\bibfield  {journal} {\bibinfo  {journal}
  {Phys. Rev. Lett.}\ }\textbf {\bibinfo {volume} {114}},\ \bibinfo {pages}
  {081103} (\bibinfo {year} {2015})}\BibitemShut {NoStop}%
\bibitem [{\citenamefont {Gerosa}\ \emph {et~al.}(2015)\citenamefont {Gerosa},
  \citenamefont {Kesden}, \citenamefont {Sperhake}, \citenamefont {Berti},\
  and\ \citenamefont {O'Shaughnessy}}]{Gerosaetal2015}%
  \BibitemOpen
  \bibfield  {author} {\bibinfo {author} {\bibfnamefont {D.}~\bibnamefont
  {Gerosa}}, \bibinfo {author} {\bibfnamefont {M.}~\bibnamefont {Kesden}},
  \bibinfo {author} {\bibfnamefont {U.}~\bibnamefont {Sperhake}}, \bibinfo
  {author} {\bibfnamefont {E.}~\bibnamefont {Berti}}, \ and\ \bibinfo {author}
  {\bibfnamefont {R.}~\bibnamefont {O'Shaughnessy}},\ }\href {\doibase
  10.1103/PhysRevD.92.064016} {\bibfield  {journal} {\bibinfo  {journal} {Phys.
  Rev. D}\ }\textbf {\bibinfo {volume} {92}},\ \bibinfo {pages} {064016}
  (\bibinfo {year} {2015})}\BibitemShut {NoStop}%
\bibitem [{\citenamefont {Pan}\ \emph {et~al.}(2014{\natexlab{b}})\citenamefont
  {Pan}, \citenamefont {Buonanno}, \citenamefont {Taracchini}, \citenamefont
  {Kidder}, \citenamefont {Mrou\'e}, \citenamefont {Pfeiffer}, \citenamefont
  {Scheel},\ and\ \citenamefont {Szil\'agyi}}]{Pan2014}%
  \BibitemOpen
  \bibfield  {author} {\bibinfo {author} {\bibfnamefont {Y.}~\bibnamefont
  {Pan}}, \bibinfo {author} {\bibfnamefont {A.}~\bibnamefont {Buonanno}},
  \bibinfo {author} {\bibfnamefont {A.}~\bibnamefont {Taracchini}}, \bibinfo
  {author} {\bibfnamefont {L.~E.}\ \bibnamefont {Kidder}}, \bibinfo {author}
  {\bibfnamefont {A.~H.}\ \bibnamefont {Mrou\'e}}, \bibinfo {author}
  {\bibfnamefont {H.~P.}\ \bibnamefont {Pfeiffer}}, \bibinfo {author}
  {\bibfnamefont {M.~A.}\ \bibnamefont {Scheel}}, \ and\ \bibinfo {author}
  {\bibfnamefont {B.}~\bibnamefont {Szil\'agyi}},\ }\href {\doibase
  10.1103/PhysRevD.89.084006} {\bibfield  {journal} {\bibinfo  {journal} {Phys.
  Rev. D}\ }\textbf {\bibinfo {volume} {89}},\ \bibinfo {pages} {084006}
  (\bibinfo {year} {2014}{\natexlab{b}})}\BibitemShut {NoStop}%
\bibitem [{\citenamefont {Taracchini}\ \emph {et~al.}(2012)\citenamefont
  {Taracchini}, \citenamefont {Pan}, \citenamefont {Buonanno}, \citenamefont
  {Barausse}, \citenamefont {Boyle}, \citenamefont {Chu}, \citenamefont
  {Lovelace}, \citenamefont {Pfeiffer},\ and\ \citenamefont
  {Scheel}}]{Taracchini:2012}%
  \BibitemOpen
  \bibfield  {author} {\bibinfo {author} {\bibfnamefont {A.}~\bibnamefont
  {Taracchini}}, \bibinfo {author} {\bibfnamefont {Y.}~\bibnamefont {Pan}},
  \bibinfo {author} {\bibfnamefont {A.}~\bibnamefont {Buonanno}}, \bibinfo
  {author} {\bibfnamefont {E.}~\bibnamefont {Barausse}}, \bibinfo {author}
  {\bibfnamefont {M.}~\bibnamefont {Boyle}}, \bibinfo {author} {\bibfnamefont
  {T.}~\bibnamefont {Chu}}, \bibinfo {author} {\bibfnamefont {G.}~\bibnamefont
  {Lovelace}}, \bibinfo {author} {\bibfnamefont {H.~P.}\ \bibnamefont
  {Pfeiffer}}, \ and\ \bibinfo {author} {\bibfnamefont {M.~A.}\ \bibnamefont
  {Scheel}},\ }\href {\doibase 10.1103/PhysRevD.86.024011} {\bibfield
  {journal} {\bibinfo  {journal} {Phys. Rev. D}\ }\textbf {\bibinfo {volume}
  {86}},\ \bibinfo {pages} {024011} (\bibinfo {year} {2012})}\BibitemShut
  {NoStop}%
\bibitem [{\citenamefont {Buonanno}\ \emph {et~al.}(2003)\citenamefont
  {Buonanno}, \citenamefont {Chen},\ and\ \citenamefont
  {Vallisneri}}]{BCV2003}%
  \BibitemOpen
  \bibfield  {author} {\bibinfo {author} {\bibfnamefont {A.}~\bibnamefont
  {Buonanno}}, \bibinfo {author} {\bibfnamefont {Y.}~\bibnamefont {Chen}}, \
  and\ \bibinfo {author} {\bibfnamefont {M.}~\bibnamefont {Vallisneri}},\
  }\href {\doibase 10.1103/PhysRevD.67.104025} {\bibfield  {journal} {\bibinfo
  {journal} {Phys. Rev. D}\ }\textbf {\bibinfo {volume} {67}},\ \bibinfo
  {pages} {104025} (\bibinfo {year} {2003})}\BibitemShut {NoStop}%
\bibitem [{\citenamefont {Buonanno}\ \emph {et~al.}(2004)\citenamefont
  {Buonanno}, \citenamefont {Chen}, \citenamefont {Pan},\ and\ \citenamefont
  {Vallisneri}}]{BYPV2004}%
  \BibitemOpen
  \bibfield  {author} {\bibinfo {author} {\bibfnamefont {A.}~\bibnamefont
  {Buonanno}}, \bibinfo {author} {\bibfnamefont {Y.}~\bibnamefont {Chen}},
  \bibinfo {author} {\bibfnamefont {Y.}~\bibnamefont {Pan}}, \ and\ \bibinfo
  {author} {\bibfnamefont {M.}~\bibnamefont {Vallisneri}},\ }\href {\doibase
  10.1103/PhysRevD.70.104003} {\bibfield  {journal} {\bibinfo  {journal} {Phys.
  Rev. D}\ }\textbf {\bibinfo {volume} {70}},\ \bibinfo {pages} {104003}
  (\bibinfo {year} {2004})}\BibitemShut {NoStop}%
\bibitem [{\citenamefont {Boyle}\ \emph {et~al.}(2011)\citenamefont {Boyle},
  \citenamefont {Owen},\ and\ \citenamefont {Pfeiffer}}]{BOP}%
  \BibitemOpen
  \bibfield  {author} {\bibinfo {author} {\bibfnamefont {M.}~\bibnamefont
  {Boyle}}, \bibinfo {author} {\bibfnamefont {R.}~\bibnamefont {Owen}}, \ and\
  \bibinfo {author} {\bibfnamefont {H.~P.}\ \bibnamefont {Pfeiffer}},\ }\href
  {\doibase 10.1103/PhysRevD.84.124011} {\bibfield  {journal} {\bibinfo
  {journal} {Phys. Rev. D}\ }\textbf {\bibinfo {volume} {84}},\ \bibinfo
  {pages} {124011} (\bibinfo {year} {2011})}\BibitemShut {NoStop}%
\bibitem [{\citenamefont {{O'Shaughnessy}}\ \emph {et~al.}(2011)\citenamefont
  {{O'Shaughnessy}}, \citenamefont {{Vaishnav}}, \citenamefont {{Healy}},
  \citenamefont {{Meeks}},\ and\ \citenamefont {{Shoemaker}}}]{OVHMS2011}%
  \BibitemOpen
  \bibfield  {author} {\bibinfo {author} {\bibfnamefont {R.}~\bibnamefont
  {{O'Shaughnessy}}}, \bibinfo {author} {\bibfnamefont {B.}~\bibnamefont
  {{Vaishnav}}}, \bibinfo {author} {\bibfnamefont {J.}~\bibnamefont {{Healy}}},
  \bibinfo {author} {\bibfnamefont {Z.}~\bibnamefont {{Meeks}}}, \ and\
  \bibinfo {author} {\bibfnamefont {D.}~\bibnamefont {{Shoemaker}}},\ }\href
  {\doibase 10.1103/PhysRevD.84.124002} {\bibfield  {journal} {\bibinfo
  {journal} {\prd}\ }\textbf {\bibinfo {volume} {84}},\ \bibinfo {eid} {124002}
  (\bibinfo {year} {2011})},\ \Eprint {http://arxiv.org/abs/1109.5224}
  {arXiv:1109.5224 [gr-qc]} \BibitemShut {NoStop}%
\bibitem [{\citenamefont {{Schmidt}}\ \emph {et~al.}(2011)\citenamefont
  {{Schmidt}}, \citenamefont {{Hannam}}, \citenamefont {{Husa}},\ and\
  \citenamefont {{Ajith}}}]{SHHA2011}%
  \BibitemOpen
  \bibfield  {author} {\bibinfo {author} {\bibfnamefont {P.}~\bibnamefont
  {{Schmidt}}}, \bibinfo {author} {\bibfnamefont {M.}~\bibnamefont {{Hannam}}},
  \bibinfo {author} {\bibfnamefont {S.}~\bibnamefont {{Husa}}}, \ and\ \bibinfo
  {author} {\bibfnamefont {P.}~\bibnamefont {{Ajith}}},\ }\href {\doibase
  10.1103/PhysRevD.84.024046} {\bibfield  {journal} {\bibinfo  {journal}
  {\prd}\ }\textbf {\bibinfo {volume} {84}},\ \bibinfo {eid} {024046} (\bibinfo
  {year} {2011})},\ \Eprint {http://arxiv.org/abs/1012.2879} {arXiv:1012.2879
  [gr-qc]} \BibitemShut {NoStop}%
\bibitem [{\citenamefont {Schmidt}\ \emph {et~al.}(2012)\citenamefont
  {Schmidt}, \citenamefont {Hannam},\ and\ \citenamefont {Husa}}]{SHH2012}%
  \BibitemOpen
  \bibfield  {author} {\bibinfo {author} {\bibfnamefont {P.}~\bibnamefont
  {Schmidt}}, \bibinfo {author} {\bibfnamefont {M.}~\bibnamefont {Hannam}}, \
  and\ \bibinfo {author} {\bibfnamefont {S.}~\bibnamefont {Husa}},\ }\href
  {\doibase 10.1103/PhysRevD.86.104063} {\bibfield  {journal} {\bibinfo
  {journal} {Phys. Rev.}\ }\textbf {\bibinfo {volume} {D86}},\ \bibinfo {pages}
  {104063} (\bibinfo {year} {2012})},\ \Eprint {http://arxiv.org/abs/1207.3088}
  {arXiv:1207.3088 [gr-qc]} \BibitemShut {NoStop}%
\bibitem [{\citenamefont {Gupta}\ and\ \citenamefont
  {Gopakumar}(2015)}]{GG2015}%
  \BibitemOpen
  \bibfield  {author} {\bibinfo {author} {\bibfnamefont {A.}~\bibnamefont
  {Gupta}}\ and\ \bibinfo {author} {\bibfnamefont {A.}~\bibnamefont
  {Gopakumar}},\ }\href {\doibase 10.1088/0264-9381/32/17/175002} {\bibfield
  {journal} {\bibinfo  {journal} {Class. Quant. Grav.}\ }\textbf {\bibinfo
  {volume} {32}},\ \bibinfo {pages} {175002} (\bibinfo {year} {2015})},\
  \Eprint {http://arxiv.org/abs/1507.00406} {arXiv:1507.00406 [gr-qc]}
  \BibitemShut {NoStop}%
\bibitem [{SpE()}]{SpECwebsite}%
  \BibitemOpen
  \href@noop {} {}\bibinfo {howpublished}
  {\url{http://www.black-holes.org/SpEC.html}}\BibitemShut {NoStop}%
\bibitem [{\citenamefont {Yo}\ \emph {et~al.}(2004)\citenamefont {Yo},
  \citenamefont {Cook}, \citenamefont {Shapiro},\ and\ \citenamefont
  {Baumgarte}}]{Yo2004}%
  \BibitemOpen
  \bibfield  {author} {\bibinfo {author} {\bibfnamefont {H.-J.}\ \bibnamefont
  {Yo}}, \bibinfo {author} {\bibfnamefont {J.~N.}\ \bibnamefont {Cook}},
  \bibinfo {author} {\bibfnamefont {S.~L.}\ \bibnamefont {Shapiro}}, \ and\
  \bibinfo {author} {\bibfnamefont {T.~W.}\ \bibnamefont {Baumgarte}},\
  }\href@noop {} {\bibfield  {journal} {\bibinfo  {journal} {Phys. Rev. D}\
  }\textbf {\bibinfo {volume} {70}},\ \bibinfo {pages} {084033} (\bibinfo
  {year} {2004})}\BibitemShut {NoStop}%
\bibitem [{\citenamefont {Cook}\ and\ \citenamefont
  {Pfeiffer}(2004)}]{Cook2004}%
  \BibitemOpen
  \bibfield  {author} {\bibinfo {author} {\bibfnamefont {G.~B.}\ \bibnamefont
  {Cook}}\ and\ \bibinfo {author} {\bibfnamefont {H.~P.}\ \bibnamefont
  {Pfeiffer}},\ }\href {\doibase 10.1103/PhysRevD.70.104016} {\bibfield
  {journal} {\bibinfo  {journal} {Phys. Rev. D}\ }\textbf {\bibinfo {volume}
  {70}},\ \bibinfo {pages} {104016} (\bibinfo {year} {2004})}\BibitemShut
  {NoStop}%
\bibitem [{\citenamefont {Pfeiffer}\ \emph {et~al.}(2003)\citenamefont
  {Pfeiffer}, \citenamefont {Kidder}, \citenamefont {Scheel},\ and\
  \citenamefont {Teukolsky}}]{Pfeiffer2003}%
  \BibitemOpen
  \bibfield  {author} {\bibinfo {author} {\bibfnamefont {H.~P.}\ \bibnamefont
  {Pfeiffer}}, \bibinfo {author} {\bibfnamefont {L.~E.}\ \bibnamefont
  {Kidder}}, \bibinfo {author} {\bibfnamefont {M.~A.}\ \bibnamefont {Scheel}},
  \ and\ \bibinfo {author} {\bibfnamefont {S.~A.}\ \bibnamefont {Teukolsky}},\
  }\href {\doibase 10.1016/S0010-4655(02)00847-0} {\bibfield  {journal}
  {\bibinfo  {journal} {Comp. Phys. Comm.}\ }\textbf {\bibinfo {volume}
  {152}},\ \bibinfo {pages} {253} (\bibinfo {year} {2003})},\ \Eprint
  {http://arxiv.org/abs/gr-qc/0202096} {gr-qc/0202096} \BibitemShut {NoStop}%
\bibitem [{\citenamefont {Lovelace}\ \emph {et~al.}(2008)\citenamefont
  {Lovelace}, \citenamefont {Owen}, \citenamefont {Pfeiffer},\ and\
  \citenamefont {Chu}}]{Lovelace2008}%
  \BibitemOpen
  \bibfield  {author} {\bibinfo {author} {\bibfnamefont {G.}~\bibnamefont
  {Lovelace}}, \bibinfo {author} {\bibfnamefont {R.}~\bibnamefont {Owen}},
  \bibinfo {author} {\bibfnamefont {H.~P.}\ \bibnamefont {Pfeiffer}}, \ and\
  \bibinfo {author} {\bibfnamefont {T.}~\bibnamefont {Chu}},\ }\href {\doibase
  10.1103/PhysRevD.78.084017} {\bibfield  {journal} {\bibinfo  {journal} {Phys.
  Rev. D}\ }\textbf {\bibinfo {volume} {78}},\ \bibinfo {pages} {084017}
  (\bibinfo {year} {2008})}\BibitemShut {NoStop}%
\bibitem [{\citenamefont {Mrou\'e}\ and\ \citenamefont
  {Pfeiffer}(2012)}]{Mroue:2012kv}%
  \BibitemOpen
  \bibfield  {author} {\bibinfo {author} {\bibfnamefont {A.~H.}\ \bibnamefont
  {Mrou\'e}}\ and\ \bibinfo {author} {\bibfnamefont {H.~P.}\ \bibnamefont
  {Pfeiffer}},\ }\href@noop {} {\  (\bibinfo {year} {2012})},\ \Eprint
  {http://arxiv.org/abs/1210.2958} {arXiv:1210.2958 [gr-qc]} \BibitemShut
  {NoStop}%
\bibitem [{\citenamefont {Buonanno}\ \emph {et~al.}(2011)\citenamefont
  {Buonanno}, \citenamefont {Kidder}, \citenamefont {Mrou\'{e}}, \citenamefont
  {Pfeiffer},\ and\ \citenamefont {Taracchini}}]{Buonanno:2010yk}%
  \BibitemOpen
  \bibfield  {author} {\bibinfo {author} {\bibfnamefont {A.}~\bibnamefont
  {Buonanno}}, \bibinfo {author} {\bibfnamefont {L.~E.}\ \bibnamefont
  {Kidder}}, \bibinfo {author} {\bibfnamefont {A.~H.}\ \bibnamefont
  {Mrou\'{e}}}, \bibinfo {author} {\bibfnamefont {H.~P.}\ \bibnamefont
  {Pfeiffer}}, \ and\ \bibinfo {author} {\bibfnamefont {A.}~\bibnamefont
  {Taracchini}},\ }\href {\doibase 10.1103/PhysRevD.83.104034} {\bibfield
  {journal} {\bibinfo  {journal} {Phys. Rev. D}\ }\textbf {\bibinfo {volume}
  {83}},\ \bibinfo {pages} {104034} (\bibinfo {year} {2011})},\ \Eprint
  {http://arxiv.org/abs/1012.1549} {arXiv:1012.1549 [gr-qc]} \BibitemShut
  {NoStop}%
\bibitem [{\citenamefont {Buchman}\ \emph {et~al.}(2012)\citenamefont
  {Buchman}, \citenamefont {Pfeiffer}, \citenamefont {Scheel},\ and\
  \citenamefont {Szil{\' a}gyi}}]{Buchman:2012dw}%
  \BibitemOpen
  \bibfield  {author} {\bibinfo {author} {\bibfnamefont {L.~T.}\ \bibnamefont
  {Buchman}}, \bibinfo {author} {\bibfnamefont {H.~P.}\ \bibnamefont
  {Pfeiffer}}, \bibinfo {author} {\bibfnamefont {M.~A.}\ \bibnamefont
  {Scheel}}, \ and\ \bibinfo {author} {\bibfnamefont {B.}~\bibnamefont {Szil{\'
  a}gyi}},\ }\href@noop {} {\bibfield  {journal} {\bibinfo  {journal} {Phys.
  Rev. D}\ }\textbf {\bibinfo {volume} {86}},\ \bibinfo {pages} {084033}
  (\bibinfo {year} {2012})},\ \Eprint {http://arxiv.org/abs/1206.3015}
  {arXiv:1206.3015 [gr-qc]} \BibitemShut {NoStop}%
\bibitem [{\citenamefont {Szil{\'a}gyi}(2014)}]{Szilagyi:2014fna}%
  \BibitemOpen
  \bibfield  {author} {\bibinfo {author} {\bibfnamefont {B.}~\bibnamefont
  {Szil{\'a}gyi}},\ }\href {\doibase 10.1142/S0218271814300146} {\bibfield
  {journal} {\bibinfo  {journal} {Int. J. Mod. Phys. D}\ }\textbf {\bibinfo
  {volume} {23}},\ \bibinfo {pages} {1430014} (\bibinfo {year} {2014})},\
  \Eprint {http://arxiv.org/abs/1405.3693} {arXiv:1405.3693 [gr-qc]}
  \BibitemShut {NoStop}%
\bibitem [{\citenamefont {Friedrich}(1985)}]{Friedrich1985}%
  \BibitemOpen
  \bibfield  {author} {\bibinfo {author} {\bibfnamefont {H.}~\bibnamefont
  {Friedrich}},\ }\href {\doibase 10.1007/BF01217728} {\bibfield  {journal}
  {\bibinfo  {journal} {Comm. Math. Phys.}\ }\textbf {\bibinfo {volume}
  {100}},\ \bibinfo {pages} {525} (\bibinfo {year} {1985})}\BibitemShut
  {NoStop}%
\bibitem [{\citenamefont {Garfinkle}(2002)}]{Garfinkle2002}%
  \BibitemOpen
  \bibfield  {author} {\bibinfo {author} {\bibfnamefont {D.}~\bibnamefont
  {Garfinkle}},\ }\href@noop {} {\bibfield  {journal} {\bibinfo  {journal}
  {Phys. Rev. D}\ }\textbf {\bibinfo {volume} {65}},\ \bibinfo {pages} {044029}
  (\bibinfo {year} {2002})}\BibitemShut {NoStop}%
\bibitem [{\citenamefont {Pretorius}(2005)}]{Pretorius2005c}%
  \BibitemOpen
  \bibfield  {author} {\bibinfo {author} {\bibfnamefont {F.}~\bibnamefont
  {Pretorius}},\ }\href {http://stacks.iop.org/0264-9381/22/425} {\bibfield
  {journal} {\bibinfo  {journal} {Class. Quant. Grav.}\ }\textbf {\bibinfo
  {volume} {22}},\ \bibinfo {pages} {425} (\bibinfo {year} {2005})}\BibitemShut
  {NoStop}%
\bibitem [{\citenamefont {Lindblom}\ \emph
  {et~al.}(2006{\natexlab{a}})\citenamefont {Lindblom}, \citenamefont {Scheel},
  \citenamefont {Kidder}, \citenamefont {Owen},\ and\ \citenamefont
  {Rinne}}]{Lindblom:2007}%
  \BibitemOpen
  \bibfield  {author} {\bibinfo {author} {\bibfnamefont {L.}~\bibnamefont
  {Lindblom}}, \bibinfo {author} {\bibfnamefont {M.~A.}\ \bibnamefont
  {Scheel}}, \bibinfo {author} {\bibfnamefont {L.~E.}\ \bibnamefont {Kidder}},
  \bibinfo {author} {\bibfnamefont {R.}~\bibnamefont {Owen}}, \ and\ \bibinfo
  {author} {\bibfnamefont {O.}~\bibnamefont {Rinne}},\ }\href {\doibase
  10.1088/0264-9381/23/16/S09} {\bibfield  {journal} {\bibinfo  {journal}
  {Class. Quant. Grav.}\ }\textbf {\bibinfo {volume} {23}},\ \bibinfo {pages}
  {S447} (\bibinfo {year} {2006}{\natexlab{a}})},\ \Eprint
  {http://arxiv.org/abs/gr-qc/0512093v3} {arXiv:gr-qc/0512093v3 [gr-qc]}
  \BibitemShut {NoStop}%
\bibitem [{\citenamefont {Szil{\' a}gyi}\ \emph {et~al.}(2009)\citenamefont
  {Szil{\' a}gyi}, \citenamefont {Lindblom},\ and\ \citenamefont
  {Scheel}}]{Szilagyi:2009qz}%
  \BibitemOpen
  \bibfield  {author} {\bibinfo {author} {\bibfnamefont {B.}~\bibnamefont
  {Szil{\' a}gyi}}, \bibinfo {author} {\bibfnamefont {L.}~\bibnamefont
  {Lindblom}}, \ and\ \bibinfo {author} {\bibfnamefont {M.~A.}\ \bibnamefont
  {Scheel}},\ }\href@noop {} {\bibfield  {journal} {\bibinfo  {journal} {Phys.
  Rev. D}\ }\textbf {\bibinfo {volume} {80}},\ \bibinfo {pages} {124010}
  (\bibinfo {year} {2009})},\ \Eprint {http://arxiv.org/abs/0909.3557}
  {arXiv:0909.3557 [gr-qc]} \BibitemShut {NoStop}%
\bibitem [{\citenamefont {{M. A. Scheel, M. Boyle, T. Chu, L. E. Kidder, K. D.
  Matthews and H. P. Pfeiffer}}(2009)}]{Scheel2009}%
  \BibitemOpen
  \bibfield  {author} {\bibinfo {author} {\bibnamefont {{M. A. Scheel, M.
  Boyle, T. Chu, L. E. Kidder, K. D. Matthews and H. P. Pfeiffer}}},\
  }\href@noop {} {\bibfield  {journal} {\bibinfo  {journal} {Phys. Rev. D}\
  }\textbf {\bibinfo {volume} {79}},\ \bibinfo {pages} {024003} (\bibinfo
  {year} {2009})},\ \Eprint {http://arxiv.org/abs/arXiv:gr-qc/0810.1767}
  {arXiv:gr-qc/0810.1767} \BibitemShut {NoStop}%
\bibitem [{\citenamefont {Hemberger}\ \emph {et~al.}(2013)\citenamefont
  {Hemberger}, \citenamefont {Scheel}, \citenamefont {Kidder}, \citenamefont
  {Szil{\'a}gyi}, \citenamefont {Lovelace}, \citenamefont {Taylor},\ and\
  \citenamefont {Teukolsky}}]{Hemberger:2012jz}%
  \BibitemOpen
  \bibfield  {author} {\bibinfo {author} {\bibfnamefont {D.~A.}\ \bibnamefont
  {Hemberger}}, \bibinfo {author} {\bibfnamefont {M.~A.}\ \bibnamefont
  {Scheel}}, \bibinfo {author} {\bibfnamefont {L.~E.}\ \bibnamefont {Kidder}},
  \bibinfo {author} {\bibfnamefont {B.}~\bibnamefont {Szil{\'a}gyi}}, \bibinfo
  {author} {\bibfnamefont {G.}~\bibnamefont {Lovelace}}, \bibinfo {author}
  {\bibfnamefont {N.~W.}\ \bibnamefont {Taylor}}, \ and\ \bibinfo {author}
  {\bibfnamefont {S.~A.}\ \bibnamefont {Teukolsky}},\ }\href {\doibase
  10.1088/0264-9381/30/11/115001} {\bibfield  {journal} {\bibinfo  {journal}
  {Class. Quant. Grav.}\ }\textbf {\bibinfo {volume} {30}},\ \bibinfo {pages}
  {115001} (\bibinfo {year} {2013})},\ \Eprint {http://arxiv.org/abs/1211.6079}
  {arXiv:1211.6079 [gr-qc]} \BibitemShut {NoStop}%
\bibitem [{\citenamefont {Lindblom}\ \emph
  {et~al.}(2006{\natexlab{b}})\citenamefont {Lindblom}, \citenamefont {Scheel},
  \citenamefont {Kidder}, \citenamefont {Owen},\ and\ \citenamefont
  {Rinne}}]{Lindblom2006}%
  \BibitemOpen
  \bibfield  {author} {\bibinfo {author} {\bibfnamefont {L.}~\bibnamefont
  {Lindblom}}, \bibinfo {author} {\bibfnamefont {M.~A.}\ \bibnamefont
  {Scheel}}, \bibinfo {author} {\bibfnamefont {L.~E.}\ \bibnamefont {Kidder}},
  \bibinfo {author} {\bibfnamefont {R.}~\bibnamefont {Owen}}, \ and\ \bibinfo
  {author} {\bibfnamefont {O.}~\bibnamefont {Rinne}},\ }\href {\doibase
  10.1088/0264-9381/23/16/S09} {\bibfield  {journal} {\bibinfo  {journal}
  {Class. Quant. Grav.}\ }\textbf {\bibinfo {volume} {23}},\ \bibinfo {pages}
  {447} (\bibinfo {year} {2006}{\natexlab{b}})},\ \Eprint
  {http://arxiv.org/abs/gr-qc/0512093} {gr-qc/0512093} \BibitemShut {NoStop}%
\bibitem [{\citenamefont {Rinne}(2006)}]{Rinne2006}%
  \BibitemOpen
  \bibfield  {author} {\bibinfo {author} {\bibfnamefont {O.}~\bibnamefont
  {Rinne}},\ }\href {http://stacks.iop.org/0264-9381/23/6275} {\bibfield
  {journal} {\bibinfo  {journal} {Class. Quant. Grav.}\ }\textbf {\bibinfo
  {volume} {23}},\ \bibinfo {pages} {6275} (\bibinfo {year}
  {2006})}\BibitemShut {NoStop}%
\bibitem [{\citenamefont {Rinne}\ \emph {et~al.}(2007)\citenamefont {Rinne},
  \citenamefont {Lindblom},\ and\ \citenamefont {Scheel}}]{Rinne2007}%
  \BibitemOpen
  \bibfield  {author} {\bibinfo {author} {\bibfnamefont {O.}~\bibnamefont
  {Rinne}}, \bibinfo {author} {\bibfnamefont {L.}~\bibnamefont {Lindblom}}, \
  and\ \bibinfo {author} {\bibfnamefont {M.~A.}\ \bibnamefont {Scheel}},\
  }\href {http://stacks.iop.org/0264-9381/24/4053} {\bibfield  {journal}
  {\bibinfo  {journal} {Class. Quant. Grav.}\ }\textbf {\bibinfo {volume}
  {24}},\ \bibinfo {pages} {4053} (\bibinfo {year} {2007})}\BibitemShut
  {NoStop}%
\bibitem [{\citenamefont {Gottlieb}\ and\ \citenamefont
  {Hesthaven}(2001)}]{Gottlieb2001}%
  \BibitemOpen
  \bibfield  {author} {\bibinfo {author} {\bibfnamefont {D.}~\bibnamefont
  {Gottlieb}}\ and\ \bibinfo {author} {\bibfnamefont {J.~S.}\ \bibnamefont
  {Hesthaven}},\ }\href {\doibase 10.1016/S0377-0427(00)00510-0} {\bibfield
  {journal} {\bibinfo  {journal} {J. Comput. Appl. Math.}\ }\textbf {\bibinfo
  {volume} {128}},\ \bibinfo {pages} {83} (\bibinfo {year} {2001})}\BibitemShut
  {NoStop}%
\bibitem [{\citenamefont {Hesthaven}(2000)}]{Hesthaven2000}%
  \BibitemOpen
  \bibfield  {author} {\bibinfo {author} {\bibfnamefont {J.~S.}\ \bibnamefont
  {Hesthaven}},\ }\href@noop {} {\bibfield  {journal} {\bibinfo  {journal}
  {Appl. Num. Math.}\ }\textbf {\bibinfo {volume} {33}},\ \bibinfo {pages} {23}
  (\bibinfo {year} {2000})}\BibitemShut {NoStop}%
\bibitem [{\citenamefont {Boyle}\ and\ \citenamefont
  {Mrou{\'{e}}}(2009)}]{Boyle-Mroue:2008}%
  \BibitemOpen
  \bibfield  {author} {\bibinfo {author} {\bibfnamefont {M.}~\bibnamefont
  {Boyle}}\ and\ \bibinfo {author} {\bibfnamefont {A.~H.}\ \bibnamefont
  {Mrou{\'{e}}}},\ }\href {\doibase 10.1103/PhysRevD.80.124045} {\bibfield
  {journal} {\bibinfo  {journal} {Phys. Rev. D}\ }\textbf {\bibinfo {volume}
  {80}},\ \bibinfo {pages} {124045} (\bibinfo {year} {2009})},\ \Eprint
  {http://arxiv.org/abs/0905.3177} {arXiv:0905.3177 [gr-qc]} \BibitemShut
  {NoStop}%
\bibitem [{\citenamefont {Taylor}\ \emph {et~al.}(2013)\citenamefont {Taylor},
  \citenamefont {Boyle}, \citenamefont {Reisswig}, \citenamefont {Scheel},
  \citenamefont {Chu}, \citenamefont {Kidder},\ and\ \citenamefont {Szil{\'
  a}gyi}}]{Taylor:2013zia}%
  \BibitemOpen
  \bibfield  {author} {\bibinfo {author} {\bibfnamefont {N.~W.}\ \bibnamefont
  {Taylor}}, \bibinfo {author} {\bibfnamefont {M.}~\bibnamefont {Boyle}},
  \bibinfo {author} {\bibfnamefont {C.}~\bibnamefont {Reisswig}}, \bibinfo
  {author} {\bibfnamefont {M.~A.}\ \bibnamefont {Scheel}}, \bibinfo {author}
  {\bibfnamefont {T.}~\bibnamefont {Chu}}, \bibinfo {author} {\bibfnamefont
  {L.~E.}\ \bibnamefont {Kidder}}, \ and\ \bibinfo {author} {\bibfnamefont
  {B.}~\bibnamefont {Szil{\' a}gyi}},\ }\href {\doibase
  10.1103/PhysRevD.88.124010} {\bibfield  {journal} {\bibinfo  {journal} {Phys.
  Rev. D}\ }\textbf {\bibinfo {volume} {88}},\ \bibinfo {pages} {124010}
  (\bibinfo {year} {2013})},\ \Eprint {http://arxiv.org/abs/1309.3605}
  {arXiv:1309.3605 [gr-qc]} \BibitemShut {NoStop}%
\bibitem [{\citenamefont {{Chu}}\ \emph {et~al.}(2016)\citenamefont {{Chu}},
  \citenamefont {{Fong}}, \citenamefont {{Kumar}}, \citenamefont {{Pfeiffer}},
  \citenamefont {{Boyle}}, \citenamefont {{Hemberger}}, \citenamefont
  {{Kidder}}, \citenamefont {{Scheel}},\ and\ \citenamefont
  {{Szilagyi}}}]{Chu:2015kft}%
  \BibitemOpen
  \bibfield  {author} {\bibinfo {author} {\bibfnamefont {T.}~\bibnamefont
  {{Chu}}}, \bibinfo {author} {\bibfnamefont {H.}~\bibnamefont {{Fong}}},
  \bibinfo {author} {\bibfnamefont {P.}~\bibnamefont {{Kumar}}}, \bibinfo
  {author} {\bibfnamefont {H.~P.}\ \bibnamefont {{Pfeiffer}}}, \bibinfo
  {author} {\bibfnamefont {M.}~\bibnamefont {{Boyle}}}, \bibinfo {author}
  {\bibfnamefont {D.~A.}\ \bibnamefont {{Hemberger}}}, \bibinfo {author}
  {\bibfnamefont {L.~E.}\ \bibnamefont {{Kidder}}}, \bibinfo {author}
  {\bibfnamefont {M.~A.}\ \bibnamefont {{Scheel}}}, \ and\ \bibinfo {author}
  {\bibfnamefont {B.}~\bibnamefont {{Szilagyi}}},\ }\href {\doibase
  10.1088/0264-9381/33/16/165001} {\bibfield  {journal} {\bibinfo  {journal}
  {Classical and Quantum Gravity}\ }\textbf {\bibinfo {volume} {33}},\ \bibinfo
  {eid} {165001} (\bibinfo {year} {2016})},\ \Eprint
  {http://arxiv.org/abs/1512.06800} {arXiv:1512.06800 [gr-qc]} \BibitemShut
  {NoStop}%
\bibitem [{\citenamefont {L{\"{o}}ffler}\ \emph {et~al.}(2012)\citenamefont
  {L{\"{o}}ffler}, \citenamefont {Faber}, \citenamefont {Bentivegna},
  \citenamefont {Bode}, \citenamefont {Diener}, \citenamefont {Haas},
  \citenamefont {Hinder}, \citenamefont {Mundim}, \citenamefont {Ott},
  \citenamefont {Schnetter}, \citenamefont {Allen}, \citenamefont
  {Campanelli},\ and\ \citenamefont {Laguna}}]{Loffler:2011}%
  \BibitemOpen
  \bibfield  {author} {\bibinfo {author} {\bibfnamefont {F.}~\bibnamefont
  {L{\"{o}}ffler}}, \bibinfo {author} {\bibfnamefont {J.}~\bibnamefont
  {Faber}}, \bibinfo {author} {\bibfnamefont {E.}~\bibnamefont {Bentivegna}},
  \bibinfo {author} {\bibfnamefont {T.}~\bibnamefont {Bode}}, \bibinfo {author}
  {\bibfnamefont {P.}~\bibnamefont {Diener}}, \bibinfo {author} {\bibfnamefont
  {R.}~\bibnamefont {Haas}}, \bibinfo {author} {\bibfnamefont {I.}~\bibnamefont
  {Hinder}}, \bibinfo {author} {\bibfnamefont {B.~C.}\ \bibnamefont {Mundim}},
  \bibinfo {author} {\bibfnamefont {C.~D.}\ \bibnamefont {Ott}}, \bibinfo
  {author} {\bibfnamefont {E.}~\bibnamefont {Schnetter}}, \bibinfo {author}
  {\bibfnamefont {G.}~\bibnamefont {Allen}}, \bibinfo {author} {\bibfnamefont
  {M.}~\bibnamefont {Campanelli}}, \ and\ \bibinfo {author} {\bibfnamefont
  {P.}~\bibnamefont {Laguna}},\ }\href {\doibase
  doi:10.1088/0264-9381/29/11/115001} {\bibfield  {journal} {\bibinfo
  {journal} {Class. Quantum Grav.}\ }\textbf {\bibinfo {volume} {29}},\
  \bibinfo {pages} {115001} (\bibinfo {year} {2012})},\ \Eprint
  {http://arxiv.org/abs/arXiv:1111.3344 [gr-qc]} {arXiv:1111.3344 [gr-qc]}
  \BibitemShut {NoStop}%
\bibitem [{\citenamefont {Goodale}\ \emph {et~al.}(2003)\citenamefont
  {Goodale}, \citenamefont {Allen}, \citenamefont {Lanfermann}, \citenamefont
  {Mass{\'o}}, \citenamefont {Radke}, \citenamefont {Seidel},\ and\
  \citenamefont {Shalf}}]{Goodale:2002a}%
  \BibitemOpen
  \bibfield  {author} {\bibinfo {author} {\bibfnamefont {T.}~\bibnamefont
  {Goodale}}, \bibinfo {author} {\bibfnamefont {G.}~\bibnamefont {Allen}},
  \bibinfo {author} {\bibfnamefont {G.}~\bibnamefont {Lanfermann}}, \bibinfo
  {author} {\bibfnamefont {J.}~\bibnamefont {Mass{\'o}}}, \bibinfo {author}
  {\bibfnamefont {T.}~\bibnamefont {Radke}}, \bibinfo {author} {\bibfnamefont
  {E.}~\bibnamefont {Seidel}}, \ and\ \bibinfo {author} {\bibfnamefont
  {J.}~\bibnamefont {Shalf}},\ }in\ \href {http://edoc.mpg.de/3341} {\emph
  {\bibinfo {booktitle} {Vector and Parallel Processing -- VECPAR'2002, 5th
  International Conference, Lecture Notes in Computer Science}}}\ (\bibinfo
  {publisher} {Springer},\ \bibinfo {address} {Berlin},\ \bibinfo {year}
  {2003})\BibitemShut {NoStop}%
\bibitem [{\citenamefont {{Schnetter}}\ \emph {et~al.}(2004)\citenamefont
  {{Schnetter}}, \citenamefont {{Hawley}},\ and\ \citenamefont
  {{Hawke}}}]{Schnetter2004}%
  \BibitemOpen
  \bibfield  {author} {\bibinfo {author} {\bibfnamefont {E.}~\bibnamefont
  {{Schnetter}}}, \bibinfo {author} {\bibfnamefont {S.~H.}\ \bibnamefont
  {{Hawley}}}, \ and\ \bibinfo {author} {\bibfnamefont {I.}~\bibnamefont
  {{Hawke}}},\ }\href {\doibase 10.1088/0264-9381/21/6/014} {\bibfield
  {journal} {\bibinfo  {journal} {Classical and Quantum Gravity}\ }\textbf
  {\bibinfo {volume} {21}},\ \bibinfo {pages} {1465} (\bibinfo {year}
  {2004})},\ \Eprint {http://arxiv.org/abs/gr-qc/0310042} {gr-qc/0310042}
  \BibitemShut {NoStop}%
\bibitem [{\citenamefont {Ansorg}\ \emph {et~al.}(2004)\citenamefont {Ansorg},
  \citenamefont {Br{\"u}gmann},\ and\ \citenamefont {Tichy}}]{Ansorg2004}%
  \BibitemOpen
  \bibfield  {author} {\bibinfo {author} {\bibfnamefont {M.}~\bibnamefont
  {Ansorg}}, \bibinfo {author} {\bibfnamefont {B.}~\bibnamefont
  {Br{\"u}gmann}}, \ and\ \bibinfo {author} {\bibfnamefont {W.}~\bibnamefont
  {Tichy}},\ }\href {\doibase 10.1103/PhysRevD.70.064011} {\bibfield  {journal}
  {\bibinfo  {journal} {Phys. Rev. D}\ }\textbf {\bibinfo {volume} {70}},\
  \bibinfo {pages} {064011} (\bibinfo {year} {2004})},\ \Eprint
  {http://arxiv.org/abs/arXiv:gr-qc/0404056} {arXiv:gr-qc/0404056} \BibitemShut
  {NoStop}%
\bibitem [{\citenamefont {Kidder}(1995)}]{kidder}%
  \BibitemOpen
  \bibfield  {author} {\bibinfo {author} {\bibfnamefont {L.~E.}\ \bibnamefont
  {Kidder}},\ }\href {\doibase 10.1103/PhysRevD.52.821} {\bibfield  {journal}
  {\bibinfo  {journal} {Phys. Rev. D}\ }\textbf {\bibinfo {volume} {52}},\
  \bibinfo {pages} {821} (\bibinfo {year} {1995})}\BibitemShut {NoStop}%
\bibitem [{\citenamefont {Tichy}\ and\ \citenamefont
  {Marronetti}(2011)}]{tichy}%
  \BibitemOpen
  \bibfield  {author} {\bibinfo {author} {\bibfnamefont {W.}~\bibnamefont
  {Tichy}}\ and\ \bibinfo {author} {\bibfnamefont {P.}~\bibnamefont
  {Marronetti}},\ }\href {\doibase 10.1103/PhysRevD.83.024012} {\bibfield
  {journal} {\bibinfo  {journal} {Phys. Rev. D}\ }\textbf {\bibinfo {volume}
  {83}},\ \bibinfo {pages} {024012} (\bibinfo {year} {2011})}\BibitemShut
  {NoStop}%
\bibitem [{\citenamefont {Brown}\ \emph {et~al.}(2009)\citenamefont {Brown},
  \citenamefont {Diener}, \citenamefont {Sarbach}, \citenamefont {Schnetter},\
  and\ \citenamefont {Tiglio}}]{Brown2008}%
  \BibitemOpen
  \bibfield  {author} {\bibinfo {author} {\bibfnamefont {D.}~\bibnamefont
  {Brown}}, \bibinfo {author} {\bibfnamefont {P.}~\bibnamefont {Diener}},
  \bibinfo {author} {\bibfnamefont {O.}~\bibnamefont {Sarbach}}, \bibinfo
  {author} {\bibfnamefont {E.}~\bibnamefont {Schnetter}}, \ and\ \bibinfo
  {author} {\bibfnamefont {M.}~\bibnamefont {Tiglio}},\ }\href {\doibase
  10.1103/PhysRevD.79.044023} {\bibfield  {journal} {\bibinfo  {journal} {Phys.
  Rev. D}\ }\textbf {\bibinfo {volume} {79}},\ \bibinfo {pages} {044023}
  (\bibinfo {year} {2009})},\ \Eprint {http://arxiv.org/abs/arXiv:0809.3533
  [gr-qc]} {arXiv:0809.3533 [gr-qc]} \BibitemShut {NoStop}%
\bibitem [{\citenamefont {Reisswig}\ and\ \citenamefont
  {Pollney}(2011)}]{Reisswig2011}%
  \BibitemOpen
  \bibfield  {author} {\bibinfo {author} {\bibfnamefont {C.}~\bibnamefont
  {Reisswig}}\ and\ \bibinfo {author} {\bibfnamefont {D.}~\bibnamefont
  {Pollney}},\ }\href {http://stacks.iop.org/0264-9381/28/i=19/a=195015}
  {\bibfield  {journal} {\bibinfo  {journal} {Classical and Quantum Gravity}\
  }\textbf {\bibinfo {volume} {28}},\ \bibinfo {pages} {195015} (\bibinfo
  {year} {2011})}\BibitemShut {NoStop}%
\bibitem [{\citenamefont {Schmidt}\ \emph {et~al.}(2017)\citenamefont
  {Schmidt}, \citenamefont {Harry},\ and\ \citenamefont
  {Pfeiffer}}]{NRInfrastructure}%
  \BibitemOpen
  \bibfield  {author} {\bibinfo {author} {\bibfnamefont {P.}~\bibnamefont
  {Schmidt}}, \bibinfo {author} {\bibfnamefont {I.~W.}\ \bibnamefont {Harry}},
  \ and\ \bibinfo {author} {\bibfnamefont {H.~P.}\ \bibnamefont {Pfeiffer}},\
  }\href@noop {} {\  (\bibinfo {year} {2017})},\ \Eprint
  {http://arxiv.org/abs/1703.01076} {arXiv:1703.01076 [gr-qc]} \BibitemShut
  {NoStop}%
\bibitem [{\citenamefont {Kidder}\ \emph {et~al.}(1993)\citenamefont {Kidder},
  \citenamefont {Will},\ and\ \citenamefont {Wiseman}}]{Kidder_Will}%
  \BibitemOpen
  \bibfield  {author} {\bibinfo {author} {\bibfnamefont {L.~E.}\ \bibnamefont
  {Kidder}}, \bibinfo {author} {\bibfnamefont {C.~M.}\ \bibnamefont {Will}}, \
  and\ \bibinfo {author} {\bibfnamefont {A.~G.}\ \bibnamefont {Wiseman}},\
  }\href {\doibase 10.1103/PhysRevD.47.R4183} {\bibfield  {journal} {\bibinfo
  {journal} {Phys. Rev. D}\ }\textbf {\bibinfo {volume} {47}},\ \bibinfo
  {pages} {R4183} (\bibinfo {year} {1993})}\BibitemShut {NoStop}%
\bibitem [{\citenamefont {P\"urrer}\ \emph {et~al.}(2013)\citenamefont
  {P\"urrer}, \citenamefont {Hannam}, \citenamefont {Ajith},\ and\
  \citenamefont {Husa}}]{Purrer:2013xma}%
  \BibitemOpen
  \bibfield  {author} {\bibinfo {author} {\bibfnamefont {M.}~\bibnamefont
  {P\"urrer}}, \bibinfo {author} {\bibfnamefont {M.}~\bibnamefont {Hannam}},
  \bibinfo {author} {\bibfnamefont {P.}~\bibnamefont {Ajith}}, \ and\ \bibinfo
  {author} {\bibfnamefont {S.}~\bibnamefont {Husa}},\ }\href {\doibase
  10.1103/PhysRevD.88.064007} {\bibfield  {journal} {\bibinfo  {journal} {Phys.
  Rev. D}\ }\textbf {\bibinfo {volume} {88}},\ \bibinfo {pages} {064007}
  (\bibinfo {year} {2013})}\BibitemShut {NoStop}%
\bibitem [{\citenamefont {Buonanno}\ and\ \citenamefont
  {Damour}(1999)}]{Buonanno99}%
  \BibitemOpen
  \bibfield  {author} {\bibinfo {author} {\bibfnamefont {A.}~\bibnamefont
  {Buonanno}}\ and\ \bibinfo {author} {\bibfnamefont {T.}~\bibnamefont
  {Damour}},\ }\href {\doibase 10.1103/PhysRevD.59.084006} {\bibfield
  {journal} {\bibinfo  {journal} {Phys. Rev. D}\ }\textbf {\bibinfo {volume}
  {59}},\ \bibinfo {pages} {084006} (\bibinfo {year} {1999})}\BibitemShut
  {NoStop}%
\bibitem [{\citenamefont {Hannam}\ \emph
  {et~al.}(2014{\natexlab{b}})\citenamefont {Hannam}, \citenamefont {Schmidt},
  \citenamefont {Boh\'e}, \citenamefont {Haegel}, \citenamefont {Husa},
  \citenamefont {Ohme}, \citenamefont {Pratten},\ and\ \citenamefont
  {P\"urrer}}]{HSBH2014}%
  \BibitemOpen
  \bibfield  {author} {\bibinfo {author} {\bibfnamefont {M.}~\bibnamefont
  {Hannam}}, \bibinfo {author} {\bibfnamefont {P.}~\bibnamefont {Schmidt}},
  \bibinfo {author} {\bibfnamefont {A.}~\bibnamefont {Boh\'e}}, \bibinfo
  {author} {\bibfnamefont {L.}~\bibnamefont {Haegel}}, \bibinfo {author}
  {\bibfnamefont {S.}~\bibnamefont {Husa}}, \bibinfo {author} {\bibfnamefont
  {F.}~\bibnamefont {Ohme}}, \bibinfo {author} {\bibfnamefont {G.}~\bibnamefont
  {Pratten}}, \ and\ \bibinfo {author} {\bibfnamefont {M.}~\bibnamefont
  {P\"urrer}},\ }\href {\doibase 10.1103/PhysRevLett.113.151101} {\bibfield
  {journal} {\bibinfo  {journal} {Phys. Rev. Lett.}\ }\textbf {\bibinfo
  {volume} {113}},\ \bibinfo {pages} {151101} (\bibinfo {year}
  {2014}{\natexlab{b}})}\BibitemShut {NoStop}%
\bibitem [{\citenamefont {{Schmidt}}\ \emph {et~al.}(2015)\citenamefont
  {{Schmidt}}, \citenamefont {{Ohme}},\ and\ \citenamefont
  {{Hannam}}}]{Schmidt2015}%
  \BibitemOpen
  \bibfield  {author} {\bibinfo {author} {\bibfnamefont {P.}~\bibnamefont
  {{Schmidt}}}, \bibinfo {author} {\bibfnamefont {F.}~\bibnamefont {{Ohme}}}, \
  and\ \bibinfo {author} {\bibfnamefont {M.}~\bibnamefont {{Hannam}}},\ }\href
  {\doibase 10.1103/PhysRevD.91.024043} {\bibfield  {journal} {\bibinfo
  {journal} {\prd}\ }\textbf {\bibinfo {volume} {91}},\ \bibinfo {eid} {024043}
  (\bibinfo {year} {2015})},\ \Eprint {http://arxiv.org/abs/1408.1810}
  {arXiv:1408.1810 [gr-qc]} \BibitemShut {NoStop}%
\bibitem [{phe()}]{phenompv2}%
  \BibitemOpen
  \href@noop {} {}\bibinfo {howpublished}
  {\url{https://dcc.ligo.org/DocDB/0122/T1500602/003/PhenomPv2_technicalnotes.pdf}}\BibitemShut
  {NoStop}%
\bibitem [{\citenamefont {Owen}(1996)}]{geometric}%
  \BibitemOpen
  \bibfield  {author} {\bibinfo {author} {\bibfnamefont {B.~J.}\ \bibnamefont
  {Owen}},\ }\href {\doibase 10.1103/PhysRevD.53.6749} {\bibfield  {journal}
  {\bibinfo  {journal} {Phys. Rev. D}\ }\textbf {\bibinfo {volume} {53}},\
  \bibinfo {pages} {6749} (\bibinfo {year} {1996})}\BibitemShut {NoStop}%
\bibitem [{\citenamefont {Harry}\ \emph {et~al.}(2009)\citenamefont {Harry},
  \citenamefont {Allen},\ and\ \citenamefont {Sathyaprakash}}]{stochastic}%
  \BibitemOpen
  \bibfield  {author} {\bibinfo {author} {\bibfnamefont {I.~W.}\ \bibnamefont
  {Harry}}, \bibinfo {author} {\bibfnamefont {B.}~\bibnamefont {Allen}}, \ and\
  \bibinfo {author} {\bibfnamefont {B.~S.}\ \bibnamefont {Sathyaprakash}},\
  }\href {\doibase 10.1103/PhysRevD.80.104014} {\bibfield  {journal} {\bibinfo
  {journal} {Phys. Rev. D}\ }\textbf {\bibinfo {volume} {80}},\ \bibinfo
  {pages} {104014} (\bibinfo {year} {2009})}\BibitemShut {NoStop}%
\bibitem [{\citenamefont {Capano}\ \emph {et~al.}(2016)\citenamefont {Capano},
  \citenamefont {Harry}, \citenamefont {Privitera},\ and\ \citenamefont
  {Buonanno}}]{stochastic2_CHPB}%
  \BibitemOpen
  \bibfield  {author} {\bibinfo {author} {\bibfnamefont {C.}~\bibnamefont
  {Capano}}, \bibinfo {author} {\bibfnamefont {I.}~\bibnamefont {Harry}},
  \bibinfo {author} {\bibfnamefont {S.}~\bibnamefont {Privitera}}, \ and\
  \bibinfo {author} {\bibfnamefont {A.}~\bibnamefont {Buonanno}},\ }\href
  {\doibase 10.1103/PhysRevD.93.124007} {\bibfield  {journal} {\bibinfo
  {journal} {Phys. Rev. D}\ }\textbf {\bibinfo {volume} {93}},\ \bibinfo
  {pages} {124007} (\bibinfo {year} {2016})}\BibitemShut {NoStop}%
\bibitem [{\citenamefont {Ajith}\ \emph {et~al.}(2014)\citenamefont {Ajith},
  \citenamefont {Fotopoulos}, \citenamefont {Privitera}, \citenamefont
  {Neunzert}, \citenamefont {Mazumder},\ and\ \citenamefont
  {Weinstein}}]{stochastic_AFPNMW}%
  \BibitemOpen
  \bibfield  {author} {\bibinfo {author} {\bibfnamefont {P.}~\bibnamefont
  {Ajith}}, \bibinfo {author} {\bibfnamefont {N.}~\bibnamefont {Fotopoulos}},
  \bibinfo {author} {\bibfnamefont {S.}~\bibnamefont {Privitera}}, \bibinfo
  {author} {\bibfnamefont {A.}~\bibnamefont {Neunzert}}, \bibinfo {author}
  {\bibfnamefont {N.}~\bibnamefont {Mazumder}}, \ and\ \bibinfo {author}
  {\bibfnamefont {A.~J.}\ \bibnamefont {Weinstein}},\ }\href {\doibase
  10.1103/PhysRevD.89.084041} {\bibfield  {journal} {\bibinfo  {journal} {Phys.
  Rev. D}\ }\textbf {\bibinfo {volume} {89}},\ \bibinfo {pages} {084041}
  (\bibinfo {year} {2014})}\BibitemShut {NoStop}%
\bibitem [{\citenamefont {Dal~Canton}\ \emph {et~al.}(2014)\citenamefont
  {Dal~Canton}, \citenamefont {Nitz}, \citenamefont {Lundgren}, \citenamefont
  {Nielsen}, \citenamefont {Brown}, \citenamefont {Dent}, \citenamefont
  {Harry}, \citenamefont {Krishnan}, \citenamefont {Miller}, \citenamefont
  {Wette}, \citenamefont {Wiesner},\ and\ \citenamefont
  {Willis}}]{pycbc_Canton_2014}%
  \BibitemOpen
  \bibfield  {author} {\bibinfo {author} {\bibfnamefont {T.}~\bibnamefont
  {Dal~Canton}}, \bibinfo {author} {\bibfnamefont {A.~H.}\ \bibnamefont
  {Nitz}}, \bibinfo {author} {\bibfnamefont {A.~P.}\ \bibnamefont {Lundgren}},
  \bibinfo {author} {\bibfnamefont {A.~B.}\ \bibnamefont {Nielsen}}, \bibinfo
  {author} {\bibfnamefont {D.~A.}\ \bibnamefont {Brown}}, \bibinfo {author}
  {\bibfnamefont {T.}~\bibnamefont {Dent}}, \bibinfo {author} {\bibfnamefont
  {I.~W.}\ \bibnamefont {Harry}}, \bibinfo {author} {\bibfnamefont
  {B.}~\bibnamefont {Krishnan}}, \bibinfo {author} {\bibfnamefont {A.~J.}\
  \bibnamefont {Miller}}, \bibinfo {author} {\bibfnamefont {K.}~\bibnamefont
  {Wette}}, \bibinfo {author} {\bibfnamefont {K.}~\bibnamefont {Wiesner}}, \
  and\ \bibinfo {author} {\bibfnamefont {J.~L.}\ \bibnamefont {Willis}},\
  }\href {\doibase 10.1103/PhysRevD.90.082004} {\bibfield  {journal} {\bibinfo
  {journal} {Phys. Rev. D}\ }\textbf {\bibinfo {volume} {90}},\ \bibinfo
  {pages} {082004} (\bibinfo {year} {2014})}\BibitemShut {NoStop}%
\bibitem [{\citenamefont {Kennedy}\ and\ \citenamefont {Eberhart}(1995)}]{pso}%
  \BibitemOpen
  \bibfield  {author} {\bibinfo {author} {\bibfnamefont {J.}~\bibnamefont
  {Kennedy}}\ and\ \bibinfo {author} {\bibfnamefont {R.}~\bibnamefont
  {Eberhart}},\ }in\ \href {\doibase 10.1109/ICNN.1995.488968} {\emph {\bibinfo
  {booktitle} {Neural Networks, 1995. Proceedings., IEEE International
  Conference on}}},\ Vol.~\bibinfo {volume} {4}\ (\bibinfo {year} {1995})\ pp.\
  \bibinfo {pages} {1942--1948 vol.4}\BibitemShut {NoStop}%
\bibitem [{\citenamefont {{Brown}}\ \emph {et~al.}(2012)\citenamefont
  {{Brown}}, \citenamefont {{Lundgren}},\ and\ \citenamefont
  {{O'Shaughnessy}}}]{Brown2012}%
  \BibitemOpen
  \bibfield  {author} {\bibinfo {author} {\bibfnamefont {D.~A.}\ \bibnamefont
  {{Brown}}}, \bibinfo {author} {\bibfnamefont {A.}~\bibnamefont {{Lundgren}}},
  \ and\ \bibinfo {author} {\bibfnamefont {R.}~\bibnamefont
  {{O'Shaughnessy}}},\ }\href {\doibase 10.1103/PhysRevD.86.064020} {\bibfield
  {journal} {\bibinfo  {journal} {\prd}\ }\textbf {\bibinfo {volume} {86}},\
  \bibinfo {eid} {064020} (\bibinfo {year} {2012})},\ \Eprint
  {http://arxiv.org/abs/1203.6060} {arXiv:1203.6060 [gr-qc]} \BibitemShut
  {NoStop}%
\bibitem [{\citenamefont {{Harry}}\ \emph {et~al.}(2014)\citenamefont
  {{Harry}}, \citenamefont {{Nitz}}, \citenamefont {{Brown}}, \citenamefont
  {{Lundgren}}, \citenamefont {{Ochsner}},\ and\ \citenamefont
  {{Keppel}}}]{Harry_Nitz_Brown_2014}%
  \BibitemOpen
  \bibfield  {author} {\bibinfo {author} {\bibfnamefont {I.~W.}\ \bibnamefont
  {{Harry}}}, \bibinfo {author} {\bibfnamefont {A.~H.}\ \bibnamefont {{Nitz}}},
  \bibinfo {author} {\bibfnamefont {D.~A.}\ \bibnamefont {{Brown}}}, \bibinfo
  {author} {\bibfnamefont {A.~P.}\ \bibnamefont {{Lundgren}}}, \bibinfo
  {author} {\bibfnamefont {E.}~\bibnamefont {{Ochsner}}}, \ and\ \bibinfo
  {author} {\bibfnamefont {D.}~\bibnamefont {{Keppel}}},\ }\href {\doibase
  10.1103/PhysRevD.89.024010} {\bibfield  {journal} {\bibinfo  {journal}
  {\prd}\ }\textbf {\bibinfo {volume} {89}},\ \bibinfo {eid} {024010} (\bibinfo
  {year} {2014})},\ \Eprint {http://arxiv.org/abs/1307.3562} {arXiv:1307.3562
  [gr-qc]} \BibitemShut {NoStop}%
\bibitem [{\citenamefont {Apostolatos}\ \emph {et~al.}(1994)\citenamefont
  {Apostolatos}, \citenamefont {Cutler}, \citenamefont {Sussman},\ and\
  \citenamefont {Thorne}}]{ACST}%
  \BibitemOpen
  \bibfield  {author} {\bibinfo {author} {\bibfnamefont {T.~A.}\ \bibnamefont
  {Apostolatos}}, \bibinfo {author} {\bibfnamefont {C.}~\bibnamefont {Cutler}},
  \bibinfo {author} {\bibfnamefont {G.~J.}\ \bibnamefont {Sussman}}, \ and\
  \bibinfo {author} {\bibfnamefont {K.~S.}\ \bibnamefont {Thorne}},\ }\href
  {\doibase 10.1103/PhysRevD.49.6274} {\bibfield  {journal} {\bibinfo
  {journal} {Phys. Rev. D}\ }\textbf {\bibinfo {volume} {49}},\ \bibinfo
  {pages} {6274} (\bibinfo {year} {1994})}\BibitemShut {NoStop}%
\bibitem [{\citenamefont {Babak}\ \emph
  {et~al.}(2017{\natexlab{b}})\citenamefont {Babak}, \citenamefont
  {Taracchini},\ and\ \citenamefont {Buonanno}}]{BTB2017}%
  \BibitemOpen
  \bibfield  {author} {\bibinfo {author} {\bibfnamefont {S.}~\bibnamefont
  {Babak}}, \bibinfo {author} {\bibfnamefont {A.}~\bibnamefont {Taracchini}}, \
  and\ \bibinfo {author} {\bibfnamefont {A.}~\bibnamefont {Buonanno}},\ }\href
  {\doibase 10.1103/PhysRevD.95.024010} {\bibfield  {journal} {\bibinfo
  {journal} {Phys. Rev. D}\ }\textbf {\bibinfo {volume} {95}},\ \bibinfo
  {pages} {024010} (\bibinfo {year} {2017}{\natexlab{b}})}\BibitemShut
  {NoStop}%
\end{thebibliography}%

\end{document}